\documentstyle[12pt]{article}
\advance \textheight \topskip
\newcommand{\beq}{\begin{equation}}
\newcommand{\eeq}{\end{equation}}
\newcommand{\bea}{\begin{eqnarray}}
\newcommand{\ena}{\end{eqnarray}}

\renewcommand{\b}{\beta}

\newcommand{\SUTH}{SU(3)$_{\mathrm{H}}$}

\begin{document}

\begin{flushright}
{\it Preprint DPFD/94/TH/46}
\end{flushright}
\vspace{12pt}
\centerline{\Large {\bf Non-Abelian Gauge Family Symmetry}}
\centerline{\Large {\bf in Rank 8 and 16 Grand Unified String Theories }}

\vspace{24pt}
\begin{center}
{\Large {\bf  A. Maslikov $^{a}$ , \\ I. Naumov $^{a}$,
  G. Volkov $^{a,b}$ }}\\
\bigskip
 $^a$ Institute for  High Energy Physics\\
142284 Protvino, Moscow Region, Russia\\
\bigskip
$^b$ INFN Sezione di Padova \\
and Dipartimento di Fisica Universit\`a di Padova \\
 Via Marzolo 8, 35100 Padua, Italy \\

\end{center}
\bigskip
\centerline{ABSTRACT}

The  one of the main points of the investigations in
high energy physics is to study
the next chain:  a law
of the quark and lepton mass spectra
$\rightarrow $ the puzzles of the quark and lepton family mixing
$\rightarrow $ a possible new family dynamics.

The new family symmetry dynamics might be connected to the
existence of some exotic gauge or matter fields or something yet.
 For this it will better to study the possibilities of the appearence
this gauge symmetry in the framework of the
Grand Unified String Theories.
In the framework of four dimensional heterotic superstring with free
fermions
we investigate the rank eight Grand Unified String Theories (GUST)
which contain
the $SU(3)_H$-gauge family symmetry.
 We explicitly construct GUST with gauge symmetry $G = SU(5)
\times U(1)\times (SU(3) \times U(1))_H$ and
 $G = SO(10)\times (SU(3) \times U(1))_H$ $\subset SO(16)$
or $E(6)\times SU(3)_H$ $\subset E(8)$
in free complex
fermion formulation. As the GUSTs originating from Kac-Moody algebras (KMA)
contain only low-dimensional representations it is usually difficult
to break
the gauge symmetry. We solve this problem taking for the observable gauge
symmetry the diagonal subgroup $G^{sym}$ of rank $16$ group
$G\times G\subset
SO(16)\times SO(16)$ or $(E(6) \times SU(3)_H)^2  $
$\subset E(8)\times E(8)$. We discuss the possible
fermion matter and Higgs sectors in these models.
In these GUST there has to exist "superweak" light chiral matter($m_H^f
< M_W$).
The understanding of quark and lepton mass spectra and family mixing leave
a possibility for the existence of an unusually low mass breaking scale of
the $SU(3)_H$ family gauge symmetry (some TeV).

\newpage

\tableofcontents

\newpage

\section{Theoretical trends beyond the SM}
\subsection{The family mixing state in Standard Model}
There are no experimental indications which would impel one to go
beyond the framework of the $SU(3)_C\times SU(2)_L\times U(1)_Y$ Standard
Model
$(SM)$ with three generations of quarks and leptons. None of the up-to-date
experiments contradict, within the limits of accuracy, the validity of the SM
predictions for low energy phenomena. The fermion mass origin and generation
mixing, CP-violation problems are among most exciting theoretical puzzles
in SM.

One has ten parameters in the quark sector of the SM with three generations:
six quark masses, three mixing angles and
the Kobayashi-Maskawa $(KM)$ CP- violation phase $(0<\delta^{KM}<\pi)$.
The CKM matrix in Wolfenstein parametrization is determined by the four
parameters- Cabibbo angle $\lambda \approx 0.22$, A, $\rho$ and $\eta$:

\begin{equation}
{\bf V_{CKM}}=
\left( \begin{array}{ccc}
V_{ud}& V_{us} & V_{ub}\\
V_{cd} & V_{cs}& V_{cb} \\
V_{td}  &V_{ts} & V_{tb} \\
\end{array} \right) =
\left( \begin{array}{ccc}
1- 1/2 {\lambda}^2 & \lambda & A {\lambda}^3 (\rho - i \eta)\\
- \lambda & 1 - 1/2{\lambda}^2 & A {\lambda}^3 \\
A {\lambda}^3 (1 - \rho - i \eta)  & -A {\lambda}^2  & 1 \\
\end{array} \right).
\end{equation}

In the complex plane the point $(\rho, \eta)$
is  a vertex of the  unitarity triangle and describes the CP- violation
in SM. The unitarity triange is constructed from
the following unitarity condition of $V_{CKM}$ :
$V_{ub}^{*} + V_{cd} \approx A {\lambda}^3$.

Recently, the interest in the CP-violation problem was excited again due
to the
data on the search for the direct CP-violation effects in neutral K-mesons
\cite{1',2'} :
\begin{equation}
Re \biggl (\frac{\varepsilon'}{\varepsilon}\biggr )
=(7.4\pm 6)\times10^{-4},\,\,\,\,\,\,
\end{equation}

\begin{equation}
Re \biggl ( \frac{\varepsilon'}{\varepsilon}\biggr )
=(23\pm 7)\times10^{-4},\,\,\,\,\,\,
\end{equation}

The major contribution to the CP-violation parameters $\varepsilon_K$ and
$\varepsilon_K'$ ($K^0$-decays), as well as to the $B_d^0-\bar B_d^0$
mixing parameter $x_d=\frac{\Delta m_{(B_d)}}{\Gamma_{(B_d)}}$  is due
to the large t-quark mass contribution. The same statement holds also
for some amplitudes of K- and B-meson rare decays.
 The CDF collaboration gives the following region for the top quark
mass: $m_t=174\pm 25$ GeV \cite{3'}. The complete fit which is based on
the low energy data as well as the latest LEP and SLC data and comparing
with the mass indicated by CDF measurements gives $m_t=162\pm9$ GeV
\cite{4'}.

The combined analysis of
the data on the $B_d^0-\bar B_d^0$ mixing parameter
$x_d=0.67\pm0.10$ \cite{5'} and $\varepsilon_K=(2.26\pm0.02)\times10^{-3}
exp(i 43.7^o)$ indicated that the top quark mass should lie in the range
of $(135\pm35)$GeV, although   a very massive top quark  in the range
of $m_t\sim 200 GeV$ is not excluded, either \cite{6'}.
Note that for $m_t\approx O(100 GeV)$ the SM predicted $\biggl (\frac
{\varepsilon'}{\varepsilon}\biggr)_K\approx(1.0\pm 0.5)\times 10^{-3}$,
and the value of $\delta_{13}=arg V_{ub}^* $  ($V_{ub}
=s_{13} \times  exp(-\delta_{13})$) in the second quadrant
$(\delta_{13}>\frac{\pi} {2})$ of the
unitary triangle is favored.
For $m_t\sim 200 GeV$ one gets the superweak- like
behaviour predicted in the SM, i.e.
$\biggl( \frac{\varepsilon'}{\varepsilon}
\biggr)$ is close to zero. In this case the value of $\delta_{13}$
is likely to  be in the first quadrant $(0<\delta_{13}<\frac{\pi}{2})$.

It is worthwhile to note that the above conclusions depend strongly on the
values
of the hadronic matrix elements $<Q_6>$,$f_K$,$f_B$, as well as on the KM
mixing angles $s_{ij}$.
For example,the determination of $V_{cb}=s_{23}=0.046\pm 0.006$
from the $\Gamma(b\longrightarrow c)$ decay rate and the determination of
the ratio $q=\left|{V_{ub}}\over {V_{cb}}\right|$
 by ARGUS and CLEO lead to the model dependent results

\begin{equation}
0.07 \leq{\Biggl ({q={{s_{13}}\over{s_{23}}}}
=\, \lambda \,\sqrt{\rho^2 \,+ \,\eta^2} \Biggr )} \leq 0.12, \label{1.3}
\end{equation}
where the parameter $A \approx 0.8- 0.9$ is defined from the
dominant decays of B- hadrons ($V_{cb}= A {\lambda}^2$).

 On the other hand, the unitarity triangle \cite{7'} constraint yields for

$|V_{td}|$, that $|V_{td}|
=|s_{23}\bigl (\lambda - q\times exp(i\delta_{13})\bigr )|
=(0.0035-0.0200)$, ($\lambda=\sin {\theta}_C$). The value of  $V_{td}$
depends crucially on the
$\delta_{13}$ phase. For the small values of this matrix element,
$\delta_{13}$ tends to be in the first quadrant, whereas the experimental
values for $\varepsilon_K, {{\varepsilon_K'}\over
{\varepsilon_K}} , x_d$ favor  larger $m_t$ values.

Another interesting possibility to check the sign of  $\cos{\delta_{13}}$
comes from the experimental observation of the $B_s^0-\bar B_s^0$ mixing.
Due to the following relation between the parameters of the
$B_d^0-\bar B_d^0$ and $B_s^0-\bar B_s^0$ mixings:

\begin{equation}
\frac{x_s}{x_d}=\frac{|V_{ts}|^2}{|V_{td}|^2}=\frac{1} {s_{12}^2
-2qs_{12}cos\delta_{13}+q^2}
=\frac{1}{{\lambda^2} \bigl [(1-\rho )^2 \,+ \,{\eta^2} \bigr] } \label{1.4}
\end{equation}

for $\delta_{13}\approx \pi$  we get ${{x_s}\over {x_d}}=10-14$ and
 $x_s=7-11$, whereas for $\delta_{13}\approx 0$ we get
${{x_s}\over {x_d}}=36-100$ and $x_s=27-70$ .

The case of the symmetric form for  the CKM-matrix \cite{8'}, which
leads to $V_{td}=V_{ub}$, $V_{ts}=V_{cb}$ and $V_{td}
={{1}\over {2}}V_{us}V_{cb}$= ${{1}\over{2}}V_{cd}V_{ts}$
(q $\simeq$ 0.1 in this case),
corresponds to  $\delta_{13}$-phase $\simeq 0$,
${x_s}/{x_d} \simeq 100$ and $x_s \simeq 70$.

As seen from the above discussion one can see that
in the quark sector of the SM
there are several experimental quantities, which are sensitive
to the values of parameters $m_t$, $\delta^{KM}$, $B_d$,...
Therefore, we need  additional experimental information to prove
the validity of the SM with three generations and to get convinced
that there are
no additional contributions to the amplitudes of flavour changing
rare processes due to new hypothetical forces beyond the SM.

\subsection{ Quark and lepton mass origin - mass ansatzes and quark mixing}

The main undrawbacks of SM now are going from our non-understanding
the generation problem, their mixing and hierarchy
of quark and lepton mass spectra. For example, for quark
masses $\mu \approx 1 GeV$
we can get approximately the following relations \cite{9'}:

\begin{eqnarray}
m_{i_k} \approx ( q_H^u )^{2k} m_0,
\,\,\,\,k = 0, 1, 2;
\,\,\, i_0= u, \, i_1 = c, \, i_2 = t,
\nonumber\\
m_{i_k} \approx ( q_H^d )^{2k} m_0,
\,\,\,\,k = 0, 1, 2;
  \,\,\, i_0= d, \, i_1 = s, \, i_2 = b,
\end{eqnarray}

where  $q_H^u \approx (q_H^d)^2$,
$q_H^d \approx 4-5 \approx 1/{\lambda}$ and $\lambda \approx
\sin {\theta}_C$.

Here we used the conventional ratios  of  the  "running"  quark
masses \cite{10'}

\begin{eqnarray}
m_d/m_s\ =0.051\pm 0.004 , \,\,\,  m_u/m_c\ =0.0038\pm 0.0012 , \,\,\,
m_s/m_b\ =0.033\pm 0.011 ,  \nonumber\\
m_c(\mu =1GeV)= (1.35\pm 0.05 ) GeV \,\,\,
and  \,\,\, m_t^{phys} \approx 0.6m_t(\mu = 1GeV) \label{5.1}.
\end{eqnarray}

This phenomenological formula (6) predicts the following value for the
$t$-quark mass:
\begin{equation}
{m_t}^{phys} \approx 180 - 200 GeV.
\end{equation}

In SM these mass matrices and mixing come from the
Yukawa sector :

\begin{eqnarray}
L_Y= Q  Y_u {\bar {q_u}}{h^*} + Q Y_d {\bar {q_d}} h + L Y_e {\bar {l_e}} h
\,\, + \,\, H.C.,
\end{eqnarray}
where $Q_i$ and $L_i$ are three quark and lepton isodoublets,
  ${q_u}_i,{q_d}_i$ and $e_i$ are three right-handed antiquark and antilepton
isosinglets, respectively. $h$ is the ordinary Higgs doublet.
In SM the $3\times 3$- family Yukawa matrices, $(Y_u)_{ij}$ and $(Y_d)_{ij}$
have no any particular symmetry.
Therefore it is necessary to reach some additional mechanisms or symmetries
beyond the SM which
could diminish the  number of the independent parameters in Yukawa
sector $L_Y$. These new structures can be
used for the determination of the mass hierarhy and family mixing.

To understand the generation mixing origin and fermion mass hierarchy
several models beyond the SM suggest special forms for the mass matrix of
"up" and "down" quarks (Fritzsch ansatz, "improved" Fritzsch ansatz,
"Democratic" ansatz, etc.\cite{11'}).
These mass matrices have less than ten independent parameters
or they could
have some matrix elements equal to zero ("texture zeroes")\cite{12'}.
This allows us
to determine the diagonalizing matrix $U_L$ and $D_L$ in terms of quark
masses:
\begin{equation}
Y_d^{diag} = D_L Y_d D_R^+,\,\\,\,\,\,\,\,
Y_u^{diag} = U_L Y_d U_R^+.
\end{equation}
For simplicity it has been taken the symmetric form of Yukawa matrices,
therefore:  $D_L=D_R^*$, $U_L=U_R^*$.
 These ansatzes or zero "textures" could be checked experimentally in
predictions for the mixing angles of the CKM matrix:$V_{CKM} = U_L D_L^+$.
For example, it can be considered the next approximate form
at $M_X$ for the symmetric "texture", using in paper \cite{12'}:

\begin{equation}
{\bf Y_u}=
\left( \begin{array}{ccc}
0 & {\lambda}^6 & 0 \\
 {\lambda}^6  & 0 &  {\lambda}^2 \\
0 &  {\lambda}^2 & 1 \\
\end{array} \right)
\,\,\,\,\,\,
{\bf Y_d}=
\left( \begin{array}{ccc}
0 & 2{\lambda}^4 & 0 \\
 2{\lambda}^4  &  2{\lambda}^3 &  2{\lambda}^3 \\
0 &  2{\lambda}^3 &  \\
\end{array} \right).
\end{equation}

Given these conditions it is possible evolve down to low energies via
the renormalizationn group equations all quantities  including the matrix
elements of Yukawa couplings $Y_{u,d}$, the values of the quark masses
and the CKM matrix elements \cite{12'}.
Also, using these relations we may compute $U_L$ (or $D_L$) in terms of CKM
matrix and/or of quark masses.

In GUT extensions of the SM with embedding the family gauge  symmetry
Yukawa matrices can acquire particular symmetry or an ansatz,
depending on the Higgs multiplets to which they couple.
The  family gauge symmetry could help us to study
by independent way  the origin of the up- ($U$) and down-
($D$) quark mixing matrices  and consequently  the structure of the
CKM matrix $V_{CKM}=UD^+$.
The possibility  a  low energy breaking scale  gives us a chance
due to  the local gauge family symmetry to define the quantum numbers of
quarks and leptons and thus establishes a link between them in families.
For considering an mass fermion ansatz in the extensions of
SM there could exist
the following types of the  $SU(3)\times SU(2_L)$ Higgs multiplets:(1,2),
(3,1), (8,1),(3,2),(8,2),(1,1),...., which could exist in spectra of the
String Models.

In the framework of the rank eight Grand Unified String
Theories we will consider an extension of SM due to local
family gauge symmetry,
 $G_H =SU(3)_H$, $SU(3)_H \times U(1)_H$ models and its developments and
the possible Higgs sector in them.
Thus,  for  understanding the quark mass spectra and
 the difference between  the origins of the up- ( or down) quark
and charged lepton  mass matrices in GUSTs we have to study  the
Higgs content of the model, which we must use from one side for breaking the
GUT- ,Quark-Lepton -,  $G_H = SU(3)_H ,...$-, $SU(2)_L \times U(1)$-
symmetries and from another side- for Yukawa matrix constructions.
The vital question arising here is
the nature of the $\nu$ mass.

\subsection{The possible ways of  E(8)-  Grand Unified
String Theories leading to the $N_G=3$ or
$N_G=3+1$ families}

For a couple of years superstring theories, and particularly the heterotic
string theory, have provided an efficient way to construct the Grand Unified
Superstring Theories ($GUST$) of all known interactions, despite the
fact that it is
still difficult to construct unique and fully realistic low energy models
resulting after decoupling of massive string modes. This is because
it is only
in 10-dimensional space-time  that there exist just two consistent (invariant
under reparametrization, superconformal, modular, Lorentz and SUSY
transformations) theories with the gauge symmetries $E(8)\times E(8)$ or
${spin(32)}/{Z_2}$ \cite{13',14'} which after compactification of
the six extra
space coordinates (into the Calabi-Yau \cite{15',16'} manifolds or into the
orbifolds) can be used for constructing GUSTs. Unfortunately, the process of
compactification to four dimensions is not unique and the number of possible
low energy models is very large. On the other hand, starting the construction
of the theory directly in $4$-dimensional space-time requires including a
considerable number of free bosons or fermions into the internal string sector
of the heterotic superstring \cite{17',18',19',20'}.
This leads to as large internal
symmetry group such as e.g. rank  $22$ group.
The way of breaking this primordial symmetry
is again not unique and leads to a huge number of possible models,
each of them
giving different low energy predictions.

On the other side, because of the presence of the affine
Kac-Moody algebra (KMA)
$\hat g$ (which is a 2-dimensional manifestation of gauge symmetries of the
string itself) on the world sheet, string constructions yield definite
predictions as to what representation of the symmetry group can be used for
low energy models building \cite{21',22'}.
Therefore the following long-standing questions
have a chance to be answered in this kind of unification schemes:

\begin{enumerate}
\item How are the chiral matter fermions assigned to the multiplets of the
unifying group?
\item How is the GUT gauge symmetry breaking realized?
\item What is the origin and the form of the fermion mass matrices?
\end{enumerate}

The first of these problems is, of course, closely connected with the
quantization of the electromagnetic charge of matter fields. In addition,
string constructions can shed some light on the questions about the number of
generation and possible existence of mirror fermions which remain unanswered
in conventional GUTs \cite{23'}.

There are not so many GUSTs describing the observable sector of String
Models. It is well known the SM gauge group, the Pati- Salam
($SU(4)\times SU(2)\times SU(2)$) gauge group, the flipped SU(5) gauge group
and SO(10) gauge group, which includes flipped SU(5) \cite{20'}.

There are good physical reasons for including the horizontal $SU(3)_H$ group
into the unification scheme. Firstly, this group naturally accommodates three
fermion families presently observed (explaining their origin) and, secondly,
can provide correct and economical description of the fermion mass spectrum
and mixing without invoking high dimensional representation of conventional
$SU(5)$, $SO(10)$ or $E(6)$ gauge groups. Construction of a string model
(GUST)
containing the horizontal gauge symmetry provides additional, strong
motivation
to this idea. Moreover, the fact that in GUSTs high dimensional
representations
are forbidden by the KMA is a very welcome feature in this context.

All this leads us naturally to consider possible forms for horizontal
symmetry
$G_H$, and $G_H$ quantum number assignments for quarks (anti-quarks) and
leptons (anti-leptons) which can be realized within GUST's framework.
To include the
horizontal interactions with three known generations in the ordinary
GUST it is
natural to consider rank eight gauge symmetry.
 We can consider
$SO(16)$ (or $E(6) \times SU(3)$) which is the maximal subgroup of $E(8)$
and which contains the rank eight
subgroup $SO(10)\times (U(1)\times SU(3))_H$ \cite{24'}. We will be,
therefore,
concerned with
the following chains (see Fig. \ref{fig1}):
\vskip 0.3cm
\small
$$
\begin{array}{cccc}
\:\:E(8) \longrightarrow SO(16) \longrightarrow
\underline {SO(10) \times (U(1) \times SU(3))_H} \longrightarrow  \\
\longrightarrow SU(5) \times U(1)_{Y_5}
\times (SU(3) \times U(1))_H  \\
\end{array}
$$
 \vskip 0.3cm
\small
or
$$
\begin{array}{cccc}
\:\:E(8) \longrightarrow E(6) \times SU(3) \longrightarrow
 (SU(3))^{ \times 4}.  \\
\end{array}
$$
 \vskip 0.3cm
\begin{figure}
\caption{The possible ways of E(8) gauge symmetry breaking leading to the
3+1 or 3 generations.}
\label{fig1}
\setlength{\unitlength}{1mm}
\begin{picture}(165,135)(10,0)
\setlength{\unitlength}{0.8mm}
{\tiny
\put(20,140){{\large E(8) } }
\put(110,140){{\large SO(16) }}
\put(6,80){{\large $ E(6)\times SU(3)_H $ }}
\put(85,80){{\large $ SO(10)\times SU(3)_H\times U(1)_H $ }}
\put(15,20){{\large $ SU(3)^{\otimes 4} $ }}
\put(73,20){{\large $ SU(5)\times U(1)\times SU(3)_H\times U(1)_H $ }}
\put(73,10){{\large $N_g = 3 $ , $N_g = 3 + 1 $ }}

\put(23,130){\vector(0,-1){33}}
\put(115,130){\vector(0,-1){33}}
\put(50,142){\vector(1,0){30}}
\put(23,70){\vector(0,-1){33}}
\put(115,70){\vector(0,-1){33}}
\put(50,82){\vector(1,0){30}}

\put(5,105){\shortstack[l]{ {$ 248 \longrightarrow $} \\ {$(78,1)\oplus $} \\
 {$ (1,8)\oplus $} \\ {$ (27,3)\oplus $} \\ {$ (\bar{27},\bar 3) $}  }}

\put(5,30){\shortstack[l]{ {$ 78 \longrightarrow $} \\ {$(8,1,1)\oplus $} \\
      {$ (1,8,1)\oplus $} \\ {$ (1,1,8)\oplus $} \\
          {$ (3,3,3)\oplus $} \\{$ (\bar 3,\bar 3,\bar 3) $} \\ \\
   {$ 27 \longrightarrow $} \\ {$(3,\bar 3,1)\oplus $} \\
      {$ (1,3,\bar 3)\oplus $} \\ {$ (\bar 3,1,3) $}    }}

\put(52,147){\shortstack[l]{ {$  248 \longrightarrow 120 \oplus 128 $} }}

\put(120,100){\shortstack[l]{ {$ 120 \longrightarrow (45,1)^0 \oplus (1,8)^0
 \oplus $} \\ {$ (1,1)^0 \oplus  (10,3)^2 \oplus (10,\bar 3)^{-2} \oplus $} \\
              {$ (1,3)^{-4} \oplus (1,3)^{+4} $} \\
 {$ 128 \longrightarrow (16,3)^{-1} \oplus (\bar{16},\bar 3)^{+1} \oplus $} \\
   {$ (16,1)^{+3} \oplus  (\bar{16},1)^{-3} $}    }}

\put(30,90){\shortstack[l]{
 {$ (78,1) \longrightarrow (45,1)^0 \oplus (1,1)^0
 \oplus  (16,1)^{+3} \oplus (\bar{16},1)^{-3} $} \\
 {$ (27,3) \longrightarrow (16,3)^{-1} \oplus (10,3)^{+2}
\oplus (1,3)^{-4} $} \\
 {$ (\bar{27},\bar{3}) \longrightarrow (\bar{16},\bar 3)^{+1}
\oplus (10,\bar 3)^{-2}
            \oplus (1,\bar 3)^{+4} $} }}

\put(120,50){\shortstack[l]{
 {$ 45 \longrightarrow (24,1) \oplus (1,1) \oplus  (10,1)
                \oplus (\bar{10},1) $} \\
 {$ 16 \longrightarrow (1)_{+5/2} \oplus (\bar 5)_{-3/2}
                        \oplus (10)_{+1/2} $} \\
 {$ \bar{16} \longrightarrow (\bar 1)_{-5/2} \oplus (5)_{+3/2}
                        \oplus (\bar{10})_{-1/2} $} }}

}
\end{picture}

\end{figure}
\normalsize

According to this scheme one can get $SU(3)_H\times U(1)_H$ gauge family
symmetry with $N_g = 3 + 1 $ (there are also other possibilities as eg.
 $E(6)\times SU(3)_H\subset E(8)$
 $N_g = 3 $ generations can be obtained due to the second way of $E(8)$
gauge symmetry breaking via $E(6)\times SU(3)_H$, see Fig.\ref{fig1}), where
the possible,
additional, fourth massive matter superfield could appear from
$\underline {78}$ as a singlet of
$SU(3)_H$ and transforms as $\underline {16}$ under the $SO(10)$ group.

In this note  starting from the rank 16 grand unified gauge group (which is
the minimal rank allowed in strings \cite{25',26'}) of the form $G\times G$
and making use of the KMA which select the possible gauge group
representations we construct the string model based on the diagonal subgroup
$G^{symm}\subset G\times G\subset SO(16)\times SO(16) (\subset E_8\times E_8)$
\cite{25'}.
We discuss and consider $G^{symm}=SU(5)\times U(1)\times (SU(3)\times U(1))_H$
$\subset SO(16)$ where the factor $(SU(3)\times U(1))_H$ is interpreted as the
horizontal gauge family symmetry. We explain how the unifying gauge symmetry
can be broken down to the Standard Model group. Furthermore, the horizontal
interaction predicted in our model can give an alternative description of the
fermion mass matrices without invoking high dimensional Higgs representations.
In contrast with other GUST constructions,
our model does not contain particles with exotic fractional electric charges
\cite{27',25'}.
This important virtue of the model is due to the symmetric construction
of the electromagnetic charge  $Q_{em}$ from $ Q^I$ and $Q^{II}$ -- the
two electric charges of each of the $U(5)$ groups \cite{25'}:

\begin{eqnarray}\label{eq102}
Q_{em} &=& Q^{II} \oplus  Q^I.
\end{eqnarray}

We consider the possible forms   of the
$G_H= SU(3)_H$ ,$SU(3)_H \times U(1)$, $G_{HL} \times G_{HR}$... - gauge family
symmetries  in the framework of Grand Unification Superstring Approach.
Also we will study the
 matter spectrum of these GUST, the possible Higgs sectors.
The form of the Higgs sector it is very important for GUST- , $G_H$- and SM
- gauge symmetries breaking
and for constructing Yukawa couplings.

\subsection{ Towards a low energy gauge family symmetry
"exactly solvable". ("Bootstrap" models.)}
The underlying analysis for  this family symmetry breaking scale
is lying on the modern experimental probability limitations
for the typical rare flavour- changing processes.  The
estimates for the family  symmetry breaking scale
have certain regularities  depending on the particular symmetry breaking
schemes and  generation mixing mechanisms (different anzatzes
for quark and lepton mass matrices with $3_H$ or $3_H + 1_H$
generations have been discussed). As  noted, the current understanding of
quark and lepton mass spectra leaves room for the existence of
an unusually  low mass breaking scale of non-abelian gauge
$SU(3)_H$ or $(SU(3) \otimes U(1))_H$
family symmetry  $\sim some\ TeV$.
Some independent experiments
for verifying the relevant hypotheses can  been considered:
light ($\pi $, K) ,
heavy (B, D) - meson and charged lepton flavour changing rare decays
\cite{28',29',30',9'},
family symmetry violation effects in $e^+ e^-$- and $ p p$ -
collider experiments (LEP, FNAL, LHC).

The introductions into the model the
Higgs fields which are
transformed
under the $SU(3)_H \times SU(2)_L$
symmetry, like as $H^a= (\underline 8,\underline 1)$
( or $ H^a_p =(\underline 8, \underline 2)$, p=1,2) and
$X^i = (\underline 3,\underline 1)$
 (or  $X^i_p = (\underline 3,\underline 2)$, p=1,2)
give the next contribution
to the family gauge boson mass matrix:
\begin{eqnarray}
(M_H^2)^{ab}_{\underline 8} =  { g_H^2}\, \sum_{d=1}^{8}\,
f^{adc} f^{bdc'} <H^c><H^{c'}> ,\label{5.6}\\
(M_H^2)^{ab}_{\underline 3} =  { g_H^2} \sum_{k=1}^{3}\,
 \frac{{\lambda}_{ik}^a}{2}\frac{{\lambda}_{kj}^b}{2}
<X^{i}>{<X^{j}>}^* ,\label{5.7}
\end{eqnarray}

The lowest bound on $M_H$ can be obtained from the analysis of the
branching ratios of $\mu$, $\pi$, K, D, B, ... rare decays
(Br$\geq 10^{-15-17}$).

In this paper we will investigate the samples of different
scenarios of $SU(3)_H$- breakings up to
the $SU(2)_H \times U(1)_{3H}$, $U(1)_{3H} \times U(1)_{8H}$ and
$U(1)_{8H}$-
subgroups, as well as the mechanism of the complete breaking of
the base group $SU(3)_H$- \cite{9'}. We will try to realize this
program conserving SUSY
on  the  scales where the relevant gauge symmetry is broken. In
the framework of these  versions of the gauge symmetry breaking, we
will search for the spectra of  horizontal gauge bosons and gauginos and
calculate the  amplitudes of some typical rare processes. Theoretical
estimates for the branching  ratios of some
rare processes obtained from these  calculations will be compared
with the experimental data on the corresponding values. Further on we
will get  some bounds on the masses of $H_{\mu}$-bosons and the appropriate
$H$-gauginos. Of particular interest is  the case of the
$SU(3)_H$ -group which   breaks completely
on  the scale $M_{H_0}$. We calculate the
splitting of eight $H$-boson masses in a model dependent  fashion.
This    splitting, depending on the quark mass spectrum, allows us
to reduce considerably  the predictive ambiguity of the  model
-"almost exactly solvable model".

   We assume that when the $SU(3)_H -$gauge symmetry of quark- lepton
generations is violated,  all the  8 gauge bosons acquire in the
eigenspectrum of horizontal interactions the same mass equal to
$M_{H_0}$. The such breaking is not difficult to get by, say, introducing
the Higgs fields transforming in accordance with the triplet
representation of the $SU(3)_H$  group. These fields are singlet
under the  Standard Model symmetries : $(z \in (3,1,1,0)$ and
${\bar z} \in (\bar 3,1,1,0) $ ,
$ <{\bar z}^{i\,\alpha}>_0=\delta^{i\,\alpha}V$ ,
$<z_i^{\alpha}>_0=\delta_i^{\alpha}V\ ,\ $,$ i,\, \alpha\, =1,2,3 $,
where $V=M_{H_0})$.  We understand that here we need in more
beautifull way to break this symmetry like by dynamical way.
But at this stage it is very important now to eastablish a link
between the spectra masses of the horizontal gauge bosons and of
till  known now  the matter fermion heavy particles like t- quark.
The degeneracy in the
masses of 8 gauge horizontal vector bosons is eliminated by using the
VEV's of the Higgs fields violating the  electroweak
symmetry and determining the mass matrix of  up- and  down-  quarks
(leptons). Thus, in the set  of  the  Higgs  fields  (see
Table 11), with $H(8,\ 2)\ ,h(8,\ 2)\ ,\ Y (\bar 3 ,\ 2),
\ X(3,\ 2)\ ,\  {\kappa}_{1,2}(1,\ 2)$  violates the
$SU(2)\times U(1)$ symmetry and determines  the
mass matrix of up-and down-quarks. On the other hand, in order to  calculate
the splitting between the masses of horizontal gauge  bosons,  one
has to take into account the VEV's of  these two sets  of
the Higgs fields.

Now we can come to constructing the    horizontal
gauge boson mass  matrix  $M_{ab}^2$ \\
(a,b=1,2,...,8):
\begin{eqnarray}
(M_H^2)_{ab}=M_{H_0}^2\delta_{ab} +({\Delta}M_d^2)_{ab} +({\Delta}M_u^2)_{ab}.
\label{5.8}
\end{eqnarray}
Here $({\Delta}M_d^2)_{ab}$ and $({\Delta}M_u^2)_{ab}$ are the "known"
functions of heavy fermions,
$({\Delta}M_{u,d}^2)_{ab} = F_{ab}(m_t, m-b,...)$ , which, mainly, get the
contributions due to the vacuum expectations of   the Higgs  bosons
that  were  used  for  construction of  the mass matrix ansatzes for
d- (u-) quarks.

For example, for the case $N_g=\underline 3 + \underline 1$ families
with Fritzch ansatz for quark mass matrices and using
$SU(3_H) \times SU(2)$ Higgs fields,
$(8,2)$, \cite{9'},
  we can  write down some rough equalities
between the masses of horizontal gauge bosons:

\begin{eqnarray}
M_{H_1}^2 &\approx& M_{H_2}^2 \approx M_{H_3}^2 \approx M_{H_0}^2 +
\frac{g_{H}^2}{4} \Bigl[\frac{1}{ \lambda^2} {{m_c m_t}
\over {1 -{m_t}/{m_{t'}}}} \Bigr]+
 ...  ,\nonumber \\
M_{H_4}^2 &\approx&  M_{H_5}^2 \approx M_{H_6}^2 \approx M_{H_7}^2
\approx M_{H_0}^2 +
\frac{g_{H}^2}{4} \Bigl[\frac{1}{\tilde {\lambda}^2} m_t m_{t'}\Bigr] +...  ,
\nonumber\\
 M_{H_8}^2
&\approx& M_{H_0}^2 +
\frac{g_{H}^2}{3} \Bigl[\frac{1}{\tilde {\lambda}^2} m_t m_{t'}\Bigr]
+...,
\end{eqnarray}
where $\lambda$ and $\tilde {\lambda}$ are Yukawa couplings.

We are interested in the dependence of the unitary compensation
for the contributions of horizontal forces  to rare processes \cite{9'}
on  different versions of the $SU(3)_H$- symmetry breaking. The investigation
of this dependence allows, first, to understand how low the horizontal
symmetry  breaking scale $M_H$ may be, and, second, how this scale is
determined by a particular  choice of a  mass matrix anzatz both for quarks
and leptons.

 We would like to consider of a possible existing
of the local family symmetry with a low energy symmetry breaking scale,
i.e. the existence of rather light H-bosons:
$m_H\geq (1-10)TeV$ \cite{9'}.
 We will analyze, in the framework of the
"minimal" horizontal
supersymmetric gauge model, the possibilities to obtain
a satisfactory
hierarchy for  quark masses and to connect it with the splitting of
horizontal gauge boson masses. We expect that due to  this approach
the horizontal model will become more definite since it will
allow to study the amplitudes of rare processes and
the CP-violation mechanism more thoroughly.
In this way  we hope to get a deeper insight into  the nature of
interdependence between  the generation mixing mechanism and
the local horizontal symmetry breaking scale.

\subsection{Estimates on the horizontal coupling constant
and the scale of unification.}

Really, the estimates on the $M_{H_0}$- scale depend on the
value of the family gauge coupling.
These estimates can be maken in GUST using the string scale
\begin{eqnarray}
M_U \approx 0,73 g_{string} \times {10}^{18} GeV
\end{eqnarray}
and the renormalization group equations (RGE) for the gauge couplings,
${\alpha}_{em}$, ${\alpha}_{3}$, ${\alpha}_{2}$, to the low energies :
${\alpha}_{em}(M_Z) \approx 1/128$,
${\alpha}_{3}(M_Z) \approx 0,11$,
${\sin}^2{\theta}_{W}(M_Z) \approx 0.233$.
The string unification scale could be contrasted with the
$SU(3^c)\times SU(2) \times U(1)$ naive unification scale,
$M_U \approx {10}^{16} GeV$, obtained by running the SM particles and their
SUSY-partners to high energies. The simplest solution to this problem
is the introduction in the spectrum of new heavy particles with SM quantum
numbers, which can be exist in string spectra  \cite{20'}d.

However there are some other ways to explain the difference between
scales of string ($M_{SU}$) and ordinary ($M_{U}$) unifications.
Thus if one uses the breaking scheme $G\times G\,\rightarrow G^{sym}$
( where $G=U(5)\times U(3)_H \subset E_8$ ) described above, then
unification scale $M_U \sim 10^{16}\,$GeV is the scale of breaking
the $G\times G$ group, and string unification do supply the equality
of coupling constant $G\times G$ on the string scale
$M_{SU}\sim 10^{18}\,$GeV. In addition if there is a symmetry
between representations of two groups $G$ then
$$g_{sym}(M_U)=\frac{1}{\sqrt{2}}\,g_G(M_U),$$
but in absence of symmetry the relation is more complicated.
Thus in this scheme knowing of scales $M_{SU}$ and $M_U$ gives
us a principal possibility to trace the evolution of coupling
constant of the original  group $G\times G$  to the low energy
and estimate the value of horizontal gauge constant $g_{3H}$.

The coincidence of $\sin^2 \theta_W$ with experiment will
show how realistic this model is. In our scheme, where
$G_{sym}=SU(5)\times U(1)\times U(3)_H$ the following equation holds:
\begin{equation}
\sin^2\theta_W(M_U)=\left.\frac{15\,k^2}{16\,k^2+24}\,\right|_{k^2=1}=
\frac{3}{8}
\label{sinW}
\end{equation}
The relation between $SU(5)_{sym}\times U(1)_{sym}$-constants
$k=g_1/g_5$ on the scale of $M_U$ is defined by set of representations of
$G^I\times G^{II}$ group.
The analysis of RG--equations under the $M_U$ scale allows to
state that horizontal coupling constant $g_{3H}$ does not exceed
electro-weak one $g_2$.

In special case \cite{9'} the difference between the $\alpha_2(\mu)$
and  $\alpha_{3H}(\mu)$
is mainly due to the particular choice of the Higgs fields leading to the
breaking of the electroweak and horizontal symmetries, respectively:

\begin{eqnarray}
\frac{1}{{\alpha}_{3H}(\mu)}\,&=&\,\frac{1}{{\alpha}_2 (\mu)}\,+\,
\frac{b_{3H}\,-\,b_2}{2\pi}\,\ln{\frac{M_X}{\mu}}\,\nonumber\\
&=&\,\frac{1}{{\alpha}_2 (\mu)}\,+\,\frac{-1\,+\,1/2\,n_3\,+\,3\,n_8\,
-\,1/2\,n_2\,}{2\pi}\,\ln{\frac{M_X}{\mu}},
\end{eqnarray}
where $n_2$and $n_3$, $n_8$ denote the nubmer of the Higgs $SU(2)$-
doublet and $SU(3)_H$ triplet and octet, respectively.
Therefore,
 for instance, when one takes into account the fields $\Phi(8_H,1_L)$,
$H(8_H,2_L)$, $h(8_H,2_L)$, the difference between ${\alpha}_{3H}(\mu)$
and ${\alpha}_2 (\mu)$ is expressed by the formula:

\begin{eqnarray}
\frac{1}{{\alpha}_{3H}(\mu)}\,-\,\frac{1}{{\alpha}_2(\mu)}\,
=\,\frac{1}{2\pi} \ln{\frac{M_X}{\mu}}
\end{eqnarray}
 whence ${{\alpha}_{3H}(\mu)}\,\leq \,{{\alpha}_2(\mu)}$.

\subsection{The N=1 SUSY character of the $SU(3_H)$- gauge family symmetry}

We will consider the supersymmetric version of the Standard Model
extended by the family (horizontal) gauge symmetry (and if one will need
we will also extend this model by the $G_R= SU(2)_R$ Right-hand gauge group
). The supersymmetric
Lagrangian of strong, electroweak and horizontal interactions,
based on the $SU(3)_C\times SU(2)_L \times U(1)_Y \times SU(3)_H$...
 (where the $G_R$)-gauge group and the
 Abelian gauge
factor $U(1)_H$ also can be taken into consideration), has the general form:

\begin{eqnarray}
  {\cal L}\ &=& \int d^2\theta \ Tr\ (\ W^k\ W^k\ )  \nonumber\\
  &+&\  \int d^4 \theta \ S_{I}^{+}\ e^{\sum_k 2g_k\hat V_k}\  S_I \nonumber\\
 &+&\  \int d^4 \theta \ Tr(\Phi ^+ \ e^{2g_H\hat V_H}\ \Phi
  \ e^{-2g_H\hat V_H}) \nonumber\\
 &+&\ \int d^4 \theta \ Tr( H^{+}_{y} \ e^{2g_2 \hat V_2+
  y2g_1 \hat V_1}  \ e^{2g_H \hat V_H } \ H_y \ e^{-2g_H \hat V_H })
\nonumber\\
 &+&  ( \int d^2\theta \ P(S_i,\Phi,H_y,\eta,\xi,...) +\ h.c.\ )  \label{2.1}
 \end{eqnarray}

In formula (\ref{2.1}) the index $k$ runs over all the gauge groups: $SU(3)_C ,
SU(2)_L , U(1)_Y$,   \SUTH{} , $\hat V={\bf T}^a V^a$ ,
 where $V^a$ are the real vector superfields ,
and ${\bf T}^a$ are the generators
of the $SU(3)_C,SU(2)_L,U(1)_Y$,\SUTH{} -groups; $S_I$ are left-chiral
superfields from fundamental  representations, and
$I=i,1,2;\ S_i=Q, u^c,d ^c, $ $L, e ^c, \nu ^c$- are matter superfields,
$ S_1=\eta,S_2=\xi$- are Higgs fundamental superfields; the Higgs left chiral
superfield  $\Phi$ is transformed
according to the adjoint representation of the
\SUTH{}-group, the Higgs left chiral superfields
$H_y:\ H_{Y=+\frac{1}{2}}=H,\ H_{Y=-\frac{1}{2}}=h$  are
transformed nontrivially under the horizontal \SUTH{}-
and electroweak $SU(2)_L$ - symmetries ( see Table 11). $P$ in formula
(\ref{2.1})
is a superpotential to be specified below. To construct it, we use the
internal $U(1)_R-$symmetry which is habitual for a simple N=1 supersymmetry.

In models with a global supersymmetry it is impossible simultaneously
to have a SUSY breaking and a vanishing cosmological term.
The reason is the semipositive definition  of the scalar potential in the rigid
supersymmetry approach
(in particular, in the case of a broken $SUSY$ we have $V_{min}>0$ ).
The problem of supersymmetry breaking, with the  cosmological
term $\Lambda =0$ vanishing, is solved in the framework of the $N=1$ $SUGRA$
models. This may be done under an appropriate choice of the
Kaehler potential, in particular, in the frames of "mini-maxi"- or
"maxi" type models \cite{31'}.
In such approaches, the spontaneous breaking of the local $SUSY$
is due to the possibility to get nonvanishing $VEV$s for the scalar fields from
the "hidden" sector of $SUGRA$ \cite{31'}. The appearance
in the observable sector
of the so-called soft
breaking terms  comes as a consequence of this effect.

 In the "flat" limit, i.e.
 neglecting gravity, one is left with lagrangian (\ref{2.1})  and soft SUSY
 breaking terms, which on the scales $\mu << M_{Pl}$ have the form:
\begin{eqnarray}
 {\cal L}_{SB}&=&\frac{1}{2}\sum_{i}m^{2}_{i}|\phi_{i}|^{2}
 +\frac{1}{2}m_{1}^{2}Tr|h|^{2}  +\frac{1}{2}m_{2}^{2}Tr|H|^{2}+\nonumber\\
 &+&\frac{1}{2}\mu_{1}^{2}|\eta |^{2}+\frac{1}{2}\mu_{2}^{2}|\xi|^{2}
 +\frac{1}{2}M^{2}Tr|\Phi |^{2}+ \label{2.2}\\
&+&\frac{1}{2}\sum_{k}M_{k}\lambda^{a}_{k}\lambda^{a}_{k}+h.c.
+ \mbox {trilinear terms},\nonumber
 \end{eqnarray}
 where $i$ runs over all the scalar matter fields $\; \tilde Q$,
 ${\tilde u}^c$, ${\tilde d}^c$,$\tilde L$, ${\tilde e}^c$,
 ${\tilde \nu}^c$ and $k$ - does over all the gauge groups:
  \SUTH{}, $SU(3)_C,\; SU(2)_L,\; U(1)_Y$. At the energies close
 to the Plank scale all the masses, as well as the gauge coupling, are
 correspondingly equal (this is true if the analytic
 kinetic function satisfies  $f_{\alpha\beta}\sim\delta_{\alpha\beta}$ )
\cite{31'},
 but at low energies they have different  values depending on the
corresponding  renormgroup equation (RGE). The squares of some masses  may
be negative,
 which permits the spontaneous gauge symmetry breaking.

Considering the SUSY version of the $SU(3)_H$-model,
it is natural to ask:
why do  we need to supersymmetrize  the model? Proceeding from our
present-day knowledge
of the nature of supersymmetry \cite{31',32'}, the answer will be:

(a) First, it  is necessary to preserve the hierarchy of the scales:
$M_{EW} <M_{SUSY}< M_H < \cdots ? \cdots < M_{GUT} $
Breaking the horizontal gauge symmetry, one has  to preserve
 SUSY on that scale.
 Another sample of hierarchy to be considered is :
$M_{EW}<M_{SUSY}\sim M_H$.
In this case, the scale $M_H$ should be rather low ($M_H\leq$ a few TeV).

(b) To use the SUSY $U(1)_{R}$ degrees of freedom
for constructing the
superpotential and forbidding undesired Yukawa couplings.

(c) Super-Higgs mechanism - it is possible to describe Higgs bosons
by means of massive gauge superfields \cite{32'}.

(d) To connect the vector- like character of the \SUTH{}-
gauge horizontal model and $N=2$ SUSY.

\subsection{The superweak-like source of CP-
 violation, the Baryon stability and neutrino mass problems.}

The existence of  horizontal interactions (\ref {2.1})
might be closely connected
with the CP-violation problem.
 This  interaction  is  described  by  the
relevant part of the SUSY $SU(3)_H$-Lagrangian and has the form

\begin{eqnarray}
{\cal L}_H=g_H\bar \psi_d {\Gamma}_{\mu}\ (\ DP_d\ \frac{  \Lambda^a } {2}\
P_d^*D^T\ ) \ \psi_d O_{ab} Z_{\mu}^b  \,\,. \label{5.9}
\end{eqnarray}

Here we have (a,b=1,2,...,8). The matrix $O_{ab}$ determines the  relationship
between the bare,  $H_{\mu}^{b}$, and physical, $Z_{\mu}^b$,  gauge
fields and  is  calculated  for
the  mass  matrix  $(M_H^2)_{ab}$   diagonalized;
$\psi_d  =(\psi_d\   ,\ \psi_s\  ,\ \psi_b\ )$ ;
$g_H$  is  the  gauge  coupling   of  the $SU(3)_H$  group.

 Here noteworthy are the  following two
points:
 The appearance of the phase in the CKM mixing matrix may be
due to new dynamics  working at short distances $(r\ll {{1}\over {M_W}})$.
Horizontal forces may be the  source of this new dynamics \cite{9'}.
Using this approach,
 we might have the CP- violation effects- both
due to electroweak and  horizontal  interactions.

(b) The CP is conserved in the
electroweak sector $(\delta^{KM}=0)$, and its breaking is provided by the
structure of the horizontal interactions. Let us think of  the
situation when $\delta^{KM}=0$. In the  SM, such a case might be
realized just accidentally. The vanishing phase of the electroweak sector
$(\delta^{KM}=0)$ might arise spontaneously due to some additional
symmetry. Again, such a situation might occur within the
horizontal extension of the electroweak model.

In particular, this model gives rise to a rather natural mechanism of superweak
-like  CP-violation due to the  $(CP=-1)$ part of the effective
Lagrangian of  horizontal interactions- $({\epsilon}^{\prime}/
{\epsilon})_K\leq \,{10}^{-4}$.
 That part of $\cal L\rm_{eff}$
includes the product of the $SU(3)_H$-currents $I_{\mu i}$ $I_{\mu j}$
(i=1,4,6,3,8; j=2,5,7 or, vice versa,  i$\longleftrightarrow$j ) \cite{9'}.
In the case of a vector- like $SU(3)_H$- gauge model the
CP- violation could
be only due to the charge symmetry breaking.

In electroweak and horizontal interactions
 we might also have two CP- violating contributions
to the amplitudes of
B-meson decays. But it is possible to construct
a scheme where CP- violation will occur only in the horizontal
interactions.
 The last fact might  lead to a very interesting CP -violation
asymmetry $A_f(t)$ for the decays of neutral $B_d^0$ - and $\bar{B}_d^0$ -
mesons to  final hadron CP -eigenstates ,for example,
to $ f \, = \,( J/{\Psi} K_S^0)$ or ( $ \pi \, \pi$)

\begin{eqnarray}
A_{f}(t) \approx \sin{ (\Delta m_{B_d} \, t ) }
Im (\frac{p}{q} \times \rho_f) , \qquad
\rho_f = {{ A (\bar{B} \rightarrow \, f)} \over
{A (B \rightarrow \,f)}}.
\end{eqnarray}

In the standard model with the Kobayashi- Maskawa mechanism of CP- violation,
the time integrated asymmetry  of $B_d^0$- and $\bar{B}_d^0$- meson decays
to  the $  J/{\Psi} K_S^0 $- final state is:

\begin{eqnarray}
A( J/{\Psi} K_S^0) \approx {\eta}_f \times
\frac{x_d}{1 + x_d^2} \times \sin {2 \phi_3 }
=-\frac{x_d}{1 + x_d^2}\times
\frac{2 \eta (1 - \rho)}{(1 - \rho)^2 + {\eta}^2}, \nonumber
\end{eqnarray}
where ${\eta}_f$=-1 for a CP-odd
${J/{\Psi} K_S^0}$- final state;
$\phi_3 = arg V_{td} $  is one of the angles: ($ \phi_i , i= 1, 2, 3 $)
of the unitary triangle.
Let us compare this asymmetry with the analogous asymmetry
of the  $B^0$ and ${\bar B}^0$-decays to  the CP- even final state
$ (\pi^+,\pi^-) $ , the latter being known to depend
 on the phase magnitudes
of $ V_{ub} $ and $V_{td}$. Then:

\begin{eqnarray}
A( \pi^+ \pi^-) \approx - {\eta}_f\times
\frac{x_d}{1 + x_d^2} \times \sin {2 \phi_2 } =
- \frac{x_d}{1 + x_d^2}
\times \frac{2{\eta}[({\rho}^2 + {\eta}^2) -\rho]}
{[(1 - \rho)^2 + {\eta}^2] [{\rho}^2 + {\eta}^2]}, \nonumber
\end{eqnarray}
where $ \phi_2 = \pi - \phi_1 - \phi_3 $ and $ \phi_1 = arg V_{ub}^*
=\delta_{13}$  ( $ \delta^{KM} = \phi_1 + \phi_3 $).

The contributions of CP-violating horizontal interactions to the
asymmetries for both $B^0$-decays are identical but the signs
differ.

The space-time structure of horizontal interactions depends on the $SU(3)_H$
quantum numbers of quark and lepton superfields and their C- conjugate
superfields. One can obtain vector
(axial)-like
horizontal interactions as far as
the $G_H$ particle quantum numbers are conjugate (equal)
 to those of antiparticles.
The question arising in these theories
is how such horizontal interactions are related with strong and
electroweak ones. All these interactions can be unified within
one gauge group, which would allow  to calculate the value of the coupling
constant of horizontal interactions.
Thus, an unification of horizontal, strong and
electroweak interactions might  rest  on the GUTs
$\tilde G \equiv G\times $\SUTH{}
 (where, for example, $\tilde G \equiv E(8)$,
$G\equiv SU(5),SO(10)$ or $E_6$), which may  be  further broken down to
\SUTH{} $\times SU(3)_C \times SU(2)_L \times U(1)_Y$.
For including "vector"- like horizontal gauge symmetry into GUT
 we have to introduce "mirror" superfields. Speaking more definitely,
if we want to construct GUTs of the
$\tilde G \equiv G\times$\SUTH{} type,
  each generation must
encompass double  $G$-matter supermultiplets, mutually conjugate under the
\SUTH-group. In this approach the first supermultiplet consists of the
superfields $f$ and $f_m^c\in 3_H$, while the second is constructed with
the help of the supermultiplets $f^c$ and $f_m\in \bar 3_H$.
In this scheme, proton decays are only possible in the case of
 mixing between ordinary and "mirror" fermions. In its turn, this mixing
must, in particular, be related
with the \SUTH{}-symmetry breaking.

The GUSTs spectra also predict the existing of the new neutral
neutrino - like particles interacting with the matter only by
"superweak"- like coupling. It is possible to estimate the masses of
these particles, and, as will be shown further, some of them have to be light
(superlight) to be observed in modern experiment.

\section{Non-Abelian Gauge Family Symmetry in Grand Unified String Models}
\subsection{World-Sheet Kac-Moody Algebra
         And Main Features of Rank Eight GUST}\label{sec2}

\subsubsection{ The representations of Kac- Moody Algebra and Vertex Operators}

Let's begin with a short review of the KMA results \cite{21',22'}.
In heterotic string the KMA is constructed by the operator product
expansion (OPE) of the fields $J^a$ of the conformal dimension $(0,1)$:
\begin{equation}
{J^a}(z) {J^b}(w) = {\frac{1}{{z-w}^2}} k {\delta}^{ab} +
{\frac{1}{z-w}} i f^{abc} J^c + ....
\end{equation}
 The structure constants $f^{abc}$ for the group $g$ are normilized so that
\begin{equation}
 f^{acd}f^{bcd} = Q_{\psi} {\delta}^{ab} =\tilde h {\psi}^2
{\delta}^{ab}
\end{equation}
, where $Q_{\psi}$ and $\psi$ are  the
quadratic Casimir and the highest weight of the adjoint representation and
$\tilde h$ is the dual Coxeter number.
The $\frac{\psi}{{\psi}^2}$ can be
expanded as in integer linear combination of the simple roots of $g$:
\begin{equation}
 \frac{\psi}{{\psi}^2} = \sum_{i=1}^{rank\,g} m_i {\alpha}_i.
\end{equation}
The dual Coxeter number can be expressed through the integers numbers $m_i$
\begin{equation}
\tilde h = 1 + \sum_{i=1}^{rank \,g} m_{i}
\end{equation}
and for the simply laced groups (all roots are equal and ${\psi}^2 =2$):
$A_n$, $D_n$, $E_6$, $E_7$, $E_8$
they are equal $n+1$, $2n-2$, $12$, $18$ and $30$, respectively.

The KMA $\hat g$ allow to grade the representations $R$ of the gauge group by
a level number $x$ (a non negative integer) and by a conformal weight $h(R)$.
An irreducible representation of the affine algebra $\hat g$ is characterized
by the vacuum representation of the algebra $g$ and the value of the central
term $k$, which is connected with the level number by the relation
$x=2 k/{\psi}^2$.
The value of the level
number of the KMA determines the possible highest weight unitary representation
which are present in the spectrum, in the following way:

\begin{equation}\label{eq1}
x=\frac{2 k}{{\psi}^2} \geq \sum_{i=1}^{rank \,g}n_{i} m_{i},
\end{equation}

\noindent where the sets of non-negative integers $\{m_i = m_1,..., m_r\}$
and $\{ n_i = n_1,...,n_r \}$ define the highest root and the highest weight
of a representation $R$ respectively \cite{21',22'}:
\begin{equation}
{\mu}_0 = \sum_{i=1}^{rank \,g} n_{i} {\alpha}_{i}
\end{equation}
In fact, the KMA on the level one is realized in the 4-dimensional heterotic
superstring theories with free world sheet fermions which allows a
complex fermion description \cite{18',19',20'}. One can obtain KMA on higher
level working with real fermions  using some tricks \cite{33'}.
For these models the level of KMA coincides with the Dynkin index of
representation $M$ to which free fermions are assigned:

\begin{equation}\label{eq2}
x = x_M = \frac{Q_M}{{\psi}^2}\frac{dim M}{dim g}
\end{equation}
($Q_M$ is a quadratic Casimir eigenvalue of representation $M$) and equals one
in cases when real fermions form vector representation $M$ of $SO(2N)$, or when
the world sheet fermions are complex and $M$ is the fundamental representation
of $U(N)$ \cite{21',22'}.

Thus, in strings with KMA on the level one realized on the world-sheet, only
very restricted set of unitary representations can arise in the spectrum:

\begin{enumerate}
\item  singlet and totally antisymmetric tensor representations of $SU(N)$
groups, for which $m_i = (1,...,1) $;
\item singlet, vector and spinor representations of $SO(2N)$ groups, with\\
$m_i = (1,2,2,...2,1,1)$;
\item singlet, $\underline{27}$, and $\bar{\underline{27}}$-plets of $E(6)$,
corresponding to $m_i = (1,2,2,3,2,1)$;
\item singlet  of $E(8)$, with $m_i = (2,3,4,6,5,4,3,2)$.
\end{enumerate}
Therefore only these representations can be used to incorporate matter
and Higgs fields in GUSTs with KMA on the level 1.

In principle it should be possible to construct explicitely an example of
a level 1 KMA-representation of the simply laced $\hat g$ algebra
(A-, D-, E - types)
from the level one representations of the Cartan subalgebra of $g$.
This construction is achieved using the vertex operator of string, where
these operators are assigned to a set of lattice point corresponding to
the roots of a simply-laced Lie algebra $g$.
In heterotic string approach the vertex operator for a gauge boson with
momentum $p$ and polarization $\zeta $ is a primary field of
conformal dimension $(1/2,1)$ and could be written in the form:
\begin{equation}
V^{a} = {\zeta}_{\mu}{{\psi}_{\mu}}({\bar z}) J^a \exp ({i p X}),
\,\,\, p_{\mu} p^{\mu} = {\zeta}_{\mu} p^{\mu} =0.
\end{equation}
$X_{\mu}$ is the string coordinate and $ {\psi}^{\mu}$ is a conformal
dimension (1/2,0) Ramond-Neveu- Schwartz fermion.

\subsubsection{The features of one level KMA in matter and Higgs
representations in
rank 8- and 16- GUST Constructions}
For example, to describe chiral matter fermions in GUST with the gauge symmetry
group $SU(5)\times U(1)\subset SO(10)$ the following sum of the level-one
complex representations: $\underline {1}(-5/2) + \underline {\bar{5}}(+3/2) +
\underline {10}(-1/2) = \underline{16}$ can be used. On the other side, as real
representations of $SU(5)\times U(1)\subset SO(10)$, from which Higgs fields
can arise, one can take for example $\underline{5} + \underline{\bar 5}$
representations arising from real representation $\underline {10}$ of $SO(10)$.
Also, real Higgs representations like $\underline {10}$(-1/2)
+ $\underline{\bar{10}}$(+1/2) of $SU(5)\times U(1)$ originating
from $\underline{16}$+$\underline{\bar{16}}$ of $SO(10)$, which has been used
in ref. \cite{10'} for further symmetry breaking, are allowed.

Another example is provided by the decomposition of $SO(16)$ representations
under $SU(8)\times U(1)\subset SO(16)$. Here only singlet,
$v=\underline{16}$,
$s = \underline{128}$ and $s^{\prime} = \underline{128}^{\prime}$
representations of $SO(16)$
are allowed by the KMA ($s = \underline{128}$ and
$s^{\prime}= \underline{128}^{\prime}$ are
the two nonequivalent, real spinor representations with the highest
weights
${\pi}_{7,8} =1/2 ({\epsilon}_1+{\epsilon}_2, + \dots +
{\epsilon}_7 \mp {\epsilon}_8)$,
${\epsilon}_{i}{\epsilon}_{j}= {\delta}_{ij}$). From the item 2. we can
obtain the following $SU(8)\times U(1)$ representations: singlet,
$\underline{8}$+$\underline{\bar 8}$ $(=\underline{16})$,
$\underline{8}+\underline{56}+\underline{\bar {56}}+\underline{\bar 8}$
$(=\underline{128})$ and $ \underline{1}+\underline{28}+\underline{70} +
+\underline{\bar{28}} + \underline{\bar {1}}$ $(=\underline{128}^{'})$.
The highest weights of $SU(8)$ representations
 ${\pi}_1 = \underline{8}$, ${\pi}_7 = \underline{\bar 8}$ and
${\pi}_3 =\underline{56}$, ${\pi}_5 =\underline{\bar {56}}$  are:
\begin{eqnarray}
{\pi}_{1}& = &1/8 (7{\epsilon}_1 - {\epsilon}_2
- {\epsilon}_3 - {\epsilon}_4 - {\epsilon}_5 - {\epsilon}_6
-{\epsilon}_7 -  {\epsilon}_8),\nonumber \\
{\pi}_{7}& = &1/8 ({\epsilon}_1 + {\epsilon}_2
+ {\epsilon}_3 + {\epsilon}_4 + {\epsilon}_5 + {\epsilon}_6
+{\epsilon}_7 - 7 {\epsilon}_8),\nonumber \\
{\pi}_{3}& = &1/8({5\epsilon}_1+ 5{\epsilon}_2 + 5{\epsilon}_3
- 3{\epsilon}_4 - 3{\epsilon}_5 - 3{\epsilon}_6 -
3{\epsilon}_7 - 3{\epsilon}_8),\nonumber\\
 {\pi}_{5}& = &1/8({-3\epsilon}_1- 3{\epsilon}_2 - 3 {\epsilon}_3
- 3{\epsilon}_4 - 3{\epsilon}_5 + 5{\epsilon}_6 +
5{\epsilon}_7  + 5{\epsilon}_8).
\end{eqnarray}
Similarly,the highest weights of $SU(8)$ representations
 ${\pi}_2 = \underline{28}$,  ${\pi}_6 =\underline{\bar {28}}$
and
${\pi}_4 =\underline{70}$ are:
\begin{eqnarray}
{\pi}_{2}& = &1/4 (3{\epsilon}_1 + 3{\epsilon}_2 -{\epsilon}_3
-{\epsilon}_4 -{\epsilon}_5 -{\epsilon}_6
-{\epsilon}_7 -  {\epsilon}_8),\nonumber \\
{\pi}_{6}& = &1/4 ({\epsilon}_1 + {\epsilon}_2
+ {\epsilon}_3 + {\epsilon}_4 + {\epsilon}_5
+  {\epsilon}_6 -
3{\epsilon}_7 - 3 {\epsilon}_8),\nonumber \\
{\pi}_{4}& = &1/2({\epsilon}_1+ {\epsilon}_2 + {\epsilon}_3
+ {\epsilon}_4 - {\epsilon}_5 - {\epsilon}_6 -
{\epsilon}_7 - {\epsilon}_8).
\end{eqnarray}
However, as we will demonstrate, in each of the string sectors
the generalized Gliozzi--Scherk--Olive projection (the $GSO$ projection
in particular guarantees modular invariance and supersymmetry of the
theory and also give some non-trivial restrictions on gauge
groups and its representations) necessarily eliminates either
$\underline{128}$ or $\underline{128}^{\prime}$. It is therefore important
that, in order to incorporate chiral matter in the model, only one spinor
representation is sufficient. Moreover, if one wants to solve the
chirality problem applying further $GSO$ projections (which break the
gauge symmetry) also the representation $\underline {\bar{10}}$ which
otherwise, together with $\underline{10}$, could form real Higgs
representation, disappears from this sector. Therefore, the existence of
$\underline{\bar{10}}_{-1/2}$+$\underline{10}_{1/2}$, needed for breaking
$SU(5)\times U(1)$ is incompatible (by our opinion) with the possible
solution of the chirality problem for the family matter fields.

Thus, in the rank eight group $SU(8) \times U(1) \subset SO(16)$ with Higgs
representations from the level-one KMA only, one can not arrange for further
symmetry breaking. Moreover, construction of the realistic
fermion mass matrices
seems to be impossible. In old-fashioned GUTs (see e.g.\cite{23'}), not
originating from strings, the representations of level-two were commonly
used to solve these problems.

The way out from this difficulty is based on the following important
observations. Firstly, all higher-dimensional representations of (simple laced)
groups like $SU(N)$, $SO(2N)$ or $E(6)$, which belong to the level-two of the
KMA (according to the equation \ref{eq1}), appear in
the direct product of the level-
one representations:
\begin{eqnarray}\label{eq3}
R_G(x=2) \subset R_G(x=1) \times {R_G}^{'}(x=1).
\end{eqnarray}

For example, the level-two representations of $SU(5)$:\\
 $\underline{15}$,
$\underline{24}$, $\underline{40}$, $\underline{45}$, $\underline{50}$,
$\underline{75}$\\
 will appear in the direct products of:\\
  $\underline{5}\times
\underline{5}$, $\underline{5}\times\underline{\bar 5}$, $\underline{5}\times
\underline{10}$, etc. respectively.

In the case of $SO(10)$ the level two
representations:\\
 $\underline{45}$, $\underline{54}$, $\underline{120}$,
$\underline{126}$, $\underline{210}$, $\underline{144}$\\
 can be obtained by the
suitable direct products:\\ $\underline{10}\times\underline{10}$,
$\underline{\bar{16}}\times\underline{10}$, $\underline{16}\times
\underline{16}$, $\underline{\bar{16}}\times\underline{16}$.\\
 The level-two
representations:\\
 $\underline{78}$, $\underline{351}$, $\underline {351}^{'}$
$\underline{650}$ \\
of $E(6)$ are factors of the decomposition of the direct
products of:\\
$\underline{\bar{27}}\times\underline{27}$ or $\underline{27}
\times\underline{27}$.\\
 The only exception from this rule is the $E(8)$
group, two level-two representations ($\underline {248}$ and
$\underline{3875}$) of which cannot be constructed as a product of level-one
representations \cite{24'}.

Secondly, the diagonal (symmetric) subgroup $G^{symm}$ of $G \times G $
effectively corresponds to the level-two KMA
$g(x=1)\oplus g(x=1)$ \cite{25',26'} because taking the $G\times G$
representations in the form $(R_G,R^{'}_G)$ of the $G\times G$,
where $R_G$ and $R_G^{'}$ belong to the level-one of G,
one obtains representations of the form $R_G\times R^{'}_G$
when one considers only the diagonal subgroup of $G\times G$.
This observation is crucial, because such a construction
allows one to obtain level-two representations. (This construction
has implicitly been used in \cite{26'} (see also \cite{25'})
where we have  constructed some examples
of GUST with gauge symmetry realized
as a diagonal subgroup of direct product
of two rank eight groups $U(8) \times U(8) \subset SO(16) \times SO(16)$.)

In strings, however, not all level-two representations can be obtained in that
way because, as we will demonstrate, some of them become massive (with masses
of order of the Planck scale). The condition ensuring that in the string
spectrum states transforming as a representation $R$ are massless reads:
\begin{equation}\label{eq4}
h(R) = \frac{Q_R}{2 k + Q_{ADJ}} =
\frac{Q_R}{2 Q_M} \leq 1 ,
\end{equation}
where $Q_i$ is the quadratic Casimir invariant of the corresponding
representations, and M has been already defined before (see eq. \ref{eq2}).
Here the conformal weight is defined by $L_0 |0> = h(R)|0>$,
\begin{eqnarray}\label{eq100}
L_0 =\frac{1}
{2k + Q_{\psi}}\times  \biggl (\sum_{a=1}^{a=dim \,g}(T_0^aT_0^a +
2 \sum_{n=1}^{n=inf}{T_{-n}^aT_n^a})\biggr),
\end{eqnarray}
where $T_n^a|0>=0$ for $n>0$.
The condition (\ref {eq4}), when combined with (\ref{eq1}), gives a
restriction at the rank of
GUT's group  ($r \leq 8$), whose representations  can accomodate chiral matter
fields. For example, for $ G =SO(16)$ or $E(6) \times SU(3)$, representations
$\underline{128}$, $(\underline{27}, \underline {3})$ ($h(\underline{128})= 1$,
$h(\underline{27}, \underline{3}) = 1$) respectively, satisfy both conditions.
Obviously, these (important for incorporation of chiral matter) representations
will exist at the level-two KMA of the symmetric subgroup of the group
$G\times G$.

In general, condition (\ref{eq4}) severely constrains  massless string states
transforming as $(R_G(x = 1), {R_G}^{'}(x = 1))$ of the direct product
$G\times G$. For example, for $SU(8)\times SU(8)$ and for $SU(5)\times SU(5)$
constructed from $SU(8)\times SU(8)$ only representations of the form
\begin{eqnarray}\label{eq5}
R_{N,N} =  \bigl((\underline {N},\underline {N}) + h.c.\bigr ),\,\,\,\,\,
\bigl((\underline {N},\underline {\bar N}) + h.c.\bigr);\,\,\,
\end{eqnarray}
with $h(R_{N,N}) = (N-1)/N$, where $N\,=\,8$ or $5$ respectively can be
massless. For $SO(2N) \times SO(2N)$ massless states are contained
only in representations
\begin{eqnarray}\label{eq6}
R_{v,v} = (\underline{2N} ,\underline{2N})
\end{eqnarray}
with $h(R_{v,v}) = 1$. Thus, for the GUSTs based on a diagonal subgroup
$G^{symm}\subset G\times G$, $G^{symm}$ - high dimensional representations,
which are embedded in $R_G(x = 1) \times {R^{'}}_G(x = 1)$ are also severely
constrained by the condition (\ref{eq4}).

For spontaneous breaking of $G\times G$ gauge symmetry down to $G^{symm}$
(rank $G^{symm}$ = rank $G$) one can use the direct product of representations
$R_G(x=1)\times R_G(x=1)$, where $R_G(x = 1)$ is the fundamental representation
of $G = SU(N)$ or vector representation of $G = SO(2N)$. Furthemore,
$G^{symm}\subset G\times G$ can subsequently be broken down to a smaller
dimension gauge group (of the same rank as $G^{symm}$) through the VEVs of the
adjoint representations which can appear as a result of $G\times G$ breaking.
Alternatively, the real Higgs superfields (\ref{eq5}) or
(\ref{eq6}) can directly break the
$G\times G $ gauge symmetry down to a $G_1^{symm}\subset G^{symm}$ (rank
$G_1^{symm}\leq$ rank $G^{symm}$). For example when $G = SU(5)\times U(1)$ or
$SO(10)\times U(1)$  $G\times G$ can directly be broken in that way down to
$SU(3^c)\times G^I_{EW}\times G^{II}_{EW}\times...$.

The above examples show clearly, that within the framework of GUSTs with the
KMA one can get interesting gauge symmetry breaking chains including the
realistic ones provided $G\times G $ gauge symmetry group is considered.
However the lack of the higher dimensional representations (which are forbidden
by \ref{eq4}) on the level-two KMA prevents the construction of the realistic
fermion mass matrices.
That is why we consider an extended grand unified string model of rank eight
{}.
$SO(16)$ or $E(6) \times SU(3)$ of $E(8)$.

The full chiral $SO(10)\times SU(3)\times U(1)$ matter multiplets can be
constructed from $SU(8)\times U(1)$--multiplets
\begin{eqnarray}\label{eq7}
(\underline {8} + \underline {56} + \underline {\bar{8}}
+ \underline {\bar{56}}) = \underline {128}
\end{eqnarray}
of $SO(16)$. In the 4-dimensional heterotic superstring with free complex
world sheet fermions, in the spectrum of the Ramond sector there
can appear also
representations which are factors in the decomposition of
$\underline{128^{'}}$.
In particular, $SU(5)$-decouplets $(\underline {10} + \underline {\bar {10}})$
from $(\underline{28}+\underline{\bar{28}})$ of $SU(8)$. However their
$U(1)_5$ hypercharge prevent using them for $SU(5)\times U(1)_5$--symmetry
breaking. Thus, in this approach we have only singlet and
$(\underline{5} + \underline{\bar5})$ Higgs fields which can break the grand
unified $SU(5)\times U(1)$ gauge symmetry. Therefore it is necessary (as we
already explained) to construct rank eight GUST based on a diagonal subgroup
$G^{symm}\subset G\times G$ primordial symmetry group, where in each
rank eight group $G$ the Higgs fields will appear only in singlets and in the
fundamental representations as in (see \ref{eq5}).

A comment concerning $U(1)$ factors can be made here. Since the available
$SU(5)\times U(1)$ decouplets have non-zero hypercharges with respect
to $U(1)_5$ and $U(1)_H$, these $U(1)$ factors may remain unbroken down to
the low energies in the model considered which seems to be very interesting.

\subsection{ GUST Constructions
            in Free Fermion For\-mu\-lation.}\label{sec3}
\subsubsection{Modular invariance and spin- basis.}\label{sbsec31}

 A Sugawara- Sommerfeld
construction of the Virasoro
algebra in terms of bilinears in the Kac- Moody generators \cite{21'},
\cite{22'} allows to get the following expression for the central Virasoro
"charge":
\begin{eqnarray}\label{eq101}
c_g = \frac {2 k dim g}{2 k + Q_{\psi}}=
\frac{x dim g}{ x + \tilde h}.
\end{eqnarray}

In heterotic string theories \cite{13',14'} ${(N=1\: SUSY)}_{LEFT}$
${(N=0\: SUSY)}_{RIGHT}$ $\oplus$ ${\cal M}_{c_{L};c_{R}}$ with $d \leq 10$,
the conformal anomalies of the space-time sector are canceled by the conformal
anomalies of the internal sector ${\cal M}_{c_L;c_R}$, where $c_L=15-3d/2$ and
$c_R=26-d$ are the conformal anomalies in the left- and right--moving string
sectors respectively.

In the fermionic formulation of the four-dimensional heterotic string theory in
addition to the two transverse bosonic coordinates $X_{\mu}$ ,$\bar{X}_{\mu}$
and their left-moving superpartners ${\psi}_{\mu}$, the internal sector
${\cal M}_{c_L;c_R}$ contains 44 right-moving ($c_R=22$) and 18 left-moving
($c_L=9$) real fermions. The model is completely defined by a set $\Xi$ of spin
boundary conditions for all these world-sheet fermions. In a diagonal basis the
vectors of $\Xi$ are determined by the values of phases $\alpha(f)$ $\in$(-1,1]
fermions $f$ acquire ($f\longrightarrow -\exp({i\pi\alpha(f)}) f $) when
parallel transported around the string. To construct the GUST according to the
scheme outlined at the end of the previous section we consider
three different basises each of them  with six elements
$B= {b_1, b_2, b_3, b_4 \equiv S, b_5, b_6 }$. (See Tables \ref{tabl1},
4 and 7.)

Following \cite{19'} we  construct the canonical basis in such a way that the
vector $\bar1$, which belongs to $\Xi$, is the first element
$b_1$ of the basis.
The basis vector $b_4=S$ is the generator of supersymmetry \cite{20'}
responsible for the conservation of the space-time $SUSY$.

We have chosen a basis in which all left movers $({\psi}_{\mu}; {\chi}_i, y_i,
{\omega}_i; i=1,...6)$ (on which the world sheet supersymmetry is realized
nonlinear\-ly) as well as 12 right movers $({\bar\varphi}_k; k=1,...12)$ are
real whereas (8 + 8) right movers $\bar{\Psi}_A$, $\bar{\Phi}_M$ are complex.
Such a construction corresponds to $SU(2)^6$ group of automorphisms of the
left supersymmetric sector of a string. Right- and left-moving real
fermions can be used for breaking $G^{comp}$ symmetry \cite{20'}. In
order to have a possibility to reduce the rank of the compactified group
$G^{comp}$, we have to select the spin boundary conditions for the maximal
possible number, $N_{LR}$ = 12, of left-moving, ${\chi}_{3,4,5,6}$,
$y_{1,2,5,6}$, ${\omega}_{1,2,3,4}$, and right-moving,
${\bar{\phi}}^{1,...12}$
(${\bar{\phi}}^p={\bar{\varphi}}_p$, $p=1,...12$) real fermions.
The KMA based on $16$ complex right moving fermions gives rise to the
"observable" gauge group, $G^{obs}$, with:
\begin{equation}\label{eq8}
rank (G^{obs})  \leq 16.
\end{equation}

The study of the Hilbert spaces of the string theories
is connected to the problem of finding all possible choices
of the GSO coefficients ${\cal C}
\left[
\begin{array}{c}
{\alpha} \\
{\beta}
\end{array}\right]$,
such that the one--loop partition function
\begin{equation}
Z=\sum_{ \alpha , \beta} {\cal C}
\left[
\begin{array}{c}
{\alpha} \\
{\beta}
\end{array}\right] \prod_f Z
\left[
\begin{array}{c}
{\alpha}_f \\
{\beta}_f
\end{array}\right]
\end{equation}
and its multiloop conterparts are all modular invariant.
In this formula ${\cal C}
\left[
\begin{array}{c}
{\alpha} \\
{\beta}
\end{array}\right]$ are GSO coefficients,
$\alpha$ and $\beta$ are $(k+l)$--component spin--vectors
$\alpha=[\alpha(f_1^r), ... , \alpha(f_k^r);
\alpha(f_1^c), ... , \alpha(f_l^c)]$,
the components $\alpha_f$, $\beta_f$ specify the spin
structure of the $f$th fermion and $Z[...]$ -- corresponding one-fermion
partition functions on torus: $Z[...]=\mbox{Tr exp}[2\pi iH_{(sect.)}]$.

The physical states in the Hilbert space of a given sector
$\alpha$ are obtained
acting on the vacuum ${|0>}_{\alpha}$ with the bosonic and fermionic operators
with frequencies
\begin{eqnarray}\label{eq9}
n(f) = 1/2 + 1/2 \alpha(f),\:\:\:   n(f^*) = 1/2  -1/2 \alpha(f^*)
\end{eqnarray}
and subsequently applying the generalized GSO projections. The physical states
satisfy the Virasoro condition:
\begin{eqnarray}\label{eq10}
M_L^2 = - 1/2 + 1/8 \:({\alpha}_L \cdot {\alpha}_L) + N_L =
-1 +1/8\: ({\alpha}_R \cdot {\alpha}_R) +N_R = M_R^2,
\end{eqnarray}
where $\alpha=({\alpha}_L,{\alpha}_R)$ is  a sector in the set $\Xi$,
$N_L = {\sum }_{L} (frequencies)$ and  $N_R = {\sum}_{R} (freq.)$.

We keep the same sign convention for the fermion number operator  $F$
as in \cite{20'}. For complex fermions we have $F_{\alpha}(f)=1$,
$F_{\alpha}(f^*)=-1$ with the exception of the periodic fermions
for which we get $F_{{\alpha}=1}(f) =-1/2(1-{\gamma}_{5f})$, where
${\gamma}_{5f}|\Omega>=|\Omega>$,
${\gamma}_{5f}b^+_o|\Omega>=-b^+_o|\Omega>$.

The full Hilbert space of the string theory is constructed as a direct sum of
different sectors ${\sum}_{i} {m_ib_i}$, ($m_i=0,1,..,N_i$), where the integers
$N_i$ define additive groups $Z(b_i)$ of the basis vectors  $b_i$.
The generalized GSO projection leaves in sectors $\alpha$ those states, whose
$b_i$-fermion number satisfies:
\begin{equation}\label{eq11}
\exp(i \pi b_i F_{\alpha})
= {\delta}_{\alpha}
{\cal C}^*
\left[
\begin{array}{c}
{\alpha} \\
b_i
\end{array}\right],
\end{equation}
where the space-time phase ${\delta}_{\alpha}=\exp(i\pi{\alpha}({\psi}_{\mu}))$
is equal $-1$ for the Ramond sector and +1 for the Neveu-Schwarz sector.

\subsubsection{$SU(5)\times U(1)\times SU(3)\times U(1)$- Model 1.}
\label{sbsec32}
Model  1 is defined by 6 basis vectors given in Table \ref{tabl1} which
generates the $Z_2\times Z_4\times Z_2\times Z_2\times Z_8\times Z_2$
group under addition.

\begin{table}[p]
\caption{\bf Basis of the boundary conditions for all world-sheet fermions.
Model 1.}
\label{tabl1}
\footnotesize
\begin{center}
\begin{tabular}{|c||c|ccc||c|cc|}
\hline
Vectors &${\psi}_{1,2} $ & ${\chi}_{1,..,6}$ & ${y}_{1,...,6}$ &
${\omega}_{1,...,6}$
& ${\bar \varphi}_{1,...,12}$ &
${\Psi}_{1,...,8} $ &
${\Phi}_{1,...,8}$ \\ \hline
\hline
$b_1$ & $1 1 $ & $1 1 1 1 1 1$ &
$1 1 1 1 1 1 $ & $1 1 1 1 1 1$ &  $1^{12} $ & $1^8 $ & $1^8$ \\
$b_2$ & $1 1$ & $1 1 1 1 1 1$ &
$0 0 0 0 0 0$ &  $0 0 0 0 0 0 $ &  $ 0^{12} $ &
${1/2}^8 $ & $0^8$  \\
$b_3$ & $1 1$ & $ 1 1 1 1 0 0 $ & $0 0 0 0 1 1 $ & $ 0 0 0 0 0 0 $ &
$0^4  1^8 $ & $ 0^8 $ &  $ 1^8$ \\
$b_4 = S $ & $1 1$ &
$1 1 0 0 0 0 $ & $ 0 0 1 1 0 0 $ & $ 0 0 0 0 1 1 $ &
$ 0^{12} $ & $ 0^8 $ & $ 0^8 $ \\
$  b_5 $ & $ 1 1 $ & $ 0 0  1 1 0 0 $ &
$0 0 0 0 0 0 $ &  $1 1  0 0 1 1$ &  $ 1^{12} $ &
${ 1/4}^5  {-3/4}^3 $ & $ {-1/4}^5\ {3/4}^3  $  \\
$  b_6 $ & $ 1 1 $ & $ 1 1  0 0  0 0 $ &
$ 0 0  0 0 1 1 $ &  $0 0  1 1 0 0$ &  $ 1^2 0^4 1^6 $ &
$ 1^8  $ & $ 0^8  $  \\
\hline \hline
\end {tabular}
\end{center}
\normalsize
\end{table}

In our approach the basis vector $b_2$ is constructed as a complex vector
with the $1/2$ spin-boundary conditions for the right-moving fermions
${\Psi}_A$, $A = 1,...8$. Initially  it generates chiral  matter fields
in the $\underline{8}+\underline{56}+\underline{\bar{56}}+\underline{\bar{8}}$
representations of $SU(8)\times U(1)$, which subsequently are decomposed under
$SU(5)\times U(1)\times SU(3)\times U(1)$ to which $SU(8)\times U(1)$ gets
broken by applying the $b_5$ $GSO$ projection.

Generalized GSO projection coefficients are originally  defined up to fifteen
signs  some of which, are fixed by the supersymmetry conditions.
Below, in Table \ref{tabl1}, we present a set of numbers
$$
\gamma\left[\begin{array}{c}b_i\\b_j\end{array}\right]=\frac{1}{i \pi}
\log{\cal C}\left[\begin{array}{c}b_i\\b_j\end{array}\right].
$$
which we use as basis for our GSO projections.

\begin{table}[b]
\caption{\bf The choice of the GSO basis $\gamma [b_i, b_j]$. Model 1.
($i$ numbers rows and $j$ -- columns)}
\label{tabl2}
\footnotesize
\begin{center}
\begin{tabular}{|c||c|c|c|c|c|c|}
\hline
& $b_1$ & $b_2$ & $b_3$ & $b_4$ & $b_5$ & $b_6$\\ \hline
\hline
$b_1$ & $0$ &    $1$ & $1$ & $1$ &    $1$ & $0$\\
$b_2$ & $1$ &  $1/2$ & $0$ & $0$ &  $1/4$ & $1$\\
$b_3$ & $1$ & $-1/2$ & $0$ & $0$ &  $1/2$ & $0$\\
$b_4$ & $1$ &    $1$ & $1$ & $1$ &    $1$ & $1$\\
$b_5$ & $0$ &    $1$ & $0$ & $0$ & $-1/2$ & $0$\\
$b_6$ & $0$ &    $0$ & $0$ & $0$ &    $1$ & $1$\\
\hline \hline
\end {tabular}
\end{center}
\normalsize
\end{table}

In our case of the ${Z_2}^4\times {Z_4}\times {Z_8}$ model, we initially have
$256\times2$ sectors. After applying the GSO-projections we
get only $49\times2$ sectors containing massless states, which depending on the
vacuum energy values, $E^{vac}_L$ and $E^{vac}_R$, can be naturally divided
into some classes  and which determine the GUST representations.

Generally RNS (Ramond -- Neveu-Schwarz) sector (built on vectors $b_1$
and $S=b_4$) has high symmetry including $N=4$ supergravity and gauge
$SO(44)$ symmetry. Corresponding gauge bosons are constructed as follows:
\begin{eqnarray}\label{eq12}
&{\psi}_{1/2}^{\mu}{|0>}_L \otimes  {\Psi}_{1/2}^I
 {\Psi}_{1/2}^{J} |0>_R ,\nonumber\\
&{\psi}_{1/2}^{\mu}{|0>}_L \otimes  {\Psi}_{1/2}^I
 {\Psi}_{1/2}^{*J} |0>_R ,\:\:I, J=1,\dots,22 &
\end{eqnarray}
 While $U(1)_J$ charges for Cartan subgruops is given by formula
 $Y=\frac{\alpha}{2}+f$ (where $F$ --- fermion number, see (\ref{eq11})),
 it is obvious that states (\ref{eq12}) generate root lattice for
 $SO(44)$:
 \begin{eqnarray}
\pm \varepsilon_I \pm \varepsilon_J \ \ (I\neq J);\qquad
  \pm \varepsilon_I \mp \varepsilon_J
\end{eqnarray}
 The others vectors breakes $N=4$ SUSY to $N=1$ and gauge group $SO(44)$
 to $SO(2)^3_{1,2,3}\times SO(6)_4\times {\left[ SU(5)\times U(1)
 \times SU(3)_H\times U(1)_H\right]}^2$, see Figure \ref{fig2}.

 Generally, additional basis vectors can generate extra vector bosons and
 extend gauge group that remains after applying GSO-projection to
 RNS-sector. In our case dangerous sectors are: $2b_2+nb_5,\ n=0,2,4,6;
 \  2b_5;6b_5$. But our choice of GSO coefficients cancels all the vector
 states in these sectors. Thus gauge bosons in this model  appear
 only from RNS-sector.

\begin{table}[t]
\caption{\bf The list of quantum numbers of the states. Model 1.}
\label{tabl3'}
\footnotesize
\noindent \begin{tabular}{|c|c||c|cccc|cccc|} \hline
N$^o$ &$ b_1 , b_2 , b_3 , b_4 , b_5 , b_6 $&
$ SO_{hid}$&$ U(5)^I $&$ U(3)^I $&$ U(5)^{II} $&$
U(3)^{II} $&$ {\tilde Y}_5^I $&$ {\tilde Y}_3^I $&$ {\tilde Y}_5^{II} $&$
{\tilde Y}_3^{II}$ \\ \hline \hline
1 & RNS &&5&$\bar 3$&1&1&--1&--1&0&0 \\
  &     &&1&1&5&$\bar 3$&0&0&--1&--1 \\
  &0\ 2\ 0\ 1\ 2(6)\ 0&&5&1&5&1&--1&0&--1&0 \\
  &&&1&3&1&3&0&1&0&1 \\
  &&&5&1&1&3&--1&0&0&1 \\
  &&&1&3&5&1&0&1&--1&0 \\ \hline \hline
2 &0\ 1\ 0\ 0\ 0\ 0&&1&3&1&1&5/2&--1/2&0&0 \\
  &&&$\bar 5$&3&1&1&--3/2&--1/2&0&0 \\
  &&&10&1&1&1&1/2&3/2&0&0 \\
  &0\ 3\ 0\ 0\ 0\ 0&&1&1&1&1&5/2&3/2&0&0 \\
  &&&$\bar 5$&1&1&1&--3/2&3/2&0&0 \\
  &&&10&3&1&1&1/2&--1/2&0&0 \\ \hline
3 &0\ 0\ 1\ 1\ 3\ 0&$-_1\ \pm_2$&1&1&1&3&0&--3/2&0&--1/2 \\
  &0\ 0\ 1\ 1\ 7\ 0&$-_1\ \pm_2$&1&$\bar 3$&1&1&0&1/2&0&3/2 \\
  &0\ 2\ 1\ 1\ 3\ 0&$+_1\ \pm_2$&1&$\bar 3$&1&3&0&1/2&0&--1/2 \\
  &0\ 0\ 1\ 1\ 7\ 0&$+_1\ \pm_2$&1&1&1&1&0&--3/2&0&--3/2 \\ \hline
4 &1\ 1\ 1\ 0\ 1\ 1&$\mp_1\ \pm_3$&1&1&1&$\bar 3$&0&--3/2&0&1/2 \\
  &1\ 1\ 1\ 0\ 5\ 1&$\mp_1\ \pm_3$&1&$\bar 3$&1&1&0&1/2&0&--3/2 \\
  &1\ 3\ 1\ 0\ 1\ 1&$\pm_1\ \pm_3$&1&$\bar 3$&1&$\bar 3$&0&1/2&0&1/2 \\
  &1\ 3\ 1\ 0\ 5\ 1&$\pm_1\ \pm_3$&1&1&1&1&0&--3/2&0&--3/2 \\ \hline
5 &0\ 1(3)\ 1\ 0\ 2(6)\ 1&$-_1\ \pm_3$&1&3($\bar 3$)&1&1&$\pm$5/4&$\pm$1/4
&$\pm$5/4&$\mp$3/4 \\
  &&$+_1\ \pm_3$&5($\bar 5$)&1&1&1&$\pm$1/4&$\mp$3/4&$\pm$5/4&$\mp$3/4 \\
  &0\ 1(3)\ 1\ 0\ 4\ 1&$-_1\ \pm_3$&1&1&1&3($\bar 3$)&$\pm$5/4&$\mp$3/4
&$\pm$5/4&$\pm$1/4 \\
  &&$+_1\ \pm_3$&1&1&5($\bar 5$)&1&$\pm$5/4&$\mp$3/4&$\pm$1/4&$\mp$3/4
\\ \hline
6 &1\ 2\ 0\ 0\ 3(5)\ 1&$\pm_1\ -_4$&1&1&1&1&$\pm$5/4&$\pm$3/4
&$\mp$5/4&$\mp$3/4 \\
  &1\ 1(3)\ 0\ 1\ 5(3)\ 1&$+_1\ \mp_4$&1&1&1&1&$\pm$5/4&$\pm$3/4
&$\pm$5/4&$\pm$3/4 \\
  &0\ 0\ 1\ 0\ 2(6)\ 0&$\mp_3\ +_4$&1&1&1&1&$\pm$5/4&$\mp$3/4
&$\pm$5/4&$\mp$3/4 \\ \hline
\end{tabular}
\normalsize
\end{table}

In NS sector the $b_3$ GSO projection leaves $(5,\bar{3})+(\bar{5},3)$ Higgs
superfields:
\begin{equation}\label{eq14}
\chi^{1,2}_{1/2}|\Omega>_L\otimes {\Psi}_{1/2}^a
{\Psi}_{1/2}^{i*};\,\, {\Psi}_{1/2}^{a*}
{\Psi}_{1/2}^{i} |\Omega>_R\ \ \mbox{and exchange}\  \Psi
 \longrightarrow\Phi, \end{equation}
where $a,\:b=1,\dots,\:5,\ \ i,\:j=1,2,3$.

Four $(3_H + 1_H)$ generations of chiral matter fields from
$({SU(5)\times SU(3)})_I$ group forming $SO(10)$--multiplets
$(\underline 1, \underline 3) + (\underline {\bar 5},\underline 3) +
(\underline {10}, \underline 3)$ ; $( \underline 1,\underline 1) +
(\underline {\bar 5},\underline 1) + (\underline {\bar {10}},\underline 1)$
are contained in $b_2$ and $3b_2$ sectors. Applying $b_3$ $GSO$
projection to the $3b_2$ sector yields the following massless states:

\begin{eqnarray}\label{eq15}
b_{\psi_{12}}^+ b_{{\chi}_{34}}^{+} b_{{\chi}_{56}}^+
|\Omega >_L& \otimes &  \Biggl \{   {\Psi}_{3/4}^{i*} ,
  {\Psi}_{1/4}^{a}   {\Psi}_{1/4}^{b}   {\Psi}_{1/4}^{c},
  {\Psi}_{1/4}^{a}   {\Psi}_{1/4}^{i}   {\Psi}_{1/4}^{j}
\Biggr \} |\Omega>_R, \nonumber \\
b_{{\chi}_{12}}^{+} b_{{\chi}_{34}}^{+} b_{{\chi}_{56}}^{+}
|\Omega >_L& \otimes &  \Biggl \{   {\Psi}_{3/4}^{a*} ,
  {\Psi}_{1/4}^{a}   {\Psi}_{1/4}^{b}   {\Psi}_{1/4}^{i},
  {\Psi}_{1/4}^{i}   {\Psi}_{1/4}^{j}   {\Psi}_{1/4}^{k}
\Biggr \} |\Omega>_R
\end{eqnarray}
with the space-time chirality  ${\gamma}_{5 {\psi}_{12}} =- 1$
and  ${\gamma}_{5 {\psi}_{12}} = 1$, respectively.
In these formulae the Ramond creation operators
$b_{\psi_{1,2}}^+$ and $b_{{\chi}_{\alpha, \beta}}^+$ of the zero modes
are built of a pair of real fermions (as indicated by double indices):
${\chi}_{\alpha, \beta}$,  $(\alpha,\beta)$ =  $(1,2)$,  $(3,4)$,  $(5,6)$.
Here, as in (\ref{eq14}) indices take values $a,b$ = 1,...,5  and
$i,j$ = 1,2,3 respectively.

We stress that without using the  $b_3$  projection we would get matter
supermultiplets belonging to real representations only i.e. "mirror"
particles would remain in the spectrum. The  $b_6$  projection instead,
eliminates all chiral matter superfields from $U(8)^{II}$ group.

Since the matter fields form the chiral multiplets of $SO(10)$, it is possible
to write down  $U(1)_{Y_5}$--hypercharges of massless states. In order to
construct the right electromagnetic charges for matter fields we must define
the hypercharges operators for the observable $U(8)^{I}$ group as

\begin{equation}\label{eq16}
Y_5=\int^\pi_0 d\sigma\sum_a \Psi^{*a}\Psi^a ,\,\,\,\,\,
Y_3=\int^\pi_0 d\sigma\sum_i \Psi^{*i}\Psi^i
\end{equation}
and analogously for the $U(8)^{II}$ group.

Then the orthogonal combinations
\begin{equation}\label{eq17}
 \tilde Y_5 = {1\over 4}(Y_5 + 5Y_3), \,\,\,\,\,
 \tilde Y_3 = {1\over 4}(Y_3 - 3Y_5),
\end{equation}
play the role of the hypercharge operators of $U(1)_{Y_5}$ and
$U(1)_{Y_H}$ groups,
respectively. In a Table \ref{tabl3'} we give the hypercharges
 $\tilde Y_5^{I},\tilde Y_3^{I}, \tilde Y_5^{II},\tilde Y_3^{II}$.

 The full list of states in this model is given in a Table \ref{tabl3'}.
 For fermion states only sectors with positive (left) chirality is
written. Superpartners arises from sectors with $S=b_4$-component
changed by 1. Chirality under hidden $SO(2)^3_{1,2,3}\times SO(6)_4$ is
defined as $\pm_1,\ \pm_2,\ \pm_3,\ \pm_4$ respectively. Low signs in item
5 and 6 correspond to sectors with components given in brackets.

 In the next section we discuss the problem of rank
eight GUST gauge symmetry breaking.  The matter is that according to the
results of section \ref{sec2} the Higgs fields
$(\underline{10}_{1/2}+\underline{\bar{10}}_{-1/2})$ do not appear.

\subsubsection{ $SU(5)\times U(1)\times SU(3)\times U(1)$ Model 2.}
Consider then another ${\left[U(5)\times U(3)\right]}^2$ model which after
breaking gauge symmetry by Higgs mechanism leads to the spectrum similar
to Model 1.

This model is defined by basis vectors given in a Table 4
with the $Z^4_2\times Z_6\times Z_{12}$ group under addition.

{\bf Table 4: Basis of the boundary conditions for Model 2. }
\begin{center}
\begin{tabular}{|c||c|ccc||c|cc|}
\hline
Vectors &${\psi}_{1,2} $ & ${\chi}_{1,..,6}$ & ${y}_{1,...,6}$ &
${\omega}_{1,...,6}$
& ${\bar \varphi}_{1,...,12}$ &
${\Psi}_{1,...,8} $ &
${\Phi}_{1,...,8}$ \\ \hline
\hline
$b_1$      & $1 1$ & $1^6$   & $1^6$   & $1^6$   &   $1^{12}$
& $1^8$          & $1^8$ \\
$b_2$      & $1 1$ &  $1^6$  & $0^6$   & $0^6$
&   $0^{12}$       & $1^5$ $1/3^3$  & $0^8$  \\
$b_3$      & $1 1$ & $1^2 0^2 1^2$ & $0^6$
& $0^2 1^2 0^2$ &  $0^8\  1^4 $ & $1/2^5\ 1/6^3$ & $-1/2^5\ 1/6^3 $ \\
$b_4 = S $ & $1 1$ & $1^2\ 0^4$ & $0^2 1^2 0^2$
& $0^4\ 1^2$ &  $0^{12}$        & $ 0^8 $        & $ 0^8 $ \\
$b_5 $     & $1 1$ & $1^4\ 0^2$ & $0^4\ 1^2$ & $0^6$
&  $1^8\ 0^4$   & $1^5\  0^3$    & $ 0^5\ 1^3  $  \\
$b_6 $     & $1 1$ & $0^2 1^2 0^2$ & $1^2\ 0^4$
& $0^4\ 1^2$ &  $1^2 0^2 1^6 0^2$     & $ 1^8  $       & $ 0^8  $  \\
\hline \hline
\end {tabular}
\end{center}

GSO coefficients are given in Table 5.\\

{\bf Table 5:The choice of the GSO basis $\gamma [b_i, b_j]$. Model 2.
($i$ numbers rows and $j$ -- columns)}
\begin{center}
\begin{tabular}{|c||c|c|c|c|c|c|}
\hline
& $b_1$ & $b_2$ & $b_3$ & $b_4$ & $b_5$ & $b_6$\\ \hline
\hline
$b_1$ & $0$ &    $1$         & $1/2$        & $0$ &  $0$ & $0$\\
$b_2$ & $0$ &  $2/3$ & $-1/6$       & $1$ &  $0$ & $1$\\
$b_3$ & $0$ & $1/3$          &$5/6$ & $1$ &  $0$ & $0$\\
$b_4$ & $0$ &    $0$         & $0$          & $0$ &  $0$ & $0$\\
$b_5$ & $0$ &    $1$         & $-1/2$       & $1$ &  $1$ & $1$\\
$b_6$ & $0$ &    $1$         & $1/2$        & $1$ &  $0$ & $1$\\
\hline \hline
\end {tabular}
\end{center}

%
The given model corresponds to the following chain of the gauge
symmetry breaking:
$$E^2_8\longrightarrow SO(16)^2\longrightarrow U(8)^2
\longrightarrow [U(5)\times U(3)]^2\ . $$
When the breaking of $U(8)^2-$group to $[U(5)\times U(3)]^2$
determined by basis vector $b_5$, and N=2 SUSY$\longrightarrow$N=1 SUSY
determined by basis vector $b_6$.

It is interesting to note how the
$U(8)^2$ gauge group restored by sectors $4b_3,\ 8b_3,\ 2b_2+c.c.$
and $4b_2+c.c.$

The full massless spectrum  for given model is given in Table 6.
By analogy with Table 3
only fermion states with positive chirality
is written and obviously vector supermultiplets are absent.
Hypercharges determines by formula:
$$ Y_n=\sum_{k=1}^{n}(\alpha_k/2 + F_k)\ . $$

The given model possesses by the hidden gauge symmetry
$SO(16)_1\times SO(2)^3_{2, 3, 4}$.
The corresponding chirality is given in column $SO_{hid.}$.
The sectors  are divided by horizontal lines and
without including the $b_5-$vector form $SU(8)-$multiplets.

For example, let us consider row No 2.
In sectors $b_2$, $5b_2$ in addition to states $(1, \bar{3})$ and
$(5, \bar{3})$ the (10, 3)--state appears, and in the sector $3b_2$
besides the $(\bar{10}, 1)-$ the states (1, 1) and $(\bar{5}, 1)$
survive too. All these states form $\bar{8}+56$ representation
of the $SU(8)^I$ group.

Analogically we can get the full structure of the theory according
$U(8)^I\times U(8)^{II}-$group.
(For correct restoration of the $SU(8)^{II}-$group we must invert
3 and $\bar{3}$ representations.)

In Model 2 matter fields appear both in $U(8)^I$ and $U(8)^{II}$ groups.
This is the main difference with comparing of the Model 1.
However, note that in the
Model 2 similary to the Model 1 all gauge fields appear in RNS--sector only
and $10 +\bar{10}$ representation (which can be the Higgs field
for gauge symmetry breaking) is absent.

{\bf Table 6: The list of quantum numbers of the states. Model 2.}\\
\noindent \begin{tabular}{|c|c||c|cccc|cccc|} \hline
N$^o$ &$ b_1 , b_2 , b_3 , b_4 , b_5 , b_6 $&$SO_{hid}$&$ U(5)^I $
&$ U(3)^I $&$ U(5)^{II} $&$
U(3)^{II} $&$  Y_5^I $&$ Y_3^I $&$ Y_5^{II} $&$
Y_3^{II}$ \\ \hline \hline
1 & RNS & $6_1\ 2_2$ & 1&1&1&1&0&0&0&0 \\
  && $2_3\ 2_4$ & 1&1&1&1&0&0&0&0 \\
  &&& 5&1&$\bar 5$&1&1&0&--1&0 \\
  &0\ 0\ 4\ 1\ 0\ 0&&1&3&1&3&0&--1&0&--1 \\
  &0\ 0\ 8\ 1\ 0\ 0&&1&$\bar 3$&1&$\bar 3$&0&1&0&1 \\ \hline\hline
2 &0\ 1\ 0\ 0\ 0\ 0&&5&$\bar 3$&1&1&--3/2&--1/2&0&0 \\
  &&&1&$\bar 3$&1&1&5/2&--1/2&0&0 \\
  &0\ 3\ 0\ 0\ 0\ 0&&$\bar {10}$&1&1&1&1/2&3/2&0&0 \\ \hline
3 &0\ 1\ 10\ 0\ 0\ 0&&1&1&$\bar {10}$&3&0&0&1/2&1/2 \\
  &0\ 3\ 6\ 0\ 0\ 0&&1&1&5&1&0&0&--3/2&--3/2 \\
  &&&1&1&1&1&0&0&5/2&--3/2 \\ \hline
4 &0\ 2\ 3\ 0\ 0\ 0&$-_3\ \pm_4$&1&3&1&1&--5/4&--1/4&5/4&3/4 \\ \hline
5 &0\ 0\ 3\ 0\ 0\ 0&$+_3\ \pm_4$&1&1&$\bar 5$&1&--5/4&3/4&1/4&3/4 \\ \hline
6 &0\ 0\ 9\ 0\ 0\ 0&$+_3\ \pm_4$&1&1&5&1&5/4&--3/4&--1/4&--3/4 \\ \hline
7 &0\ 4\ 9\ 0\ 0\ 0&$-_3\ \pm_4$&1&$\bar 3$&1&1&5/4&1/4&--5/4&--3/4 \\ \hline
8,9 &0\ 5\ 0\ 1\ 0\ 1&$-_1\ \pm_3$&1&3&1&1&0&--1&0&0 \\
  &0\ 3\ 0\ 1\ 0\ 1&$+_1\ +_3$&5&1&1&1&1&0&0&0 \\
  &&$+_1\ -_3$&$\bar 5$&1&1&1&--1&0&0&0 \\
  &&$-_1\ +_3$&1&1&5&1&0&0&1&0 \\
  &&$-_1\ -_3$&1&1&$\bar 5$&1&0&0&--1&0 \\
  &0\ 5\ 8\ 1\ 0\ 1&$+_1\ +_3$&1&1&1&$\bar 3$&0&0&0&1 \\ \hline
10 &0\ 3\ 3\ 0\ 0\ 1&$+_1\ \pm_4$&1&1&1&1&--5/4&3/4&5/4&3/4 \\ \hline
11 &1\ 0\ 3\ 0\ 0\ 1&$\pm_2\ -_3$&1&1&5&1&--1/4&3/4&--5/4&--3/4 \\
   &1\ 2\ 11\ 0\ 0\ 1&$\pm_2\ -_3$&1&1&1&$\bar 3$&--5/4&3/4&--5/4&1/4 \\ \hline
12 &1\ 0\ 9\ 0\ 0\ 1&$\pm_2\ +_3$&$\bar 5$&1&1&1&1/4&--3/4&5/4&3/4 \\
   &1\ 4\ 9\ 0\ 0\ 1&$\pm_2\ +_3$&1&$\bar 3$&1&1&5/4&1/4&5/4&3/4 \\ \hline
13 &0\ 0\ 0\ 1\ 1\ 1&$\pm_2\ +_3$&1&1&1&1&0&--3/2&0&3/2 \\
   &0\ 2\ 0\ 1\ 1\ 1&$\pm_2\ -_3$&1&3&1&1&0&1/2&0&3/2 \\
   &0\ 2\ 8\ 1\ 1\ 1&$\pm_2\ -_3$&1&1&1&$\bar 3$&0&--3/2&0&--1/2 \\
   &0\ 4\ 8\ 1\ 1\ 1&$\pm_2\ +_3$&1&3&1&$\bar 3$&0&1/2&0&--1/2 \\
   &1\ 0\ 3\ 1\ 1\ 1&$+_1\ +_3$&1&1&1&1&5/4&3/4&--5/4&3/4 \\
   &1\ 0\ 9\ 1\ 1\ 1&$+_1\ +_3$&1&1&1&1&--5/4&--3/4&5/4&--3/4 \\
   &1\ 3\ 3\ 0\ 1\ 1&$-_1\ -_3$&1&1&1&1&--5/4&--3/4&--5/4&3/4 \\
   &1\ 3\ 9\ 0\ 1\ 1&$-_1\ +_3$&1&1&1&1&5/4&3/4&5/4&--3/4 \\ \hline
\end{tabular}

\subsubsection{ $SO(10) \times SU(3)\times U(1)$ Model 3.}
 As an illustration we can consider the GUST construction involving $SO(10)$ as
GUT gauge group.  We consider the set  consists of six vectors $B=
{b_1, b_2, b_3, b_4 \equiv S, b_5, b_6 }$ given in Table 7.\\

{\bf Table 7: Basis of the boundary conditions for the Model 3.}
\begin{center}
\begin{tabular}{|c||c|ccc||c|cc|}
\hline
Vectors &${\psi}_{1,2} $ & ${\chi}_{1,..,6}$ & ${y}_{1,...,6}$ &
${\omega}_{1,...,6}$
& ${\bar \varphi}_{1,...,12}$ &
${\Psi}_{1,...,8} $ &
${\Phi}_{1,...,8}$ \\ \hline
\hline
$b_1$ & $1 1 $ & $1 1 1 1 1 1$ &
$1 1 1 1 1 1 $ & $1 1 1 1 1 1$ &  $1^{12} $ & $1^8 $ & $1^8$ \\
$b_2$ & $1 1$ & $1 1 1 1 1 1$ &
$0 0 0 0 0 0$ &  $0 0 0 0 0 0 $ &  $ 0^{12} $ &
$1^5 {1/3}^3 $ & $0^8$  \\
$b_3$ & $1 1$ & $ 0 0 0 0 0 0 $ & $1 1 1 1 1 1 $ & $ 0 0 0 0 0 0 $ &
$0^8 1^4$ & $ 0^5 1^3 $ &  $ 0^5 1^3$ \\
$b_4 = S $ & $1 1$ &
$1 1 0 0 0 0 $ & $ 0 0 1 1 0 0 $ & $ 0 0 0 0 1 1 $ &
$ 0^{12} $ & $ 0^8 $ & $ 0^8 $ \\
$  b_5 $ & $ 1 1 $ & $ 1 1 1 1 1 1 $ &
$0 0 0 0 0 0 $ &  $0 0 0 0 0 0$ &  $ 0^{12} $ &
$ 0^8 $ & $ 1^5 {1/3}^3  $  \\
$  b_6 $ & $ 1 1 $ & $ 0 0  1 1  0 0 $ &
$ 1 1  0 0 0 0 $ &  $0 0  0 0 1 1$ &  $ 1^2 0^2 1^6 0^2 $ &
$ 1^8  $ & $ 0^8  $  \\
\hline \hline
\end {tabular}
\end{center}

GSO projections are given in Table 8.
It is interesting to note that in this model the horizontal gauge symmetry
$U(3)$ extends to $SU(4)$. Vector bosons which are needed for this appear
in sectors $2b_2\ (4b_2)$ and $2b_5\ (4b_5)$. For further breaking $SU(4)$
to $SU(3) \times U(1)$ we need an additional basis spin-vector.

So, the given model possesses gauge group
$G^{comp.}\times [SO(10)\times SU(4)]^2 $
and matter fields appear both in first and in second group symmetricaly.
Sectors $3b_2$ and $5b_2 +c.c.$ give the matter fields
$(\underline{16}, \underline{4}; \underline{1}, \underline{1})$
(first group) and sectors $3b_5$ and $5b_5 +c.c.$ give the matter fields
$(\underline{1}, \underline{1}; \underline{16}, \underline{4})$
(second group).

Of course for getting a realistic model we must add some basis vectors
which give addition GSO--projections.\\
{\bf Table 8:The choice of the GSO basis $\gamma [b_i, b_j]$. Model 3.
($i$ numbers rows and $j$ -- columns)}
\begin{center}
\begin{tabular}{|c||c|c|c|c|c|c|}
\hline
& $b_1$ & $b_2$ & $b_3$ & $b_4$ & $b_5$ & $b_6$ \\ \hline
\hline
$b_1$ & $0$ &    $1$ & $0$ & $0$ &    $1$ & $0$ \\
$b_2$ & $0$ &  $2/3$ & $1$ & $1$ &    $1$ & $1$ \\
$b_3$ & $0$ &    $1$ & $0$ & $1$ &    $1$ & $1$ \\
$b_4$ & $0$ &    $0$ & $0$ & $0$ &    $0$ & $0$ \\
$b_5$ & $0$ &    $1$ & $1$ & $1$ &  $2/3$ & $0$ \\
$b_6$ & $0$ &    $1$ & $0$ & $1$ &    $1$ & $1$ \\
\hline \hline
\end {tabular}
\end{center}

The condition of generation chirality in this model results in choice
of Higgs fields as a vector representations of SO(10)
 ($\underline{16}+\underline{\bar{16}}$ are absent). According to
conclusion (\ref{eq6}) the only Higgs fields $(\underline{10}, \underline{1};
\underline{10}, \underline{1})$ of $(SO(10)\times SU(4))^{\times 2}$
appear in model (from RNS--sector) which can be used for GUT gauge symmetry.

\newcommand{\thr}{\frac{1}{3}}
\newcommand{\tthr}{\frac{2}{3}}
\subsubsection{ $E_6 \times SU(3)$ tree generations model (Model 4).}
 This model illustrates a branch of $E_8$ breaking
$E_8\rightarrow E_6\times SU(3)$ and is an interesting result on a way to
obtain three generations with gauge horizontal symmetry. Basis of the boundary
conditions (see Table 9) is rather simple but there are some subtle points.
In \cite{Lop} the possible left parts of basis vectors were worked out,
see it for details. We just use the notation given in \cite{Lop}
( hat on left part means complex fermion, other fermions on the left
sector are real, all of the right movers are complex)
 and an example of commuting set of vectors.

{\bf Table 9: Basis of the boundary conditions for the Model 4.}

\footnotesize
\begin{center}
\begin{tabular}{|c||c|cc||c|cc|}
\hline
Vectors &${\psi}_{1,2} $ & ${\chi}_{1,..,9}$ &
${\omega}_{1,...,9}$
& ${\bar \varphi}_{1,...,6}$ &
${\Psi}_{1,...,8} $ &
${\Phi}_{1,...,8}$ \\ \hline
\hline
$b_1$ & $1 1 $ & $1^9$ &
$1^9$ & $1^{6} $ & $1^8 $ & $1^8$ \\
$b_2$ & $1 1$ & $\widehat\thr,1;-\widehat\tthr,0,0,\widehat\thr$ &
$\widehat\thr,1;-\widehat\tthr,0,0,\widehat\thr$ &  $ \tthr^3\:-\tthr^3$ &
$0^2\:-\tthr^6 $ & $1^2\:\thr^6$  \\
$b_3$ & $0 0$ & $0^9$ &
$0^9$ & $ 0^6$ &
$1^8$&$0^8$  \\
$b_4$ & $1 1$ & $\widehat{1},1;\widehat0,0,0,\widehat0$ &
$\widehat1,1;\widehat0,0,0,\widehat0$ &$0^6$ & $ 0^8 $ & $0^8$ \\
\hline \hline
\end {tabular}
\end{center}
\normalsize

A construction of an $E_6\times SU(3)$ group caused us to use rational
for left boundary conditions. It seems that it is the only way to
obtain such a gauge group with appropriate matter contents.

The model has $N=2$ SUSY. We can also construct model with
$N=0$ but according to \cite{Lop} using vectors that can give rise
to $E_6\times SU(3)$ (with realistic matter fields)
one cannot obtain $N=1$ SUSY.

{\bf Table 10: The choice of the GSO basis $\gamma [b_i, b_j]$. Model 4.
($i$ numbers rows and $j$ -- columns).}

\footnotesize
\begin{center}
\begin{tabular}{|c||c|c|c|c|}
\hline
& $b_1$ & $b_2$ & $b_3$ & $b_4$   \\ \hline
\hline
$b_1$ & $0$ &  $1/3$ & $1  $ & $1$ \\
$b_2$ & $1$ &  $  1$ & $1  $ & $1$ \\
$b_3$ & $1$ &  $1$ & $1  $ & $0$ \\
$b_4$ & $1$ &  $1/3$ & $1  $ & $1$ \\
\hline \hline
\end {tabular}
\end{center}
\normalsize

Let us give a brief review of the model contents. First notice
that all superpartners of states in sector $\alpha$ are found in
sector $\alpha+b_4$ as in all previous models. Although the same sector
may contain, say, matter fields and gauginos simultaneously.

The observable gauge group
 $(SU(3)^I_H\times E^I_6)\times (SU(3)^{II}_H\times E^{II}_6)$ and hidden
group $SU(6)\times U(1)$ are rising up from sectors NS, $b_3$ and $3b_2+b_4$.
Matter fields in representations $({\bf 3},{\bf 27})+
(\overline{\bf 3},\overline{\bf 27})$
for each $SU(3)_H\times E_6$ group are found in sectors $3b_2$, $b_3+b_4$
and $b_4$. Also there are some interesting states in sectors $b_2,\:b_2+b3,\:
2b_2+b_3+b_4,\:2b_2+b_4$ and $5b_2,\:5b_2+b3,\:4b_2+b_3+b_4,\:4b_2+b_4$
that form representations $(\overline{\bf 3},{\bf 3})$
and $({\bf 3},\overline{\bf 3})$
of the $SU(3)^I_H\times SU(3)^{II}_H$ group. This states are singlets
under both $E_6$ groups.

We suppose that the model permits further breaking of $E_6$ down to other
grand unification groups, but problem with breaking supersymmetry ${N=2}
\rightarrow N=1$ is a great obstacle on this way.

\subsection{Gauge Symmetry Breaking and GUST Spectrum}
Let us consider the Model 1 in details.
In the Model 1 there exists a
possibility to break the GUST group $(U(5)\times U(3))^I
\times (U(5)\times U(3))^{II}$
down to the symmetric group by the ordinary Higgs mechanism \cite{13'}:
\begin{equation}\label{eq18}
G^I\times G^{II} \rightarrow {G}^{symm}
\rightarrow ...
\end{equation}
To achieve such breaking one can use nonzero vacuum expectation values of the
tensor Higgs fields (see Table \ref{tabl3'}, row No 1),
contained in the $2b_2 + 2(6)b_5 (+S)$ sectors which transform
under the $(SU(5)\times U(1)\times SU(3)\times U(1))^{symm}$ group in the
following way:
\begin{eqnarray}\label{eq19}
\begin{array}{lll}
(\underline 5,\underline 1;\underline 5,\underline 1)_{(-1,0;-1,0)}
 &\rightarrow  &  (\underline {24},\underline 1)_{(0,0)}  +
(\underline 1,\underline 1)_{(0,0)}; \\
(\underline 1,\underline 3;\underline 1,\underline 3)_{(0,1;0,1)}\,
 &\rightarrow  &
(\underline 1,\underline 8)_{(0,0)}  +  (\underline 1,\underline 1)_{(0,0)},
\end{array}
\end{eqnarray}

\begin{eqnarray}\label{eq20}
\begin{array}{lll}
(\underline 5,\underline 1;\underline 1,\underline 3)_{(-1,0;0,1)}
 &\rightarrow  &
 (\underline {\bar 5},\underline 3)_{(1,1)}; \\
 (\underline 1,\underline 3;\underline 5,\underline 1)_{(0,1;-1,0)}
 &\rightarrow  &
(\underline 5,\underline {\bar3})_{(-1,-1)}.
\end{array}
\end{eqnarray}

The diagonal vacuum expectation values for the Higgs fields
(\ref{eq19}) break the
the GUST group $(U(5)\times U(3))^I \times (U(5)\times U(3))^{II}$ down to
the "skew"-symmetric group with the generators $\triangle_{symm}$ of the form:
\begin{equation}\label{eq21}
\triangle_{symm} (t) = -t^* \times 1 + 1\times t,
\end{equation}
The corresponding  hypercharge of the symmetric group reads:
\begin{equation}\label{eq22}
\bar Y = \tilde {Y}^{II} - \tilde {Y}^{I}.
\end{equation}
Similarily, for the electromagnetic charge we get:
\begin{eqnarray}\label{eq23}
Q_{em} &=& Q^{II} - Q^I = \nonumber\\
&=& (T^{II}_5 - T^{I}_5) + \frac{2}{5}(\tilde Y^{II}_5 - \tilde Y^{I}_5) =
\bar T_5 + \frac{2}{5}\bar Y_5,
\end{eqnarray}
where $ T_5 = diag (\frac{1}{15},\frac{1}{15},\frac{1}{15}, \frac{2}{5},
-\frac{3}{5})$. Note, that this charge quantization does not lead to exotic
states with fractional electromagnetic charges \\(e.g. $Q_{em} =\pm 1/2,
\pm 1/6$).

Thus, in the breaking scheme (\ref{eq21})
it is possible to avoid colour singlet
states with fractional electromagnetic charges, to achieve
desired GUT breaking and moreover to get the usual value for the weak mixing
angle at the unification scale (see (\ref{sinW})).

Adjoint  representations which appear on the $\it rhs$ of
(\ref{eq19}) can be used for
further breaking of the symmetric group. This can lead to the final physical
symmetry
\begin{equation}\label{eq24}
(SU(3^c)\times SU(2_{EW})\times U(1_Y)\times U(1)^{'})
\times (SU(3_H)\times U(1_H))
\end{equation}
with  low-energy gauge symmetry of the quark -- lepton generations with an
additional $U(1)^{'}$--factor.

Note, that using the same  Higgs fields as in (\ref{eq19}), there exists also
another, interesting way of breaking the $G^I \times G^{II}$ gauge symmetry:
\begin{eqnarray}\label{eq25}
 G^I\times G^{II} \rightarrow SU(3^c) \times SU(2)^I_{EW}
\times SU(2)^{II}_{EW} \times U(1_{\bar Y}) \times \nonumber\\
\times SU(3_H)^I \times SU(3_H)^{II} \times U(1_{{\bar Y}_H})
\rightarrow ....
\end{eqnarray}
It is attractive because it naturally solves the Higgs doublet--triplet mass
splitting problem with rather low energy scale of GUST symmetry breaking \
\cite{34'}.

In turn, the Higgs fields ${\hat h}_{(\Gamma, N)}$ from the NS sector
\begin{eqnarray}\label{eq26}
(\underline 5, \underline {\bar 3})_{(-1,-1)} +
(\underline {\bar 5}, \underline 3)_{(1,1)}
\end{eqnarray}
originates from N=2 SUSY vector representation  $\underline{63}$
of $SU(8)^{I}$ (or $SU(8)^{II}$) by applying the $b_5$ GSO projection
 (see Fig. \ref{fig2}).
These Higgs fields (and fields (\ref{eq20})) can be used for constructing
chiral fermion (see Table \ref{tabl3'}, row No 2) mass matrices.

The $b$ spin boundary conditions (Tabl.\ref{tabl1})
generate chiral matter and Higgs
fields with the $GUST$ gauge symmetry $G_{comp}\times (G^I\times G^{II})_{obs}$
(where $G_{comp} = {U(1)}^3\times SO(6) $ and $G^{I,II}$ have been already
defined). The chiral matter spectrum, which we denote
${\hat \Psi}_{(\Gamma, N)}$   with ($\Gamma = \underline 1,\underline {\bar 5},
\underline{10};   N=\underline3, \underline1$),  consists of
$N_g = 3_H + 1_H $   families. See Table \ref{tabl3'}, row No 2 for
the $((SU(5)\times U(1))\times (SU(3)\times U(1))_H)^{symm}$ quantum numbers.

The $SU(3_H)$ anomalies of the matter fields (row No 2) are naturally canceled
by
the chiral "horizontal" superfields  forming two sets:
${\hat \Psi}^H_{(1,N;1,N)}$ and ${\hat \Phi}^H_{(1, N;1, N)}$,
 $\Gamma = \underline 1$,  $N = \underline 1, \,  \underline 3$,
(with both ${SO(2)}_2$ chiralities, see Table \ref{tabl3'}, row No 3, 4).

The horizontal fields (No 3, 4) compensate all  $SU(3)^{I}$ anomalies
introduced by the chiral matter spectrum (No 2) of the  $(U(5)\times U(3))^{I}$
group (due to $b_6$ GSO  projection the chiral fields of
the $(U(5)\times U(3))^{II}$ group disappear from the final string spectrum).
Performing the decomposition of fields (No 3, 4) under
$(SU(5)\times SU(3))^{symm}$
we get (among other) three "horizontal" fields:
\begin{eqnarray}\label{eq27}
(\underline 1,\underline {\bar 3})_{(0,-1)},
(\underline 1,\underline 1)_{(0,-3)},
(\underline 1,\underline {\bar 6})_{(0,1)},
\end{eqnarray}
coming from
${\hat \Psi}^H_{(\underline1,\underline3;\underline1,\underline1)}$, (or
${\hat \Psi}^H_{(\underline1,\underline1;\underline1,\underline {\bar3})}$),
${\hat \Psi}^H_{(\underline1,\underline1;\underline1,\underline1)}$ and
${\hat \Psi}^H_{(\underline1,\underline3;\underline1,\underline {\bar3})}$
respectively  which
make the low energy spectrum of the resulting model
(\ref{eq25}) ${SU(3_H)}^{symm}$-
anomaly free. The other fields arising from
(rows No 3, 4, Table \ref{tabl3'})  form
anomaly-free representations of $(SU(3_H) \times U(1_H))^{symm}$:
\begin{eqnarray}\label{eq28}
2(\underline 1,\underline 1)_{(0,0)},
(\underline 1,\underline {\bar3})_{(0,-1(2))} +
(\underline 1,\underline 3)_{(0,1(-2))},
(\underline 1,\underline {8})_{(0,0)}.
\end{eqnarray}

The  superfields  ${\hat \phi}_{(\Gamma, N)} + h.c.$, where
($\Gamma = \underline 1, \underline 5$; $N = \underline 1,\underline 3$), from
the Table \ref{tabl3'}, row No 5
forming
representations of $(U(5) \times U(3))^{I.II}$ have either $Q^I$ or $Q^{II}$
exotic fractional charges.
Because of the strong $G^{comp}$ gauge forces these fields may
develop the double scalar condensate $ {<\hat \phi \hat \phi>}$, which can also
serve for $U(5)\times U(5)$ gauge symmetry breaking. For example, the composite
condensate ${<{\hat \phi}_{(5,1;1,1)} {\hat \phi}_{(1,1;\bar 5,1)}>}$ can
break the $U(5)\times U(5)$ gauge symmetry down to the symmetric diagonal
subgroup with generators of the form
\begin{eqnarray}\label{eq29}
{\triangle}_{symm} (t) = t \times 1 + 1 \times t,
\end{eqnarray}
so for the electromagnetic charges  we would have the form
\begin{eqnarray}\label{eq30}
Q_{em} = Q^{II} + Q^I.
\end{eqnarray}
leading again to no exotic, fractionally charged states
in the low-energy string spectrum.

The superfields which transform nontrivially under the compactified group
$G^{comp} = SO(6)\times {SO(2)}^{\times 3}$,
(denoted as $\hat{\sigma}+ h.c.$),
and which are singlets of $(SU(5)\times SU(3))\times
(SU(5)\times SU(3))$, arise
in three sectors, see Table \ref{tabl3'}, row No 6.
The superfields $\hat\sigma$ form the spinor representations $\underline4+
\underline {\bar4}$ of $SO(6)$ and they are also spinors of one of the $SO(2)$
groups. They have following hypercharges ${\tilde Y}_5^{I,II}$,
${\tilde Y}_3^{I,II}$:
\begin{eqnarray}\label{eq31}
\tilde Y = (5/4,\mp 3/4;  5/4,\mp 3/4),
\tilde Y = (5/4, 3/4; -5/4,-3/4).
\end{eqnarray}
With respect to the diagonal
$G^{symm}$ group with generators given by (\ref{eq21}) or
(\ref{eq29}), the fields $\hat {\sigma}$ from sets a), b)  or the set c),
are of zero hypercharges and can, therefore, be used for breaking the
$SO(6)\times {SO(2)}^{\times 3}$ group.

Note, that for the fields $\hat {\phi}$ and for the fields $\hat {\sigma}$
any other electromagnetic charge quantization diffrent than (\ref{eq23}) or
(\ref{eq30}) would lead to "quarks" and "leptons"
with the exotic fractional charges, for example,
for the $\underline 5$- and $\underline 1$- multiplets according to the values
of hypercharges (see eqs.\ref{eq31})the generator
$Q^{II}$  (or $Q^{I}$) has the
eigenvalues \\
$(\pm 1/6,\pm 1/6,\pm 1/6,\pm 1/2,\mp1/2)$ or $\pm 1/2$, respectively.

Scheme of the breaking of the gauge group to the symmetric subgroup,
which is like scheme of the Model 1, works for the Model 2 too.
In this case vector-like multiplets
$(\underline5,\ \underline1;\ \underline{\bar 5},\ \underline1)$
from RNS--sector and
$(\underline1,\ \underline3;\ \underline1,\ \underline3)$
from $4b_3$ $(8b_3)$ play the role of Higgs fields.
Then generators of the symmetric subgroup and electromagnetic
charges of particles are determined by formulas:
\begin{eqnarray}
\Delta^{(5)}_{sym}&=&t^{(5)}\times 1\ \oplus\ 1\times t^{(5)} \nonumber \\
\Delta^{(3)}_{sym}&=&(-t^{(3)})\times 1\ \oplus\ 1\times t^{(3)} \nonumber \\
Q_{em}=t^{(5)}_5-2/5\,Y^5&,&\mbox{where}
\ t^{(5)}_5=(1/15,\ 1/15,\ 1/15,\ 2/5,\ -3/5) \label{eq}
\end{eqnarray}

After this symmetry breaking matter fields (see Table 6)
rows No 2, 3) standardly for flip models take place in representations
of the $U(5)-$group and form four generations
$(\underline1 +\underline5 +\underline{\bar{10}};\ \underline{\bar3}
+\underline1)_{sym}$.
And Higgs fields form adjoint representation of the symmetric group,
similar to Model 1, which is necessary for breaking of the gauge
group to the Standard group. Besides, quantization of the
electromagnetic charge according to the formula (\ref{eq})
does not lead to appearance of exotic charges in lowenergy
spectrum for this model too.

\subsection{Superpotential and Non-renormalizable Contributions}

The ability to correctly describe the fermion masses and mixings will, of
course, constitute the decisive criterion for selection of a model of this
kind. Therefore, within our approach one has to
\begin{enumerate}
\item study the possible nature of the $G_H$ horizontal gauge symmetry
($N_g=3_H$ or $3_H+1_H$),
\item investigate the possible cases for
$G_H$-quantum numbers for quarks (anti-quarks) and leptons (anti-leptons),
i.e. whether one can obtain vector-like or axial-like structure (or even
chiral $G_{HL}\times G_{HR}$ structure) for the horizontal interactions.
\item the structure of the sector of the matter fields which are needed for
the $SU(3)_H$ anomaly cancelation (chiral neutral "horizontal" or "mirror"
fermions),
\item write down all possible renormalizable and relevant non-renormalizable
contributions to the superpotential $W$ and their consequences for fermion mass
matrices.
\end{enumerate}
All these questions are currently under investigation.
Here we restrict ourselves to some general remarks only.

With the chiral matter and "horizontal" Higgs fields available in the Model 1
constructed in this paper, the possible form of the renormalizable (trilinear)
part of the superpotential responsible for fermion mass matrices is well
restricted by the gauge symmetry:
\begin{eqnarray}\label{eq32}
W_1&=& g\sqrt{2} \biggl[ {\hat \Psi}_{(1,3)}
{\hat \Psi}_{({\bar 5},1)}
{\hat h}_{(5,{\bar 3})} +
 {\hat \Psi}_{(1,1)} {\hat \Psi}_{({\bar 5},3)}
{\hat h}_{(5,{\bar 3})} + \nonumber\\
&+& {\hat \Psi}_{(10,3)} {\hat \Psi}_{({\bar 5},3)}
{\hat h}_{({\bar5},3)} +
{\hat \Psi}_{(10,3)} {\hat \Psi}_{(10,1)}
{\hat h}_{(5,{\bar 3})} \biggr]
\end{eqnarray}
{}From the above form of the Yukawa couplings
follows that two (chiral) generations
have to be very light (comparing to $M_W$ scale).
The construction of realistic quarks and leptons mass matrices depends, of
course, on the nature of the horizontal interactions.
In the construction described in Sec.\ref{sec3}
there is a freedom of choosing spin
boundary conditions for $N_{LR}$=12 left and right fermions in the basis
vectors $b_3$, $b_5$, $b_6$,...,
which in the Ramond sector $2b_2$, may yield another
Higgs fields, denoted as ${\tilde h}_{(\Gamma,N)}$ and transforming as
$(\underline 5,\underline 3)_{(-1,1)}$ +
$(\underline {\bar 5},\underline {\bar 3})_{(1,-1)}$
$\subset \underline {28}$ + $\underline {\bar {28}}$ of $SU(8)$.
Using these Higgs fields we get the following alternative form of the
renormalizable part of the superpotential $W$:
\begin{eqnarray}\label{eq33}
{W'}_1&=& g\sqrt{2} \biggl[ {\hat \Psi}_{(1,3)}
{\hat \Psi}_{({\bar 5},3)} {\tilde h}_{(5,3)} +
{\hat \Psi}_{(10,1)} {\hat \Psi}_{({\bar 5},3)}
{\tilde h}_{({\bar 5},{\bar 3})} + \nonumber\\
&+& {\hat \Psi}_{(10,3)} {\hat \Psi}_{(10,3)} {\tilde h}_{(5,3)} +
{\hat \Psi}_{(10,3)} {\hat \Psi}_{({\bar 5},1)}
{\tilde h}_{({\bar5},{\bar3})} \biggr]
\end{eqnarray}
To construct the realistic fermion mass matrices one has to also use the
Higgs fields  (\ref{eq19}, \ref{eq20}) and (Table \ref{tabl3'}, No 5)
and also to take into account
all relevant non-renormalizable contributions \cite{20'}.

The Higgs fields (\ref{eq19}) can be used for constructing Yukawa couplings
of the horizontal superfields (No 3 and 4). The most general contribution
of these fields to the superpotential is:
\begin{eqnarray}\label{eq34}
W_2&=& g\sqrt{2} \biggl[
{\hat \Phi}^H_{(1,1;1,\bar3)}{\hat \Phi}^H_{(1,\bar 3;1,1)}
{\hat \Phi}_{(1,3;1,3)} +
{\hat \Phi}^H_{(1,1;1,1)}{\hat \Phi}^H_{(1,\bar 3;1,\bar 3)}
{\hat \Phi}_{(1,3;1,3)} + \nonumber\\
&+& {\hat \Phi}^H_{(1,\bar 3;1,\bar 3)}{\hat \Phi}^H_{(1,\bar 3;1,\bar 3)}
{\hat \Phi}_{(1,\bar 3;1,\bar 3)} +
 {\hat \Psi}^H_{(1,3;1,1)}{\hat \Psi}^H_{(1,3;1,\bar 3)}
{\hat \Phi}_{(1,3;1,3)} +\nonumber\\
&+&{\hat \Psi}^H_{(1,1;1,\bar 3)}
{\hat \Psi}^H_{(1,3;1,\bar 3)}
{\hat \Phi}_{(1,\bar 3;1,\bar 3)}  \biggr]
\end{eqnarray}
{}From this expression it follows that some of the horizontal fields in
(\ref{eq28}) (No 3, 4) remain massless at the tree-level.
This is a remarkable prediction: fields (\ref{eq28}) interact with the ordinary
chiral
matter fields only through the $U(1_H)$ and $SU(3_H)$
gauge boson and therefore are very interesting in the context of the
experimental searches for the new gauge bosons.

The superfields ${\hat \Phi}^H_{(1,3;1,1)}$
and ${\hat\Psi}^H_{(1,\bar3;1,1)}$ (see No 3, 4)
can be used to construct the non-renormalizable contributions to the
superpotential $W$. For example, the term
\begin{eqnarray}\label{eq35}
\Delta W_1 =  \frac {cg^3}{M_{Pl}^2} {\Psi}_{(10,1)} {\Psi}_{(10,1)}
{\Phi}_{(1,3;5,1)} {\hat \Phi}^H_{(1,3;1,1)}{\hat \Phi}^H_{(1,3;1,1)}
\end{eqnarray}
can give contribution to the mass to the fourth generation down--type quark
($c = {\cal O}(1)$, see \cite{20'}). To get a reasonable value of the mass for
this quark we must arrange for the $SU(3_H)^I$ gauge symmetry breaking at the
energy scale near the Planck scale, i.e. $ < {\hat \Phi}^H_{(1,3;1,1)}> $ =
$<{\hat \Phi}^H_{(1,\bar 3;1,1)}> \sim M_{Pl}$. In this case one can get the
$SU(3_H)^{II}$ -family gauge group with a low energy breaking symmetry scale.
Finally, we remark that the Higgs sector of our GUST allows for conservation of
the $G_H$ gauge family symmetry down to the low energies ($\sim {\cal O}(1TeV)$
\cite{9'}). Thus we can expect at this energy region new interesting physics
(new gauge bosons, new chiral matter fermions, superweak-like CP--violation
in $K$,- $B$,- $D$-meson decays with ${\delta}_{KM} < {10}^{-4}$ \cite{9'}).

\noindent
\begin{figure}[p]
\caption{Supersymmetry breaking.}
\label{fig2}
\bigskip
\centering\begin{tabular}{ccccccc} \bf \large
 N=2  SUSY : & \it V&=&(1,$\frac{1}{2}$)&+&($\frac{1}{2}$,0)&$SU(8)$ \\
$\Downarrow$ & & & & & & $\Downarrow$ \\
\bf\large N=1  SUSY : &  $V_{N=2}$ &$\rightarrow$&$V_{N=1}$&+&$S_{N=1}$&
  $\qquad SU(5)\times SU(3)\times U(1)$ \\
\end{tabular}

\bigskip

\raggedleft\begin{tabular}{|p{35mm}||c|c|c|c||}
\hline
   & \large \bf \hspace*{9mm}J=1\hspace*{9mm}
   & \large \bf \hspace*{8.5mm}J=1/2\hspace*{8mm}
   & \large \bf \hspace*{4.5mm}J=1/2\hspace*{4mm}
   & \large \bf \hspace*{4mm}J=0\hspace*{4mm} \\
\hline
{\large \bf $E_{\it vac}=[-1/2;-1]$  NS sector } & (63)
& --- & --- & (63) \\ \hline
{\large \bf $E_{\it vac}=[0;-1]$ SUSY sector }
& --- & $ (63)\times 2 $ & $ (63)\times 2 $ & --- \\
\hline \hline
\multicolumn{1}{|c|}{ } &
\multicolumn{4}{|c||}{ \large\bf Gauge multiplets } \\
\hline \hline
\end{tabular}

\bigskip
\begin{flushleft}
\hspace{60mm} $ {\bf \Downarrow} \quad b_5 $ projection GSO
\end{flushleft}

\bigskip

\raggedleft\begin{tabular}{|p{35mm}||c|c||c|c||}
\hline
   & \large \bf J=1
   & \large \bf J=1/2
   & \large \bf J=1/2
   & \large \bf J=0 \\
\hline
{\large \bf $E_{\it vac}=[-1/2;-1]$  NS sector } &
 \small (24,1)+(1,1)+(1,8) & --- & --- & \small (5,\=3)+(\=5,3) \\ \hline
{\large \bf $E_{\it vac}=[0;-1]$ SUSY sector } &
 --- & \scriptsize ((24,1)+(1,1)+(1,8))$\times$ 2  &
       \scriptsize ((5,\=3)+(\=5,3))$\times$ 2  & --- \\
\hline \hline
\multicolumn{1}{|c|}{} &
\multicolumn{2}{|c|}{ \large\bf Gauge multiplets } &
\multicolumn{2}{|c||}{ \large\bf Higgs multiplets } \\
\hline \hline
\end{tabular}

\end{figure}

\section{Low Energy Construction of the \SUTH{} model }

\subsection{The spontaneous breaking of  SUSY \SUTH{}
horizontal gauge symmetry.}

Since the expected scale of the horizontal symmetry breaking is sufficiently
large:
$M_H>>M_{EW}$ , $M_H>>M_{SUSY}$ (where $M_{EW}$ is the scale of the electroweak
symmetry breaking,
and $M_{SUSY}$ is the value of the splitting into ordinary particles and their
superpartners),
it is reasonable to search for the SUSY-preserving stationary vacuum solutions.

Let us construct the gauge invariant superpotential $P$ of  Lagrangian
(\ref{2.1}).
With the fields given in Table 11, the most general superpotential will
have the form
\begin{eqnarray}
P&=& \lambda_0 \biggl[\; {{1}\over {3}}Tr\bigl({\hat \Phi}^3 \bigr)+
{{1}\over {2}}
M_ITr\bigl({\hat \Phi}^2 \bigr)\;\biggr]+
\lambda_1 \biggl[\; \eta {\hat \Phi }\xi
 +M'\eta \xi\; \biggr]
+ \lambda_2 Tr\bigl(\hat h {\hat \Phi}\hat H \bigr)+\nonumber\\
&+& \mbox { (Yukawa couplings)} +
\mbox { ( Majorana terms $\nu^c$ ),} \label{2.3}
\end{eqnarray}
where  Yukawa Couplings could be constructed, for example,
using the Higgs fields, H and h, transforming under $SU(3)_H \times SU(2_L)$,
like (8,2):
\begin{eqnarray}
P_Y=\lambda_3 Q{\hat H}d^c + \lambda_4 L{\hat H}e^c +
\lambda_5 Q{\hat h}u^c.
\end{eqnarray}
Also, one can consider another types of superpotential $P_Y$, using
the  Higgs fields from Table 11.

{\bf Table 11.} The Higgs Superfields with their
 $SU(3_H), SU(3)_C,\ SU(2)_L,\ U(1)_Y$ (and possible
$U(1)_H$- factor) Quantum Numbers
\begin{center}
\begin{tabular}{||c|ccccc|} \hline
&H&C&L&Y&$Y_H$ \\ \hline
$\Phi$& 8 & 1 & 1 & 0 & 0 \\
H & 8 & 1 & 2 & $-1/2$ & $-y_{H1}$ \\
h & 8 & 1 & 2 & 1/2 &$y_{H1}$ \\
$\xi$ & $\bar 3$ & 1 & 1 & 0 & 0 \\
$\eta$ & 3 & 1 & 1 & 0 & 0 \\
$Y$&$\bar 3$&  1&  2&  1/2 &$ -y_{H2}$ \\
$X$&$ 3$&  1&  2&$ -1/2$& $y_{H2}$   \\
$\kappa_1$&1&  1&  1 &  0 &$ -y_{H3}$ \\
$\kappa_2$&      1 & 1 & 1 &  0 & $y_{H3}$    \\ \hline
\end{tabular}
\end{center}
Note, that Higgs fields $X$ and $Y$ are very important in
models with forth SU(3$)_H$-singlet generation.

The spontaneous horizontal symmetry breaking may be constructed via
different scenarios -both with intermediate scale, and without it:

\begin{eqnarray}
&&(i)\ \ \ \ \ {SU(3)_H} \stackrel{M_I}{\longrightarrow}
{SU(2)_H \times U(1)_H} \stackrel{M_H}{\longrightarrow}
{c.b.}  \nonumber\\
&&(ii)\ \ \ \ {SU(3)_H} \stackrel{M_I}{\longrightarrow}
{U(1)_H\times U(1)_H}
\stackrel {M_H}{\longrightarrow} {c.b.} \nonumber\\
&&(iii)\ \ \ \ {SU(3)_H} \stackrel {M_I}{\longrightarrow} {U(1)_H}
\stackrel {M_H}{\longrightarrow} {c.b.} \nonumber\\
&&(iv)\ \ \ \ {SU(3)_H} \stackrel {M_{H_0}}{\longrightarrow}
\mbox{complet breaking.} \label{2.4}
\end{eqnarray}

If we assume that the soft breaking mass parameters
in formula (\ref{2.2}) should not be more than
0(1 TeV), then the soft breaking terms on the scale $M_I$ of
the \SUTH{}-              intermediate breaking may be neglected, and
it is possible to go on working in the approximation
of conserved SUSY. The SUSY preserving stationary vacuum solutions
are degenerate in the models with global SUSY.
In the construction of the stationary solutions,
only the following contributions of the scalar potential are taken
into account:
\begin{eqnarray}
V &=& \sum_i |F_i|^2+\sum_{a} |D^{a}|^2=V_F+V_D \geq 0 \label{2.5}\\
\mbox {where    }V_F &=& \sum {\biggl|{{\partial P_F}\over {\partial F_i}}
\biggr|}^2=
{\biggl|{{\partial P_F}\over {\partial F_{\Phi^a}}}\biggr|}^2+
{\biggl|{{\partial P_F}\over {\partial F_{\xi_i}}}\biggr|}^2+
{\biggl|{{\partial P_F}\over {\partial F_{\eta_i}}}\biggr|}^2\label{2.6}
\end{eqnarray}

The case $<V>=0$
of supersymmetric vacuum can be realized within different gauge
scenarios (\ref{2.4}).
By switching on the SUGRA, the vanishing scalar potential is
no more required to conserve the supersymmetry with the necessity.
Hence, different gauge
breaking scenarios (\ref{2.4}) do not result in obligatory
vacuum degeneracy, as in the case of the global SUSY version.
Let us write down each of the terms of formula (\ref{2.6}):
\begin{eqnarray}
P_F(\Phi ,\xi ,\eta )&=& \lambda_0{\biggl[\;{\frac{i}{4\times 3}}\;f^{abc}
\Phi^a \Phi^b \Phi^c+{\frac{1}{4\times 3}}
\;d^{abc}\Phi^a \Phi^b \Phi^c+{\frac{1}
{4}}\;M_I\Phi^c \Phi^c\; \biggr]}_F+\nonumber\\
&+& \lambda_1 \biggl[\; \eta_i \;{(T^c)_j^i}\;
\xi^j \Phi^c+M'\eta_i \xi^i\; \biggr]_F+ \label{2.7}\\
&+& \lambda_2 \biggl[\;{\frac{i}{4}}f^{abc}\; h^a_i \Phi^b H^c_j \epsilon^{ij}+
\frac{d^{abc}}{4} \; h^a_i \Phi^b H^c_j \epsilon^{ij}\; \biggr]_F+h.c.\nonumber
\end{eqnarray}
The contribution of $D$-terms into the scalar potential will be :
\begin{eqnarray}
V_D&=&g_H^2|\eta^+ T^a\eta -\xi^+ T^a\xi +i/2\ f^{abc} \Phi^b{\Phi^c}^++
i/2\ f^{abc} h^b{h^c}^++i/2\ f^{abc} H^b{H^c}^+|^2 \nonumber\\
&+&g_2^2|h^+ \tau^i /2\ h +\ H^+ \tau^i /2\ H|^2+
(g')^2 | 1/2\ h^+ h -\ 1/2\  H^+ H |^2 \label{2.8}
\end{eqnarray}
The SUSY-preserving condition for scalar potential (\ref{2.5}) is
determined by the flat $F_i-$
and $D^{a}$ directions: $<F_i>_0=<D^{a}>_0=0$.
It is possible to remove the degeneracy of the
supersymmetric vacuum solutions taking into account the interaction with
supergravity, which was endeavored in SUSY GUT's, e.g. in the $SU(5)$ one
\cite{31'}
$(SU(5)\rightarrow SU(5),$
$SU(4)\times U(1),$ $SU(3) \times SU(2) \times U(1))$.

The horizontal symmetry spontaneous breaking to
the intermediate subgroups in the
first three cases of (\ref{2.4})
can be realized, using the scalar components of the
chiral complex superfields $\Phi$,
which are singlet under the standard gauge group.
The $\Phi$-superfield transforms as
the adjoint representation of $SU(3)_H$. The
intermediate scale $M_I$ can be sufficiently large: $M_I > 10^5-10^6$GeV.
 The complete breaking of the remnant symmetry group $V_H$ on the scale
$M_H$ will  occur
due to the nonvanishing VEV's of the
scalars from the chiral superfields $\eta (3_H)$
and $\xi(\bar 3_H)$. The $V_{min}$, again, corresponds to the flat directions:
$<F_{\eta,\xi}>_0=0$. The version (iv) corresponds to the minimum of the scalar
potential in the case when  $<\Phi>_0=0$.

As for the electroweak breaking,
it is due to the VEV's of the fields $h$ and $H$,
providing masses for quarks and leptons.
Note that VEV's of the fields $h$ and $H$
must be of the order of $M_W$ as they
determine the quark and lepton mass matrices. On the
other hand, the masses of physical Higgs fields $h$ and $H$, which mix
generations, must be some orders higher than $M_W$, so as not to contradict the
experimental restrictions on FCNC. As a careful search for the Higgs
potential shows, this is the picture that can be attained.

\subsection{The intermediate horizontal symmetry breaking }

As  noted in the previous Section, the spontaneous
horizontal gauge symmetry breaking takes place when the
 fields $\phi,\  \eta$ and $\xi$ get  nonvanishing VEVs. We are
interested in the possibility of realizing the structure, when some of the
 horizontal gauge bosons (and the corresponding gauginos) may have
relatively small masses $(M_H \sim 1-10 \mbox{TeV})$ \cite{9'}.
Our consideration of the family symmetry breaking
will be done in two steps. To this end, we look for the
SUSY stationary vacuum solutions, such as $<\Phi>_{0}\ \gg\ <\eta>_{0},
<\xi>_{0}$.
So, the degeneracy of the corresponding $H$-gauge bosons is assumed
near one or two scales.
The complete breaking of the \SUTH{}- group corresponds to the "condensation"
 of all eight bosons near the $M_H$ scale. For intermediate \SUTH{}-
breakings, some of the gauge massive superfields will have the masses around
the scale $M_I$, while the other superfields
from the remnant symmetry group will be condensed on the scale
$M_H(M_H << M_I)$.
We will analyze several subgroups of \SUTH{}- and check if the low scale $M_H$
is consistent with the experimental data for these models. Such  analysis
will allow us to get a deeper insight into the dynamics of horizontal forces
 and investigate the effects of their compensation, especially in pure leptonic
and pure quark processes. At first stage, due to the nonvanishing VEV
of $\Phi$, the horizontal symmetry group breaks down to some subgroup $V$
satisfying $[V,\ <\Phi>_0]=0$. At the second stage, the remnant group
$V$ is broken down completely, as fields $\eta$ and $\xi$ will acquire
nonzero VEVs. Let us consider several cases of this breaking.

Case (i): $V=SU(2)_H\times U(1)_{8H}$.
As has already been mentioned, in the gauge model with
the global SUSY stationary supersymmetry conserving vacuum solutions  are
degenerate: $V_{min}=0$. Let us recall that the superinvariance condition
for the model on the scale $M_I$ requires the existence of flat $D^a-$ and
$F_{\Phi}^a-$ directions: ${<D^a>}_0={<F^a_{\Phi}>}_0=0\ \  (a=1,2,3,8)$.
Equations (\ref{2.6}-\ref{2.8}) give the following form of these constraints:

\begin{eqnarray}
1/2\ d^{abc}(\Phi_1^a \Phi_1^b -\Phi_2^a \Phi_2^b)+M_I\Phi_1^c =0\nonumber\\
& & (<F_{\Phi}>_0=0) \label{3.1}\\
d^{abc}\Phi_1^a\Phi_2^b +M_I\Phi_2^c =0 \nonumber\\
\,\,\,\nonumber\\
 if^{abc}\Phi^b{\Phi^c}^+ =0  & & (<D^a>_0=0), \label{3.2}
\end{eqnarray}

where $\Phi^a=\Phi^a_1 + i\Phi^a_2$ , $d^{abc}$ and $f^{abc}$ are the
$SU(3)$
structure constants. From equations (\ref{3.1}) and (\ref{3.2}) it is easy to
verify that the SUSY \SUTH{}- group can be broken down to the SUSY
$SU(2)_H\times U(1)_{8H}$
if, for example, the 8-th component of the field $\Phi$ acquires a
nonvanishing VEV:
\begin{equation}
<\Phi^8>_0=\frac{\sqrt{3} a_8}{2} =\frac{\sqrt{3} M_I}{2}\label{3.3}
\end{equation}

In this case of the gauge symmetry breaking the supersymmetry conservation
allows to describe the mass spectrum of new massive $N=1$ supermultiplets in a
rather simply  way.
We start with  eight vector massless superfields $V_H^a(1,\ 1/2)$
($4\times 8^a =32$ degrees of freedom) and
eight chiral massless superfields $\Phi^a(1/2\ ;\ 0,\ 0)$
($4\times 8^a =32$ degrees of freedom). As a
result of the super-Higgs effect, we get four massive vector
supermultiplets $(1\ ,\ \frac{1}{2})+(\frac{1}{2}\ ,\ 0+0)=
(1\ ,\ \frac{1}{2}+\frac{1}{2}\ ,\ 0)_{massive}$
with $8\times 4^a=32$ degrees of freedom  and with the same universal mass.
The  formula for the gauge boson mass is
\begin{eqnarray}
(M^2)_{ab}&=&1/2\ g_H^2 f^{8ac} f^{8bc} a_8^2=\ 3/8\ g_H^2 a_8^2\delta^{ab}
\nonumber\\
a,b&=&4,5,6,7\ \ \ \ or \label{3.4}\\
M_{4,5,6,7}^2&=&3/8\ g_H^2 M_I^2\ ,\ \ M_{1,2,3,8}^2=0\nonumber
\end{eqnarray}
The mass term of $\lambda$-gauginos is expressed as follows:
\begin{eqnarray}
{\cal L}_M&=&1/{\sqrt{2}}\ g_H f^{8bc} \psi_{\Phi}^b\lambda^c a_8=
\frac{\sqrt{3}}{2} \frac{g_H}{\sqrt{2}}\ M_I[\psi_{\Phi}^4\lambda^5 -
\psi_{\Phi}^5\lambda^4 +\psi_{\Phi}^6\lambda^7 -\psi_{\Phi}^7\lambda^6 ]
\nonumber\\
&-&{\lambda}_0 M_I\ 3/4\ (\psi_{\Phi}^1
\psi_{\Phi}^1 +\psi_{\Phi}^2\psi_{\Phi}^2+
\psi_{\Phi}^3\psi_{\Phi}^3 -\ 1/3\
\psi_{\Phi}^8\psi_{\Phi}^8 )+h.c. \label{3.5}
\end{eqnarray}
So the gauginos $\lambda^4,
\lambda^5, \lambda^6,\lambda^7$ combining with fermions
$\psi_{\Phi}^4,
\psi_{\Phi}^5,\psi_{\Phi}^6,\psi_{\Phi}^7$ give the Dirac gauginos
with the masses $M=\frac{1}{\sqrt{2}} \frac{\sqrt{3}}{2} g_H M_I$.
Four real scalar states from the supermultiplets $\Phi^{4,5,6,7}$
transform into the longitudinal components
of four corresponding massive vector bosons, while the remaining four scalar
states contribute to
four massive $N=1$ supermultipets
$(1\ ,\ \frac{1}{2}+\frac{1}{2}\ ,\ 0)_{massive}$.
There are also four massless
vector superfields $V_H^{1,2,3,8}$
(16 degrees of freedom) and four massive chiral
superfields
$(\psi^a_{\Phi}, \Phi^a)\ \ (a=1,2,3,8)$ at this stage of breaking. So, due
to the super-Higgs mechanism of SUSY breaking four massless
vector superfields have absorbed
four massless chiral superfields and formed four massive vector superfields.
The chiral superfields $\Phi^{4,5,6,7}$ play the role of Higgs superfields and
they all have been absorbed completely.

At the second stage of the $SU(2)_H\times U(1)_{8H}$ gauge symmetry
breaking with a simultaneous supersymmetry conservation, one can use the
chiral superfields $\eta_{\alpha},
\xi_{\alpha}$ with  equal VEV's: ${<\eta_{\alpha i}>}_{0}=
{<\xi^i_{\alpha}>}_{0}=
\delta_{\alpha}^i\gamma\ \  (i,\alpha=1,2,3,)$.
As a result of this breaking, four massive
vector supermultiplets $V_H^{1,2,3,8}(1\ ,\ \frac{1}{2}+\frac{1}{2}\ ,\ 0)$
acquire the
universal mass $M^2_H=2g^2_H \gamma^2$. A detailed analysis shows that the
Majorana higgsinos from the supermultiplets $\eta_{\alpha}, \xi_{\alpha}$
participate in the formation of four Dirac gauginos, whose upper components are
$\lambda^1,\lambda^2,
\lambda^3$ and $\lambda^8$. The degenerate mass of these Dirac
gauginos will be $\sqrt{2}g_H \gamma$:
\begin{eqnarray}
{\cal L}_M&=&i/\sqrt{2}\ \ g_H\gamma\biggl\{\lambda^3 [(\eta_{11} -\xi_1^1)
-(\eta_{22} -\xi_2^2 ) ]+\ 1/\sqrt{3}\
\ \lambda^8 [(\eta_{11} -\xi_1^1)+\nonumber\\
&+&(\eta_{22} -\xi_2^2) -2(\eta_{33} -\xi_3^3)]+
\lambda^1 [ (\eta_{21} -\xi_1^2) +(\eta_{12} -\xi_2^1)] \nonumber\\
&-&i\lambda^2
[ (\eta_{12} +\xi_2^1) + (\eta_{21} +\xi_1^2) ] \biggr\}+h.c. \label{3.6}
\end{eqnarray}
It is easy now to rewrite the Lagrangian of the interactions in
terms of physical states
(remembering that for the matter fields $\psi_{mi}=U_{ij}\psi_{oj}$ and
$A_{mi}=\tilde U_{ij}A_{oj}$, where $A$ denotes the scalar
partner of $\psi-$fermions).
The gauge boson interactions with matter fields have the form:
\begin{eqnarray}
{\cal L}=g_H H_a^{\mu}&\biggl\{&{\bar \psi}_u\gamma_{\mu} U_L T^a U_L^+
\ \frac{1+\gamma_5}{2}\ \psi_u +{\bar \psi}_u \gamma_{\mu} U_R T^a U_R^+
\ \frac{1-\gamma_5}{2}\ \psi_u +\nonumber \\
&+&(u\rightarrow d,l,\nu)\biggr\} \label{3.7}
\end{eqnarray}
Let us consider now the gaugino interactions. The initial
Lagrangian has the form:
\begin{equation}
{\cal L}=ig_H\sqrt {2}(A_i^+ T_{ij}^a \lambda^a \psi_j -h.c.)\ \ \ a=1,2,3,8
\label{3.8}
\end{equation}
Consider only the interaction between left "up" quarks, left "up"
squarks and gauginos.
The generalization of this Lagrangian to all leptons and quarks will be
obtained by simply adding similar terms to  $u_R,\ d,\ d_R,\ \nu,\ l$
and $l_R$.
Expression (\ref{3.8}) for "up" quarks looks like:
\begin{eqnarray}
{\cal L}&=&ig_H
\biggl (A_1^*,\ A_2^*,\ A_3^*\biggr )_L \left( \begin{array}{ccc}
\lambda^3+\frac{1}{\sqrt{3}}\lambda^8 & \lambda^+ & 0 \\
\lambda^- & -\lambda^3+\frac{1}{\sqrt{3}}\lambda^8 & 0 \\
0 & 0 & -\frac{2}{\sqrt{3}}\lambda^8
\end{array} \right) \left( \begin{array}{c}
\psi_1 \\ \psi_2 \\ \psi_3
\end{array} \right) +h.c.=\nonumber \\
&=&-g_H\biggl [A_{iL}^* {\tilde U}_{ik}(\Lambda' )_{bk}U_{jb}^*\
\frac{1+\gamma_5}{2}\ \psi_j \biggr ]+h.c. \label{3.9}
\end{eqnarray}
where $A_{iL}={\tilde U}_{ij} {A_{0j}}_L$ , $i,j,...,b=1,\; 2,\; 3$.

Case (ii).
To realize this version of the intermediate \SUTH{}- symmetry breaking, one
has to use the pair of the chiral superfields $\Phi$, $\tilde \Phi$
with different $U(1)_R$ quantum
numbers. Then one easily verifies that the stationary supersymmetric vacuum
solutions will be realized in
accordance with equations (\ref{2.5}-\ref{2.7}). These solutions
will look like
\begin {equation}
<\Phi_{1}^{3}>_0\ =\ <\Phi_1^8>_0\ =\ -\sqrt{3}M\ ,\ \ \ <\tilde\Phi_1^8>_0\ =\
2\sqrt{3}M \label{3.10}
\end {equation}

When these fields are applied
simultaneously with the above VEVs (\ref{3.10}), the following gauge
boson mass spectrum is obtained: $M_{H_4}=M_{H_5}=3\sqrt{2}M_I$,
$M_{H_1}=M_{H_2}=M_{H_6}=M_{H_7}=3M_I$, $M_{H_3}=M_{H_8}=0$. The remnant
group in this case is the $SUSY\ U(1)_{3H}\times U(1)_{8H}$ - group.
As one would expect,
the rank of the group did not change, whereas the remnant group
was broken by the chiral superfields $\eta$, $\xi$ on the scale $M_H$. Here it
makes no difficulty to get the mass degeneracy of the superfields $V_3$ and
$V_8$ conserving SUSY while doing this. Again, the super-Higgs mechanism is
applied leading to
the formation of the massive superfields with the universal mass
$M_H$.  In this connection, a rather simple way may
be proposed to estimate the
bound on $M_H$ from the comparison with the data on rare processes.

And, finally, let us consider case (iii) when $V=U(1)_{8H}$.
We confine ourselves to the case when the
scalar components of the complex chiral superfield
$\bar\Phi$ have the nonzero VEVs: ${<\Phi_1^1>}_0 \not = 0 \ ,\ {<\Phi_1^2>}_0
\not = 0$ , ${<\Phi_1^3>}_0\not =0 \ ,\ {<\Phi_1^8>}_0\not = 0 \ .$

Although this choice of VEV's fulfils the equations for the flat $F_{\Phi}$
directions with the solutions ${<\Phi_1^8>}_0=-\sqrt{3} M_I$ ,
${<\Phi_1^1>}_0^2 +{<\Phi_1^2>}_0^2 -{<\Phi_2^3>}_0^2 =9M_I^2$ ,
this solution does not determine the vacuum of the theory
as might be expected. The corresponding solutions for $D^{1,2}$ are
incompatible with the $F_{\Phi}$-flat solutions. As in the previous case (ii),
in order to overcome this difficulty one has to introduce a new Higgs
superfield to compensate for the nonvanishing contributions of $D$-terms to
the scalar potential
of the theory. This compensation requires a specific choice of the vacuum
expectations for the second Higgs superfield $\tilde \Phi$.
In this case, only one vector supermultiplet $V_H^8$ is left on the
intermediate scale.

The abovementioned examples are enough to
 research further into the regularity of the behavior
of the violation scale $M_H$ by comparing model predictions with the
experiment. Here we just
note that the SUSY stationary solutions with CP violation in the horizontal
sector are available
both for the scales $M_I$ and $M_H$. Indeed, for instance, in
case (iii) the CP violation occurs on the scale $M_I$. Another supersymmetric
vacuum, ${<\Phi_1^8>}_0=-\sqrt{3}M_I$ , ${<\Phi_1^1>}_0^2+{<\Phi_1^2>}_0^2
+{<\Phi_1^3>}_0^2 =9M_I^2$ ,
corresponding to the intermediate symmetry group $SU(2)_H\times
U(1)_{8H}$,
results in the CP violation
in the neutral $K$-meson decays due to only horizontally acting forces on the
scale $M_H$.
In the last case, in the electroweak sector of CP violation one may have
$\delta_{KM}=0$. That is the very case outlined in our introduction.

\subsection{The role of horizontal interactions with intermediate symmetry
breaking scale in rare processes}

Let us analyze the contribution of horizontal interactions to rare processes.
We will consider first the oscillations of $K^0,\ B^0_d$ and $B^0_s$ mesons.
The experimental data on
$K^0\longleftrightarrow {\bar K}^0$ oscillations are as follows:
\begin {equation}
\biggl( \frac{\Delta m_K}{m_K} \biggr)_H \, < \, \biggl( \frac{\Delta m_K}
{m_K} \biggr)_{exp} \,\approx \,7\times 10^{-15} \label{4.1}
\end {equation}
The theoretical expression for the $m(K_L)-m(K_S)=\Delta m_K$
mass difference is given by the equation:
\begin {eqnarray}
\Delta m_K = \frac{g_H^2}{M_H^2}\ Re(C_K^0) f^2_Km_K \biggl[
 \frac{1}{6}\ +\
\frac{1}{3}\frac{m_K^2}{(m_s-m_d)^2} \biggr]  ,\label{4.2}
\end {eqnarray}
where $C_K^0=\sum_{a}^{\prime}(DT^aD^+)_{21}(DT^aD^+)_{21}$,
 $C_K^0$ being a unitary coefficient showing the contributions of the Feynman
diagrams with the exchange of the horizontal bosons from
the considered gauge groups. The symbol "$\prime$" denotes that the sum is
over the definite set of indexes "a", but it should be noted that the
sum over the complete set ( a=1,2,...8 ) is equal to zero.
"D" is the orthogonal matrix diagonalizing the Fritzsch-like
mass matrix for "down" quarks.
One-particle contributions
to the vector, axial, scalar and pseudoscalar currents
might be calculated like in Ref. \cite{35'}.

{}From formula (\ref{4.2}), using the experimental data  and the values for
$\alpha_H=\ g^2_H/4\pi\ \approx 1.9\cdot 10^{-2}$ on the scale
$M_H$ \cite{9'}, one can obtain the lower limits on the  light
H-boson masses . These values are given in Table 12 together with the
$C_K^0$ values.

Analogous expressions can be obtained for the $B_d$ and $B_s$ mesons.
The mixing elements for $B_d$ and $B_s$ mesons are as follows: $C_{B_d}^0=
\sum_{a}^{\prime}{(DT^aD^+)_{31}(DT^aD^+)_{31}}$ and
$C_{B_s}^0=\sum_{a}^{\prime}(DT^aD^+)_{32}(DT^aD^+)_{32}$. Their values
are given in Table 12. Using the $H$-boson mass limits from Table 12 and the
value $f_{B_d}\approx 150$ MeV,
 one can calculate the $B_d$ meson mass difference (see Table 12).
 The one-particle contribution ($R_1$) to the $B_d$ (and $B_s$)
meson amplitudes is unknown. But, assuming that it is not much
greater than the vacuum contribution, one can see that the
$\Delta m_{B_d}/m_{B_d}$ values given in Table 12 are very close to those
obtained from ARGUS \cite{5'} (except for case (ii)):
\begin {equation}
\biggl(\frac{\Delta m_{B_d}}{m_{B_d}}\biggr)_H \,<\,
\biggl(\frac{\Delta m_{B_d}}
{m_{B_d}}\biggr)_{exp}=(0.73\pm 0.14)\times 10^{-13} \label{4.3}
\end {equation}

{\bf Table 12.} The $M_H, C_K^0, C_{B_d}^0, C_{B_s}^0,\ \
\Delta m_{B_d}/m_{B_d}$ and \ $\Delta m_{B_s}/\Delta m_{B_d}$
Values for Different Models.

\footnotesize

\begin{center}
\begin{tabular}{|l|ccccc|}
\hline
Models&$C_K^0$ & $M_H$ (TeV)& $C_{B_d}^0$& $\frac{\Delta m(B_d)}{m(B_d)}$ &
$\frac{\Delta m(B_s)}{\Delta m(B_d)}$ \\ \hline
\hline
$SU(2)_H\times U(1)_{8H}$ & $3.8\times 10^{-5}$ & $8\div 9$ & $1.5\times
10^{-3}$ &
$(1.1\div 0.8)\times 10^{-13} k $ & $m_s/m_d$ \\
$U(1)_{3H}\times U(1)_{8H}$ & $1.6\times 10^{-2}$ & $170\div 200$ &
$1.5\times 10^{-3}$ & $(2.3\div 1.8)\times 10^{-16}k$ & $m_s/m_d$ \\
$U(1)_{8H}$ & $2.8\times 10^{-5}$ & $7\div 8$ & $1.1\times 10^{-3}$ &
$(1\div 0.8)\times 10^{-13}k$ & $m_s/m_d $ \\
\SUTH{} & $<(10^{-5}\div 10^{-6})$ &
$O(1\ TeV)$ & $-$ & $-$ &
$-$ \\ \hline
\end {tabular}
\end{center}
\normalsize
$k=(R_{1P} +\ 1/2\ )\ \ ,\ \ g_H\approx g_{EW}\ \ .$

Let us calculate now the relative mass difference of mesons  $B_d$ and $B_s$.
Assuming that $f_{B_s} \approx f_{B_d}$, one obtains the values given in
Table 12.
So the oscillations of $B_s$ mesons in such models must be
stronger than those of
$B_d$ mesons. Let us consider now the decay $\mu \rightarrow 3e$.
The branching ratio of this process will be

\begin {equation}
B(\mu\rightarrow 3e)=\ 12\ \frac{g_H^4}{g_W^4}\ \frac{m_W^4}{M_H^4}\
|C_{(\mu)}^0|^2 , \label{4.4}
\end {equation}
where $C_{(\mu)}^0=\sum_{a}^{\prime}{(LT^aL^+)_{21}(LT^aL^+)_{11}}
=\sum_{a}^{\prime}{L_{21}^aL_{11}^a}$;
 L- is the orthogonal matrix diagonalizing the real Fritzsch- like mass matrix
for "down" leptons. Using the experimental value \cite{36'}
$B(\mu\rightarrow 3e) \,<\, 10^{-12}$, it is easy to
obtain the limits for the horizontal boson masses:
\begin {eqnarray}
(i)\ \ M_H\,> \, 7.5\ TeV\ \ \ \ &(ii)&\ M_H\,> \, 30\ TeV\nonumber \\
(iii)\ \ M_H\,>\, 5\ TeV\ \ \ \ &(iv)&\ M_H\,>\, O(1\ TeV) . \label{4.5}
\end {eqnarray}

Let us turn next to the process of the muon-to-electron conversion in
the presence of a nucleus.
The branching ratio of this
process for the nucleus with equal numbers of protons
and neutrons and large $Z$ is \cite{28'}:
\begin {equation}
\frac{\Gamma (\mu N\rightarrow eN)}{\Gamma (\mu N\rightarrow \nu N)}=\ 432\
\frac{{|\sum_a^{\prime} L_{21}^a [{U_{11}^a}^* +{D_{11}^a}^*] |}^2}
{1/4\ (1+3g_A^2)}\
\frac{m_W^4}{M_H^4}\ \frac{g_H^4}{g_W^4}. \label{4.6}
\end {equation}

In eq.(\ref{4.6}),
$L^a_{21}, U^a_{11}$ and $D^a_{11}$ are the mixing elements for
leptons, up- and down-quarks. Using the recent experimental value
for the $\mu$ to $e$
conversion : ${\Gamma} (\mu N \rightarrow eN)
\, < \, {\Gamma}(\mu N\rightarrow \nu N) \times 5\cdot 10^{-12}$ \cite{37'},
from eq. (\ref{4.6}) one can obtain the limits for the horizontal gauge bosons
masses.

 We consider the choice for the forms of the quark and lepton mass
matrices, for instance, the "improved" Fritzsch ansatz like Matumoto
\cite{11'}, the corresponding estimates
will do not change much (except for $U(1)_{8H}$)

\begin {eqnarray}
(i)\ \ M_H\,>\, 60 TeV\ \ \ \
&(ii)&\ M_H\,>\, 65  TeV\nonumber \\
(iii)\ \ M_H\,>\, 3 TeV\ \ \ \
&(iv)&\ M_H\,>\, 60 TeV. \label{4.8}
\end {eqnarray}
This fact can easily be explained by the coincidence of the values of the
mass matrix element $(M_d)_{12}$ in these two ansatzes and its
dominant role in the definition of the $V_{us}$-CKM matrix element.

{}From another very important quark-lepton rare decay $ K^+\longrightarrow
{\pi}^+ {\mu}^+  e^-$, whose partial width is now experimentally estimated as

\begin{eqnarray}
 Br(  K^+\longrightarrow
{\pi}^+ {\mu}^+ e^-) \,<\,  2.1 \times 10^{-10}, BNL-E777, \label{4.9}
\end{eqnarray}
the constraints on $M_{H_0}$ are also rather large (except in (iii)):

\begin{eqnarray}
M_{H_0} > {g_H}/{g_W} \times 35 TeV .\label{4.10}
\end{eqnarray}
Let us compare it with the bounds on the pure quark or lepton rare
processes. For the $U(1)_{8H}$- group, the corresponding bound
on the scale
$M_{H_0} $ is  approximately some TeVs.

Finally, let us consider the decay $\mu \rightarrow e\gamma$.
The one-loop contribution
with the  $H$-boson exchange  is suppressed against the $\mu \rightarrow 3e$
decay: $\Gamma (\mu \rightarrow e\gamma) \ll \Gamma (\mu \rightarrow 3e)$. So,
 the major contribution to the $\mu \rightarrow e\gamma$ decay width will
come from the one-loop diagram with the exchange of horizontal gauginos and
scalar charged leptons.
The branching ratio of this decay is :

\begin {eqnarray}
B(\mu\rightarrow e\gamma)=\ \frac{48\pi^2}{G_F^2 m_{\mu}^2}\ F^2 \label{4.11}
\end {eqnarray},
where formfactor
$F^2$ is given in ref.\cite{28'}. Using the experimental value \
cite{38'}:
$B(\mu \rightarrow e\gamma)\,<\, 4.9\cdot 10^{-11}$, one can easily obtain the
bounds on the gaugino masses:
\begin {eqnarray}
(i)\ \ {\tilde M}_H\,>\, 0.6\ TeV\ \ \ \ &(ii)&\ {\tilde M}_H\,>\, 0.25\
TeV\nonumber \\
(iii)\ \ {\tilde M}_H\,>\, 0.3\ TeV\ \ \ \ &(iv)&\ {\tilde M}_H\,>\,
O(100\ GeV), \label{4.12}
\end {eqnarray}
where the scalar lepton mass is 80 GeV.

To conclude, let us note that the analysis of the supersymmetric horizontal
model shows that in several schemes of $H$-symmetry breaking (cases (i), (iii)
and (iv)) the limits
for the lower bounds of some $H$-bosons from the experimental results on
the amplitudes of pure quark and pure leptonic rare processes
($|{\Delta}H |  \neq  0 $)  can be relatively low (${\leq} 10 TeV $).
In this case the contribution of $H$-interaction to $B^0_d$ meson oscillations
may turn out to be the major contribution and  explain, in principle, the
experimental value of the $B^0_{d1}-B^0_{d2}$ mass difference.
However,  similar bounds on  $M_H$,
derived from some
quark-lepton rare reactions $ (|\Delta H | = 0)$, may turn out
to be  much more than the above estimates, except for  case (iii).

Really, we should look closer at this situation  : in particular,
we should   clear up whether our understanding of
the origin of quark and lepton generations is correct, i.e. that
we see one
and the same quark and lepton mixing mechanism in operation.
Indeed, by now  no reliable evidences to the lepton mixing have been
obtained.
It is natural to think that the problem of the nature of mixings  ascends
to the main question of the SM, i.e. the origin of quark and lepton masses.
So we state that we   know nothing  about how the mixings of quark
and lepton families are correlated and can only hypothesize it.

Before  considering a particular model of such a nontrivial  correlation and
discussing  some of its consequences for the breaking scale  of the horizontal
gauge interaction $M_H$, we have to investigate the breaking of the
\SUTH-model
without the intermediate scale $M_{H_I}$, trying to connect the splitting
of its global breaking scale $M_{H_0}$ with  known heavy quark masses.
This  will enable us to get  more information about the 8-gauge boson
masses and make our estimates more predictive.

\section{The  SUSY   \SUTH-gauge model with correllation
3- family mixing and 8-gauge boson mass splitting.}

 In this Section we will confine ourselves to the consideration of two
types of Hermitian fermion mass matrices and, going on with the previous
analysis, estimate the local H- symmetry breaking scale (see
subsection 1.4)

As the 1st approach, we consider
a  modified "calculable" Fritzsch anzatz for 3 generations, with nonzero
antidiagonal mass matrix elements and the upper bound on the t-quark
from the experimental data on the matrix element $V_{cb}$ will be in
according to the experimental values.

\begin{equation}
{\bf M_f^{MF}}=
\left( \begin{array}{ccc}
0 & A e^{-i \alpha} & E e^{-i \gamma} \\
A e^{ i \alpha} & F & B e^{-i \beta} \\
E e^{ i \gamma} & B e^{ i \beta} & C \\
\end{array} \right) , \label{6.1}
\end{equation}

where $( {A >> or \sim E})  << ({B << or \sim F} ) << C $.
We will not  study now the most general form of
hermitian quark mass matrices of the three families. It will be enough,
and more useful, for us
to consider the specific forms of quark ansatzes which would fit well
the CKM- mixing matrix elements
with the experimental accuracy attainable today.

Another (democratic) ansatz is noteworthy for the possibility to single out
the t-quark mass value. And besides, within this approach the mass
 matrices of the
above form can, at least, be more correctly interpreted in physical
terms - e.g., via  the compositeness  of quarks.

\subsection{The {\it modified} \,Fritzsch \, ansatz}

Let us consider some of the "calculable" ansatzes for $3\times\,3 $
 "$up$" and "$down$" quark mixing matrices with nonzero antidiagonal
elements (\ref{6.1}) and consistent with the modern values of the
$CKM$-matrix for charged currents \cite{11'}c, where the matrix elements of the
up- and down mass matrices were taken like as:

\begin{eqnarray}
 A_u &=& E_u = \sqrt{|m_u|m_c} , F_u = -B_u = m_c , C = m_t ,\nonumber\\
{\alpha}_u &=& {\beta}_u = {\gamma}_u = 0; \nonumber\\
 A_d &=& E_d = \sqrt{|m_d|m_s} , F_d = B_d = m_s ,
 C_d = m_b , \nonumber \\
 {\alpha}_d &=& -\frac{\pi}{2}, {\beta}_d = {\gamma}_d = 0 . \label{6.2}
\end{eqnarray}

In the leading approximation the unitary matrices $U$ and
$D$ ($O_{CKM}=U\,D^+$), diagonalizing these mass states
(\ref{6.1}) (\ref{6.2}),
 have the following form:

\begin{equation}
{\bf U }\approx
\left ( \begin{array}{ccc}
1 & -\sqrt{\frac{m_u}{m_c}} & -2\sqrt{\frac{m_u}{m_c}}\frac{m_c}{m_t} \\
\sqrt{ \frac{m_u}{m_c}} & 1 & \frac{m_c}{m_t} \\
\sqrt{ \frac{m_u}{m_c}} \frac{m_c}{m_t} & -\frac{m_c}{m_t} & 1 \\
\end{array} \right) \label{6.3}
\end{equation}

and
\begin{equation}
{\bf D}\approx
\left ( \begin{array}{ccc}
e^{-i\frac{\pi}{4}} & \sqrt{ \frac{m_d}{m_s}}e^{-i\frac{3 \pi}{4}} &
\frac{ \sqrt{2m_dm_s}}{m_b} e^{i \frac{\pi}{2}}\nonumber \\
\sqrt{ \frac{m_d}{m_s}} & e^{i \frac{\pi}{2}} &
\frac{m_s}{m_b} e^{-i \frac{\pi}{2}} \nonumber\\
\frac{ \sqrt{m_dm_s}}{m_b} e^{i \frac{\pi}{2}} &
\frac{m_s}{m_b} e^{i \frac{ \pi}{2}} & e^{i \frac{ \pi}{2}} \\
\end{array} \right). \label{6.4}
\end{equation}

Using the above ansatz for the $b \rightarrow c$ transition, one can get a
higher restriction for the upper bound on the t- quark mass:
$ V_{cb} \approx\frac{m_s}{m_b} + \frac{m_c}{m_t}$ .

The experimental precision of the measurements of $V_{cb}$  could
indicate, and only to a
certain extent,
the magnitude  of the $d_{23}$.
Unfortunately, our  knowledge of  the $d_{13}$ -element  is still not
sufficient because of big experimental uncertainties in  q or
$ \sqrt{{\rho}^2 \,+\, {\eta}^2}$ values ($q \leq 0.2$).
In  scheme \cite{11'}c  $ q= |V_{ub}|/|V_{cb}| \approx 0.1 $ and
the Wolfenstein parameters  are
$ \rho \approx - 0.15 $ and $ \eta \approx 0.49 $
($\sqrt{\rho^2 + \eta^2 }\approx \frac{1}{2}$ ), so that the
$(V_{ub}-V_{td})$- intersection of the unitary triangle lies
in the second (left) quadrant of the $\rho,\eta$- complex plane.

To obtain the form of these mass matrices,
it's necessary to consider, besides  $H$ - and
 $h$ -Higgs superfields (see (\ref{2.3})),  the additional superfields
$H_0\,(\,1_H,\,2_L, \,-\frac{1}{2})$ and
$h_0\,(\,1_H,\,2_L,\,\frac{1}{2})$, which are  $SU(3)_H$ -singlets.
Then the corresponding addenda  $k_1\,Q\,h_0\,u^c$  and
$k_2\,Q\,H_0\,d^c$  will appear in the superpotential (\ref{2.3}).

The splitting between the
horizontal gauge boson masses will be determined only by the nonvanishing
VEV's of the $H(8_H,\,2_L,\,-\frac{1}{2})$ and $h(8_H,\,2_L,\,+\frac{1}{2})$
Higgs fields: $<\hat H>_0=\varphi^a (\frac{\lambda^a}{2})$  and
$<\hat h>_0= \tilde{\varphi^a}(\frac{\lambda^a}{2})$, where

\begin{eqnarray}
\varphi_2&=&-2 \sqrt{m_dm_s}/{\lambda_3};\,\,
\varphi_4=2 \sqrt{m_dm_s}/{\lambda_e}; \,\,
\varphi_6=2 m_s/{\lambda_3};\nonumber \\
\varphi_3&=&-2 m_s/{\lambda_3}  ;\,\,\,\,
\varphi_8=-\frac{2}{\sqrt{3}}(m_b - \frac{m_s}{2})/{\lambda_3} . \label{6.5}
\end{eqnarray}
and similarly for the nonvanishing VEV's of the $h$-Higgs fields:

\begin{eqnarray}
\tilde{\varphi_1}&=&\tilde{\varphi_4}=2 \sqrt{m_um_c}/{\lambda_5};\,\,\,
\tilde{\varphi_6}=-2m_c/{\lambda_5}; \nonumber\\
\tilde{\varphi_3}&=&-m_c/{\lambda_5};\,\,\,\,
\tilde{\varphi_8}=-\frac{2}{\sqrt{3}}(m_t-\frac{m_c}{2})/{\lambda_5}.
\end{eqnarray}

Now we have the possibility to calculate the contributions of the horizontal
interactions to the amplitudes of following rare processes. \\
a) pure quark processes:
$K^0\leftrightarrow \bar K^0$ , $B^0_d\leftrightarrow \bar B^0_d$ ,
$B^0_s\leftrightarrow \bar B^0_s$ oscillations
and CP-violating effects in K-, B-,  D-meson decays;\\

b) purely lepton processes: $\mu\rightarrow e\gamma\ ,\ \mu\rightarrow 3e\
,\ \tau\rightarrow\mu \gamma$ , etc.\\

c) quark-lepton processes: $K^{\pm}\rightarrow\pi^{\pm} \mu^{\pm} e^{\pm}$
, $(\mu -e)-$conversion  on
nuclei, $K^0\rightarrow\pi^0  \mu e$,
$K^{\pm}\rightarrow\pi^{\pm} \nu_i \nu_j$, etc.;\\

After the calculations the expressions for
the ($K_L^0\,-\,K_S^0$)- and ($D_L^0\, -\, D_S^0$)-
meson mass differences (processes a) )take the following general forms:

\begin{eqnarray}
\biggl[\frac{(M_{12})_{12}^K}{m_K}\biggr]_H &=&
\frac{1}{2}\frac{g_H^4}{M_{H_0}^4}  \Biggl\{
\biggl[\tilde{\varphi_a}(D\,\frac{\lambda^a}{2}\,D^+)_{12}\biggr]^2
+\biggl[\varphi_a(D\,\frac{\lambda^a}{2}\,D^+)_{12}\biggr]^2 \Biggr\}
{f_K^2}R_K, \nonumber\\
\biggl[\frac{(M_{12})_{12}^D}{m_D}\biggr]_H &=&
\frac{1}{2}\frac{g_H^4}{M_{H_0}^4}  \Biggl\{
\biggl[\tilde{\varphi_a}(U\,\frac{\lambda^a}{2}\,U^+)_{12}\biggr]^2
+\biggl[\varphi_a(U\,\frac{\lambda^a}{2}\,U^+)_{12}\biggr]^2 \Biggr\}
{f_D^2}R_D. \label{6.7}
\end{eqnarray}

Putting into formula (\ref{6.7}) the expressions for $\varphi$,
$\tilde{\varphi}$ (really the second term in the $(M_{12})_{12}^K$
and the first in the $(M_{12})_{12}^D$ are equal to zero)
 and the elements $d_{ij}$ of the $D$- mixing matrix,
we can obtain the lower limit for the value $M_{H_0}$. So, we analyze
the ratios:
\begin{equation}
\biggl[\frac{\Delta m_K}{m_K}\biggr]_H=
\frac{g_H^2}{M_{H_0}^2}Re[C_K]f_K^2R_K < 7\cdot 10^{-15} \label{6.8}
\end{equation}
and
\begin{equation}
\biggl[\frac{Im{M}_{12}}{m_K}\biggr]_H=
\frac{1}{2}
\frac{g_H^2}{M_{H_0}^2}Im[C_K]f_K^2R_K < 2\cdot 10^{-17}.\label{6.9}
\end{equation}

In formulas (\ref{6.8}) and
(\ref{6.9}) the expression for $C_K$ is as follows :
\begin{eqnarray}
C_K=&-& \frac{g_H^2}{2 \lambda_5^2} \frac{m_t^2}{M_{H_0}^2}
\Biggl[\,\frac{m_c}{m_t}
\Biggl(\sqrt{\frac{m_u}{m_c}} +\sqrt{ \frac{m_d}{m_s}} \frac{m_s}{m_b}
\Biggr) \nonumber\\
&+&\,i\,\Biggl( \sqrt{\frac{m_u}{m_c}}
\frac{m_c}{m_t}+3 \sqrt{\frac{m_d}{m_s}} \frac{m_c}{m_t} \frac{m_s}{m_b}
+ 2\sqrt{\frac{m_d}{m_s}} \frac{m_s^2}{m_b^2}\Biggr)
\Biggr]^2.
\end{eqnarray}

For getting the lower bounds for $M_{H_0}$
 from  formulas (\ref{6.8}) and (\ref{6.9}) we can take    for the value
of $R_K={1}/{6}+
{1}/{3}({m_K^2}/{(m_s-m_d)^2})$ ,$f_K=0.163\,GeV$ , $m_t=150GeV$ and
$g_H\simeq 0.488$ \cite{9'}.

In quite an analogous way, we should also write the expression for
$M_{12}(B_d)_H$  ($M_{12}(B_d)_H$):

\begin{equation}\label{6.12}
\biggl[\frac{M_{12}(B_d) } {m_{B_d}}\biggr]_H=\frac{1}{2}
\frac{g_H^2}{M_{H_0}^2}C_{B_d}f_{B_d}^2R_{B_d}\,.
\end{equation}

The unitary  suppression coefficient will take the following form:

\begin{displaymath}
C_{B_d} \approx \frac{g_H^2}{2 \lambda_5^2} \frac{m_t^2}{M_{H_0}^2}
 \Biggl[e^{-i\frac{3\pi}{4}}\sqrt{\frac{m_u}{m_c}}\frac{m_c}{m_t}
 -e^{-i\frac{5\pi}{4}}\sqrt{\frac{m_d}{m_s}} \frac{m_c}{m_t}
 + \sqrt{\frac{2m_d}{m_s}} \frac{m_s}{m_b}\Biggr]^2 .\label{6.14'}
\end{displaymath}

{}From these  formulas
and assuming that $(\Delta m(B_d)/ m(B_d))|_H
  \leq (0.73\pm 0.14)10^{-13}$,
$R_{B_d}\simeq$ $ {1}/{6}+{1}/{3}({m_{B_d}^2}/
{(m_b-m_d)^2})$ , $f_{B_d}\simeq\,0.14 GeV $, we also can obtained the
lower limits on $M_{H_0}$.

 Note that, if we take the value  $f_{B_D}=0.2 \,GeV$, the lower
bounds on the horizontal symmetry breaking scales following from
(\ref{6.12}) and (\ref{6.14'}) are
approximately equal ( see (\ref{6.14}) ).

In this ansatz for the CP- violation parameter
$\frac{1}{2}\times Im(\frac{p}{q})_H$  we have a well-defined magnitude:

\begin{eqnarray}
\frac{1}{2}\times Im (\frac{p}{q})_{B_d} \approx
\frac{ImM_{12}(B_{d})_H}{ReM_{12}(B_d)_H} \, \approx 0.3.
\end{eqnarray}

In SM with this quark mass ansatz the asymmetries will be equal:
$A(J/\Psi) \approx - 0.34 $ and $A({\pi}^+ {\pi}^-) \approx -0.44$,
respectively.
The contributions of CP-violating horizontal interactions to the
asymmetries for both $B^0$-decays are identical but the signs
differ (in this approach $max |A_f| \approx 0.17$) .

Finally, let us give the useful estimate:

\begin{equation}
\biggl[\frac{\Delta m_{B_s}}{\Delta m_{B_d}}\biggr]_H\approx
\biggl[\frac{V_{ts}}{V_{td}}\biggr]^2
\approx\frac{1}{2}\frac{m_s}{m_d}
\times (1\, +\, \frac{m_c}{m_t} \frac{m_b}{m_s})^2
\approx17 \div 20.
\end{equation}

As follows from the expression (\ref{6.7}) in the approach with the ansatz
(\ref{6.2}) the value of the ($D_L^0 - D_S^0$) - mass difference will
be considerably suppressed by the unitarity coefficient
$ (U {\lambda}^8 U^+)_{12}$ ( see (\ref{6.3})  comparing with the similar
coefficient $ (D {\lambda}^8 D^+)_{12} $) in (\ref{6.4})).
But it is very important to note that in our approach with the symmetric
ansatz \cite{8'} this suppression  for the  $\Delta m_D $ - mass difference
will be absent. From formulas (\ref{6.7}) it's possible to get in this ansatz
the next approximate expression for the
$D_L^0 - D_S^0 $ and $K_L^0- K_S^0 $ - mass difference ratio:

 \begin{eqnarray}
\frac{ \Delta m_D }{\Delta m_K } \approx
\frac{m_D}{m_K} \frac{f_D^2 R_D}{f_K^2 R_K}
\frac{[\tilde{\varphi_8}^{sym}]^2}
{[{\varphi_8}^{sym}]^2} .
\end{eqnarray}

{}From this expression it follows that for the corresponding values of
the VEV's  ratio -$\tilde{\varphi_8} / \varphi_8 $ -
the magnitude for $D_L^0 - D_S^0$ - mass difference could be
considerable so that

\begin{eqnarray}
\rho_D \longrightarrow \rho_{exp} = 5\cdot 10^{-3}.
\end{eqnarray}

This is a characteristic feature of the horizontal model with symmetric
ansatz in which we could get the considerable magnitude for
$\bar D^0 - D^0 $- mixing and this is very intrigueing
for the future experiments.

After all, in this section we calculate the branching ratio for the
$\mu \rightarrow\,\,3e $-decay (process b) ). For this we have proposed
the  form of the charged
lepton mass matrix - the one that  was used for down- quarks.
Using  this ansatz for a charged lepton mixing, we find out that:
\begin{eqnarray}
Br(\mu\rightarrow3e)&&\simeq
12\frac{g_H^4}{g_W^4}\frac{M_W^4}{M_{H_0}^4}
\Biggl\{f^{abc}\Biggr(L \frac{\lambda^a}{2}L^+\Biggr)_{21}(g_H{\phi^b})
f^{a^{\prime}b^{\prime}c} \Biggl(L \frac{\lambda^{a^{\prime}}}{2}L^
+ \Biggr)_{11}(g_H{\phi^{b^{\prime}}})
\Biggr\}^{2}\nonumber\\
&&\simeq \frac{3}{2}\frac{g_H^4}{g_W^4}\frac{M_W^4}{M_{H_0}^4}
\frac{g_H^4}{\lambda_5^4}\frac{m_t^4}{M_{H_0}^4}
\Biggl[-\frac{1}{\sqrt{2}}\frac{m_c^2}{m_t^2}\sqrt{\frac{m_u}{m_c}}
+\frac{m_c}{m_t}\Biggl(\sqrt{\frac{m_u}{m_c}}
- \sqrt{\frac{2m_e}{m_{\mu}}} \Biggr) \frac{m_{\mu}}{m_{\tau}}
\Biggr]^2 ,\label{6.19}
\end{eqnarray}
where$(\phi=\tilde{\varphi^a},{\varphi^a})$ are the  nonvanishing
VEV's of the $h^a$- and $H^a$-Higgs fields, respectively.

Using the experimental information for the  rare processes a) and b)
and the formulas (\ref{6.8}), (\ref{6.9}), (\ref{6.12}), (\ref{6.14}) and
(\ref{6.19}) depending on the Yukawa coupling we can get   very small
values for breaking scale, $M_{H_0}$, of  the  $SU(3_H)$ gauge symmetry:

\begin{eqnarray}
&&M_{H_0}\,>\,\,  0.2 TeV \,-\, 0.5 TeV, \bigl(\,\lambda_5\simeq\,
O(1)\bigr) ,\nonumber\\
&&M_{H_0}\,>\, \,0.6 TeV\,-\, 1.5 TeV\,\bigl(\,\lambda_5\simeq\,O(0.1)
\bigr). \label{6.14}
\end{eqnarray}

\subsection{The  SUSY \SUTH-gauge model with "Democratic" ansatz for
quark and lepton families}

In the electroweak $SU(2) \times U(1)$- model it is impossible to define
 separately
the "mixings" in up-  and down-quark mass matrices
("absolute mixing").
The SM still provides
a
certain freedom in  choosing  the primary  mass matrices for quarks
in such a way as to get   large mixings both for up-,  $U$, and  down-quarks
$D$,
provided  $U D^+ = V_{CKM}$.
To get  information on this "absolute mixing", one  should  investigate
 rare processes in the framework of the horizontal gauge model.
Therefore, it would be very interesting to consider a
scheme, in which  these mixings are large, and  to study  possible constraints
on the horizontal symmetry breaking scale $M_{H_0}$.
So, we consider the "up"-  and "down"- fermion mass matrices
of the following ("democratic") form, which has been used for explaining
the special  role of the t- quark mass ($m_t\sim \Lambda_{EW} \gg m_c, m_u $)
 \cite{39',40'}:

\begin{eqnarray}
{\bf M_f^0} = \frac{1}{3}m_f
\left( \begin{array}{ccc}
 1 & 1 & 1 \\
 1 & 1 & 1 \\
 1 & 1 & 1 \\
\end{array} \right). \label{6.21}
\end{eqnarray}%

A BCS theory of quark generation could  explain  such
form of mass matrix.  The term BCS mechanism is used to refer to
Cooper pair formation through attractive forces between some
constituent  ur-fermions \cite{39',40'}.

To obtain the democratic form of quark  mass matrices in the model
with the family gauge symmetry , it is necessary to consider , besides
$H$- and $h$- superfields (see (\ref{2.3})), the additional superfields
$H_0(1_H, 2_L, - \frac{1}{2})$ and $h_0(1_H, 2_L, \frac{1}{2})$ and
to conserve the $S(3)_L \times S(3)_R$ vacuum symmetry.
The diagonalization of the democratic matrices  yields a mass gap,
i.e. the masses of
$t$- or $b$-  quarks are split far apart from all other degenerate masses
of $c-\,,\,u$- or $s-\,,\,d$-quarks. The complete mass matrices are
$M_f\,=M_f^0\,+\Delta{M}_f$. As a result of the diagonalization,
they yield the physical fermion mass matrices $M^D$
for "up" and "down" quarks: $M_f^D\,=\,V_f^+\,M_f\,V_f\,$, where
$\,V_f\,=V_{f0}\,V_{f1}\,$ and $\,V_{di}\,=\,D_i\,$, $V_{ui}\,=\,U_i\,$,
$i=(0\,,1)\,$. Here $V_{f0} $( $f=d\,or\,u$) are  unitary matrices
constructed from the eigenvectors of the $M_f^0$-matrix and  equal to :

\begin{eqnarray}
\qquad{\bf (U^0)^T} ={\bf (D^0)^T}=
\left( \begin{array}{ccc}
\frac{1}{\sqrt{2}} &-\frac{1}{\sqrt{6}}&\frac{1}{\sqrt{3}} \\
0&\frac{2}{\sqrt{6}} &\frac{1}{\sqrt{3}} \\
-\frac{1}{\sqrt{2}} & -\frac{1}{\sqrt{6}}&\frac{1}{\sqrt{3}} \\
\end{array} \right). \label{6.22}
\end{eqnarray}

In the first approximation, there is a conservation of the isotopic symmetry
of the mixing mechanism in the up" and "down" quark mass matrices.
The matrices $U_1$ and $D_1$  ($U_1 \,\neq\,D_1$) are  small corrections
to produce the correct form of the $V_{CKM}$-matrix. Using the $U_1$- and
$D_1$- correction matrices, we can construct the mass matrices $M_f$
differing from  $M_f^0$ by a small correction factor.

 Let us construct explicitely
the corresponding splitting of the horizontal gauge boson mass
matrix. In this case, each of the $8\times8$-dimensional mass matrices:
$\biggl[\Delta{M}_H^2\biggr]_d^{ab}$ and $\biggl[\Delta{M}_H^2\biggr]_u^{ab}$ ,
$a\,,\,b\,=1,2...8$, is broken into $3\times3$- and $5\times5$-dimensional
matrices. The additional contributions to the mass spectra of
the $H^{\mu}_{2,5,7}$- and $H^{\mu}_{1,4,6,3,8}$- horizontal  gauge bosons
take, correspondingly, the following forms:

\begin{eqnarray}
{\bf\biggl[\Delta{M}_H^2\biggr]_f^{{a,b}={2,5,7}}} =
\frac{g_H^2m_f^2}{12\lambda_f^2}
\left( \begin{array}{ccc}
 \,2 &\, 1 & -1 \\
 \,1 &\, 2 &\,1 \\
  -1 &\, 1 &\,2 \\
\end{array} \right), \label{6.23}
\end{eqnarray}

\begin{eqnarray}
{\bf\biggl[\Delta{M}_H^2\biggr]_f^{{a,b}={1,4,6,3,8}}} =
\frac{g_H^2m_f^2}{36\lambda_f^2}
\left( \begin{array}{ccccc}
 2 & -1 & -1& 0 & 2\sqrt{3} \\
 -1 & 2 & -1 & 3& -\sqrt{3} \\
-1 & -1 &  2 & -3& -\sqrt{3} \\
 0 & 3 & -3 & 6 & 0 \\
2\sqrt{3} & -\sqrt{3} &-\sqrt{3}& 0 & 6 \\
\end{array} \right). \label{6.24}
\end{eqnarray}
The diagonalization of these mass matrices can  easily be realized
by the orthogonal matrices $O^{(-)}$ and $O^{(+)}$:
so that $Z_a^{\mu(-)}=O_{ab}^{(-)}H_b^{\mu(-)}$  ( ${a,b}={2,5,7}$) and
 $Z_a^{\mu(+)}=O_{ab}^{(+)}H_b^{\mu(+)}$
(${a,b}={1,4,6,3,8}$).
In accordance with  expressions (\ref{6.23}) and (\ref{6.24}), let us
write down the forms of the $O_{ab}^{(-)}$- and $O_{ab}^{(+)}$-
diagonalizing matrices:

\begin{eqnarray}
\qquad{\bf O_{ab}^{(-)T}} =
\left( \begin{array}{ccc}
\frac{1}{\sqrt{6}} & \frac{1}{\sqrt{2}}&\frac{1}{\sqrt{3}} \\
\frac{2}{\sqrt{6}} & 0 &-\frac{1}{\sqrt{3}} \\
\frac{1}{\sqrt{6}} & -\frac{1}{\sqrt{2}}&\frac{1}{\sqrt{3}} \\
\end{array} \right),
\end{eqnarray}

\begin{eqnarray}
\qquad{\bf  O_{ab}^{(+)T}} =
\left( \begin{array}{ccccc}
\frac{1}{\sqrt{3}}&0&-\frac{2}{3}&0&\frac{\sqrt{2}}{3}\\
\frac{1}{\sqrt{3}}&-
\frac{1}{\sqrt{3}}&\frac{1}{3}&\frac{1}{\sqrt{6}}&-\frac{1}{3\sqrt{2}}\\
\frac{1}{\sqrt{3}}&
\frac{1}{\sqrt{3}}&\frac{1}{3}&-\frac{1}{\sqrt{6}}&-\frac{1}{3\sqrt{2}}\\
0&\frac{1}{\sqrt{3}}&0&\sqrt{\frac{2}{3}}&0\\
0&0&\frac{\sqrt{3}}{3}&0&\sqrt{\frac{2}{3}}\\
\end{array} \right). \label{6.28}
\end{eqnarray}

Note that the signs $(-)$ and $(+)$ indicate  the opposite CP-
transformation properties of the $ J^{\mu}_{H_{2,5,7}}$- and
 $ J^{\mu}_{H_{1,4,6,3,8}}$- gauge horizontal currents for  each value of the
index
$\mu$  $(\mu= 0 or 1,2,3)$. Until  there is
 no mixing between these currents,
there may be   no  CP- violation in the gauge sector of the horizontal
interactions.

As a result of this approach, we get a very simple splitting between
8- gauge $Z^{\mu}_a$- bosons:

\begin{eqnarray}
M_{Z_1}^{2}&=&M_{Z_4}^{2}=M_{Z_6}^{2}=M_{Z_7}^{2}=M_{H_0}^{2} \nonumber\\
M_{Z_3}^{2}&=&M_{Z_8}^{2}=M_{Z_2}^{2}=M_{Z_5}^{2}=M_{H_0}^{2}+
\frac{g_H^2}{4} {\sum}_{f}{\frac{m_f^2}{\lambda_f^2}}\,.\label{6.29}
\end{eqnarray}

The mass spectra of $ Z^{\mu}_{1,4,6,7}$- gauge bosons correspond to
the global $SU(2)_H \times U(1)_H $-  symmetry in the gauge sector,
which was  considered  in sections 3 and 4.

If we use the family mixing
(\ref{6.1}), the Lagrangian for the quark- gauge boson
interactions will be

\begin{eqnarray}
{\cal L}_{Q}&=&\frac{g_H}{2}\bar{Q} \gamma_{\mu} \Biggl(
\Biggl[-\,D_1 {\lambda^8}D_1^{+}\Biggr]\tilde{Z}_1^{\mu}\nonumber\\
&+&\Biggl[
\frac{\sqrt{3}}{2}D_1{\lambda^3}D_1^{+}
-\frac{1}{2}D_1{\lambda^1}D_1^{+}\Biggr]\tilde{Z}_4^{\mu}
-\Biggl[\frac{1}{2}D_1{\lambda^3}D_1^{+}
+\frac{\sqrt{3}}{2}D_1{\lambda^1}D_1^{+}\Biggr]\tilde{Z}_6^{\mu}\nonumber\\
&+&\Biggl[\frac{1}{2}D_1{\lambda^4}D_1^{+}
-\frac{\sqrt{3}}{2}D_1{\lambda^6}D_1^{+}\Biggr]\tilde{Z}_3^{\mu}
+\Biggl[\frac{1}{2}D_1{\lambda^6}D_1^{+}
+\frac{\sqrt{3}}{2}D_1{\lambda^4}D_1^{+}\Biggr]\tilde{Z}_8^{\mu}\nonumber\\
&+&\Biggl[D_1{\lambda^5}D_1^{+}\Biggr]\tilde{Z}_2^{\mu}
-\Biggl[D_1{\lambda^7}D_1^{+}\Biggr]\tilde{Z}_5^{\mu}
-\Biggl[D_1{\lambda^2}D_1^{+}\Biggr]\tilde{Z}_7^{\mu}
\Biggr) Q \,, \label{6.30}
\end{eqnarray}
where  one has ${Q}=Q_d=(d,s,b)$, or ${Q}=Q_u=(u,c,t)$.

At this expression we take a certain small quark mixing
( the "democracy" is broken) but we will not consider
  the additional gauge boson mixing and
just  assume that $M_{Z_a} \approx M_{\tilde{Z}_a}$,
a=1,2,3..8.
For our purpose, it will suffice to   take into account only a
new small correction to the quark family mixing -$D_{1}$ and  $U_{1}$
matrices, leading
to the correct form of the CKM- matrix for charged EW currents.
We may  consider the chain of symmetry breaking from the original
$ U(3)_L \times U(3)_R $ of the massless quarks and leptons:

\begin{equation}
{ U(3)_L \times U(3)_R} \stackrel{m_{t,b}\not=0}
 {\longrightarrow} {S(3)_L \times S(3)_R } \stackrel{m_{c,s}\not=0}
{\longrightarrow} {S(2)_L \times S(2)_R} \stackrel{m_{u,d}\not=0}
{\longrightarrow} {1} \label{6.31}
\end{equation}

At the first stage only the third family is massive and the other
two are massless. At the second stage only one generation remains
massless. At last, at the third stage the first generation also
gets mass. For instance, we may consider the up-  and down-  quark
mass matrices  like  (\ref{6.1}) and (\ref{6.2}):
\begin{eqnarray}
{\bf M_u^{dem}} \longrightarrow \frac{1}{3}m_t
\left( \begin{array}{ccc}
 1 & 1 & 1 \\
 1 & 1 & 1 \\
 1 & 1 & 1 \\
\end{array} \right) +  \frac{m_c}{6} u_1
\left( \begin{array}{ccc}
 1 & 1 & u_2 \\
 1 & 1 & u_2 \\
 u_2 & u_2 & -2 \\
\end{array} \right) +  \sqrt{\frac{m_um_c}{3}}u_3
\left( \begin{array}{ccc}
 1 & 0 & u_4 \\
 0 & -1 & -u_4 \\
 u_4 & - u_4 & 0 \\
\end{array} \right) \label{6.32}
\end{eqnarray}%
and
\begin{eqnarray}
{\bf M_d^{dem}} \longrightarrow \frac{1}{3}m_b
\left( \begin{array}{ccc}
 1 & 1 & 1 \\
 1 & 1 & 1 \\
 1 & 1 & 1 \\
\end{array} \right) +  \frac{m_s}{6} d_1
\left( \begin{array}{ccc}
 1 & 1 & d_2 \\
 1 & 1 & d_2 \\
 d_2 & d_2 & -2 \\
\end{array} \right) +  \sqrt{\frac{m_dm_s}{3}} d_3
\left( \begin{array}{ccc}
 1 & -\frac{i}{\sqrt{2}} & d_4 \\
 \frac{i}{\sqrt{2}} & -1 & -d_4 \\
 d_4^* & - d_4^* & 0 \\
\end{array} \right)
\end{eqnarray}%
where
\begin{eqnarray}
 u_1 &=& 2\sqrt{2} -1 ,\,
 u_2 = -\frac{4 +\sqrt{2}}{2\sqrt{2}-1 } ,\,
 u_3 = \sqrt{2} - 1 ,\,
u_4 = \frac{\sqrt{2} + 1}{ 2 - \sqrt{2}} ;\nonumber\\
d_1 &=& 1 + 2 \sqrt{2} , \,
d_2 = \frac{4- \sqrt{2}}{1 + 2 \sqrt{2}},\,
d_3 = \sqrt{2}, \,u_4 = \frac{1 + i \sqrt{2}}{2}.
\end{eqnarray}

Now,  for the further  estimates of the $SU(3)_H$- symmetry breaking
scale $M_{H_0}$ we  use the results from  section 4 and
 the following  useful relation

\begin{eqnarray}
\sum_{a}{(DT^aD^+)_{ik}(DT^aD^+)_{mn}}=
\frac{1}{2}\biggl( \delta_{in} \delta_{km} -
\frac{1}{3} \delta_{ik} \delta_{mn} \biggr) \,. \label{6.35}
\end{eqnarray}

Then,   the expression
for the $K_L^0$- $K_S^0$ meson mass difference (\ref{4.2}), derived
from formulas (\ref{6.31}), (\ref{6.28}) and (\ref{6.30}),  will change only
due to the new suppression coefficient $\sim {m_f^2}/{M_{H_0}^2}$,
 e.g.

\begin{eqnarray}
\Biggl\{\frac{\Delta m_K}{m_K}\Biggr\}_H= \frac{g_H^2}{M_{H_0}^2}
\Biggl\{\frac{g_H^2}{4 M_{H_0}^2} {\sum}_{f}{\frac{m_f^2}{\lambda_f^2}}\Biggr\}
Re(C_K^0) f^2_K R_K \,. \label{6.36}
\end{eqnarray}
where $C_K^0=\sum_{a}^{\prime}(D_1T^aD_1^+)_{21}(D_1T^aD_1^+)_{21}$,
and  the  index $(\prime)$ indicates that summation is only  over  the
diagrams with the exchanges of
$ Z^{\mu}_1, Z^{\mu}_4, Z^{\mu}_6, Z^{\mu}_7$-
gauge horizontal bosons. In  this approximation, the lower bound
on the local horizontal symmetry breaking scale may be smaller than in case (i)
from sbsection 3.4 (the $ SU(2)_H\times U(1)_H$ symmetry). For instance, if we
assume that $m_f=m_t$  and make our usual
assumption for the relation ${g_H}/{\lambda_f}$,  we can get:

\begin{equation}
M_{H_0}\, \approx\,
\sqrt{\frac{g_H}{2 \lambda_f}} \sqrt{\frac{m_t}{M_H}}\times M_H\,
> O(0.8) TeV\,. \label{6.37}
\end{equation}
In the last inequality we use the estimate for $ M_H \,>\, 8-9 TeV$
taken from Table 12.
Note, that the consequences for the $B_{d,s}^0 \longleftrightarrow
\bar{B}_{d,s}^0 $
oscillations  remain  as in Table 12- e.g.,  this value
for the gauge symmetry breaking scale  corresponds to the
present quantity of the $B_{d_1}^0 $-  $B_{d_2}^0 $- meson
mass difference.

Really, one could  expect   a very low horizontal local symmetry
breaking scale- $M_{H_0}$ in  pure quark  (or pure lepton) rare processes
due to the changes  of the  quantum numbers of generations
 therein:
$|\Delta H| \neq 0$.
{}From the experimental limits on the amplitudes of quark- lepton rare
processes,
where $\Delta H = 0 $, we  obtain considerably  larger  values for this scale
 (\ref{4.8}),(\ref{4.10}). What are the consequences of the studies  of
the lower bound on the horizontal local symmetry breaking scale ?
Here are some of them:

1.The most pronounced  processes promoting  the   discovery of a new hypothetic
interaction  are quark- lepton rare processes like
 $ K\,  \longrightarrow \pi\,+\, \mu\,+\,e\,$-,  or the
$ \mu/e $- conversion on nuclei. Within this class,
the decay $ K \, \longrightarrow \, \pi \,+\, {\nu}_i \,+\,{\nu}_j\, $
may also turn out to be very important.
In this case,  the "traditional" construction for the
quark- lepton families has been assumed:

$ C_1  =  ( Q_1  [u,d];  L_1  [{\nu}_e, e] ),
  C_2  =  ( Q_2  [c,s];  L_2  [{\nu}_{\mu},  \mu] ),
 C_3  =  ( Q_3  [t,b];  L_3  [{\nu}_{\tau}, \tau] ) $

and  here the mixings  between
$Q_i$-  and  $L_j$- families are not very large.
Then we should think that  the local horizontal symmetry breaking  scale
is very large, as follows from the limits (\ref{4.8}) - e.g., it may be
more than 60 Tev.
For this  large enough scale, the contributions to the pure quark
rare reactions (meson mixings), or pure lepton rare decays
 ($ \mu \rightarrow 3 e $  etc.), from these forces will be very small.
In particular, if the splitting  between the masses of 8-  gauge horizontal
 bosons is as in our previous examples:
 $ |(\Delta M_H^2)_a| \ll M_H^2 $.
 In the case of the large splitting  $ |(\Delta M_H^2)_a| \gg
min M_{H^a}^2 $,
 we may use , for practical purposes, the results of the
 ${U(1)}_{T_3} \times {U(1)}_{T_8} $-gauge group,
 where it was  established that the  lower limits on the  $M_H$-scale are:
 $ M_H > 170 - 195 TeV$ ( $\Delta m_K$ ,Table 12);
 $M_H > 60-100 TeV$  (from the $\mu/e$- conversion on nuclei);
and $ M_H > 35 TeV $ (from
  $ K\,  \longrightarrow
\pi\,+\, \mu\,+\,e\,$ if  $B\leq 10^{-10}$ (\ref{4.10})), or
  $M_H > 100 TeV$ ( if the limit $ 10^{-12}$ is reached in the nearest future
 in BNL- experiment).
 The lower bound on $M_H$ obtained  from the modern experimental limit
 on  a  pure lepton rare decay,  like
 $ \mu \rightarrow 3 e$, is compatible with  the bounds resulting from  the
$ K\,  \longrightarrow \pi\,+\, \mu\,+\,e\ $- experiment. So,
 we have $M_H > 28 TeV $ (\ref{4.5}).

2. An alternative scenario  we have to consider is connected
with searching for another  possible  mechanism  of
the   (q-l) - mixing
 to diminish the scale $M_H$  to the values approaching
the region of (1-10) Tev. For this purpose,  we may also use  an
indefinite correlation both between
 the $Q_i$- and $L_j$-
family mixings, and,
within $L_j$- lepton families, -  between charged lepton and neutrino mixings,
so far as  the experimental situation allows us to do so.
These  explicit differences in the origin of quark and
lepton mass spectra make one  also suppose that
 leptonic families might mix by  quite a different mechanism, different from
 the  above example of  quark mixing. One should also
remember that in the SM it is impossible, in principle, to
establish a correlation between the $Q_i$-quark and $L_j$- lepton
 mixings.
 Due to electroweak interactions, we can only get  information on the
correlations between up- and down- quark mixing.
 But now there is  still  a certain freedom in the choice of the mixing models
for charged leptons or neutrino, especially in the case of  very
small neutrino masses.

{}From the analysis of   pure quark
(lepton) rare processes in the gauge horizontal model  we may get
complete information  about  separate "absolute" mixings of
up- quarks (neutrinos) and down- quarks (charged leptons).
 And from quark- lepton rare reactions in the frames of
gauge horizontal interactions we may define  correlations
between $[d, s, b] $ (  [ $u, c, t $] )  quark-    and $[e, \mu, \tau] $
or $ [\nu_e, \nu_{\mu}, \nu_{\tau} ]$- lepton bases.
 In the above  examples (see
section 4),  the supposition of the absence of   correlation
between down- quark and charged lepton mixings resulted in rather high
limits for  the
$M_H$-scale, obtained from quark- lepton processes,    compared to those
from  pure "q", or pure "l" -processes. Now let us
 consider
the scheme when the magnitude of  correlation between quark-  and
charged lepton
mixings  is  large.

\begin{eqnarray}
{\cal L}_{l}&=&\frac{g_H}{2} \bar{\Psi}_l \gamma_{\mu}\Biggl(
\Biggl[\frac{\sqrt{3}}{2}{L_1 \lambda^3 L_1^+}
+\frac{1}{2}{L_1 \lambda^8 L_1^+}\Biggr]Z_1^{\mu}\nonumber\\
&+&\Biggl[\frac{\sqrt{3}}{4}{ L_1\lambda^3 L_1^+}
-\frac{3}{4}{L_1 \lambda^8 L_1^+}
-\frac{1}{2}{L_1 \lambda^6 L_1^+}\Biggr]Z_4^{\mu}\nonumber\\
&-&\Biggl[\frac{1}{4}{L_1 \lambda^3 L_1^+}
-\frac{\sqrt{3}}{4}{L_1 \lambda^8 L_1^+}
+\frac{\sqrt{3}}{2}{L_1 \lambda^6 L_1^+}\Biggr]Z_6^{\mu}\nonumber\\
&+&\Biggl[-\frac{\sqrt{3}}{2}{L_1 \lambda^1 L_1^+}
+\frac{1}{2}{ L_1 \lambda^4 L_1^+}\Biggr]Z_3^{\mu}
+\Biggl[\frac{1}{2}{L_1 \lambda^1 L_1^+}
+\frac{\sqrt{3}}{2}{L_1 \lambda^4 L_1^+}\Biggr]Z_8^{\mu}\nonumber\\
&-&\Biggl[{L_1 \lambda^5 L_1^+}\Biggr]Z_2^{\mu}
+\Biggl[{L_1 \lambda^2 L_1^+}\Biggr]Z_5^{\mu}
-\Biggl[{L_1 \lambda^7 L_1^+}\Biggr]Z_7^{\mu}
\Biggr) {\Psi}_l \,, \label{6.38}
\end{eqnarray}
where ${\Psi}_l =(e,\mu,\tau)$.

It is obvious  that the lower bound on  $M_{H_0}$
can also  be very small ($\sim O(1) TeV$ ) as far as pure charge lepton
rare processes are considered
(e.g.,  the modern high experimental limit for the
$ \mu^+ \rightarrow e^+\, + e^+\, + e^-\, $- decay). Again, in this approach
there appears a similar additional suppression  factor
 $\sim {m_f^2}/{M_H^2}$
for the pa[Brtial width of this process.
So, assuming that $m_f\approx m_t$ and
$ (L_1T^3L^T_1)_{21} (L_1T^3L^T_1)_{11}  =  L^3 \approx
\sqrt{{m_e}/{m_{\mu}}}$,
as was accepted in formula (\ref{4.5}) (Fritzsch ansatz for lepton mixing),
we have: $ M_{H_0}> \sqrt{{g_H}/(2\lambda_f)} \sqrt{m_f M_H}> O(1) TeV$
for $M_H > 7.5 TeV$.

Besides, in this model  we could obtain   lower limits for  the
horizontal symmetry breaking scale by analyzing quark- lepton rare processes
like  $ K\,  \longrightarrow \pi\,+\, \mu\,+\,e\,$-, or the
$ \mu/e $- conversion. For example, for process of the first type
the estimate (\ref{4.10}) is:

\begin{eqnarray}
M_{H_0} \approx \sqrt{2 |d_{13}|}\,M_H \,\,\,\,
( K^+\,  \longrightarrow \pi^+\,+\, \mu^+\,+\,e^-\,) \,, \label{6.39}\\
M_{H_0} \approx \sqrt{2 |d_{12}d_{23}|}\,M_H
(  K^+\,  \longrightarrow \pi^+\,+\, \mu^-\,+\,e^+\,) \,. 
\nonumber
\end{eqnarray}

If we take the values for $(D_1)_{ij}=d_{12},d_{23},d_{13}$ from all the three
ansatzes, we may verify  (\ref{4.10}) that the scale
$M_{H_0}$   will be rather low:
$M_{H_0}\,>\, O(1 TeV) $ for $ M_H \,>\, 35 TeV$.
The last estimate is conditioned by the model-dependent  value
of $d_{13}$, the latter being not very precisely defined from
the comparison  with the $V_{cb}$- matrix element.
Here we  assume that $ |d_{13}|\ll |V_{cb}|$, which  does not contradict
 the experiment. Note,  in this scheme we make a  very interesting
prediction for the heavy quark-,
 or heavy $\tau$- lepton rare decay. For example,
for this horizontal symmetry  breaking scale ($M_{H_0}\sim 1 TeV )$
the partial width for the $\tau$- lepton decay - $\tau \rightarrow
\mu + d + \bar{s} $
may be rather large $\sim {10}^{-5}$.

By analogy, we find that the rate for $\mu$ to $e$- conversion
(\ref{4.6}) is reduced
by the factor $2 |d_{12}d_{13}|$. Combining  this factor and the expression
(\ref{4.6}), we come to the following limit for $M_{H_0}$:

\begin{eqnarray}
M_{H_{0}} \, \approx \sqrt{2|d_{12}d_{13}|}\,M_H \,>\, O(1-2)\, TeV\,.
\label{6.41}
\end{eqnarray}

In this scheme the bounds (\ref{6.39}) and (\ref{6.41}) differ
by the factor of $\sqrt{|d_{12}}|$. Note, that our earlier limits
for $M_H$ (\ref{4.10})
 from for these two very important processes
also differ by the same  factor $\approx 2-2.5$.

Finally,  we have to consider the decay
$ K \, \longrightarrow \, \pi \,+\, \nu \,+\,\nu\, $.
Now the experimental lower limit for  the partial width
of this process is :

\begin{eqnarray}
Br\biggl( K \, \longrightarrow \, \pi \,+\, {\nu}_i \,+\,{\nu}_j\, \biggr)\,\,
<\,\, 5\times {10}^{-9}. BNL \,. \label{6.42}
\end{eqnarray}

According to  (\ref{6.42}),
the immediate  estimate of  this decay in our approach
gives us the following constraint:
$M_{H_0} > 10  TeV. $  To lower this limit,
it's necessary to elucidate the origin of the neutrino mass spectrum.
Clearly,  to achieve this one has to consider
an extension of the   fermion matter spectrum of new particles-
first of all,  new neutral neutrino- like
particles ($T_{SU(2)}=\frac{1}{2}, SU(3)_H$-singlet).
This could result in an efficient decrease in the value
of the coupling constant $g_H$ in the neutrino horizontal interaction.
The studies of the regularities in the observed mass spectrum of
ordinary particles might indicate possible existence of new partcles
like those occurring in GUT's (E(6)) or in other earlier extensiions
of the SM (Left- Right models with mirror particles
or with the fermion spectrum doubled).

\section{Discussion. The quark-lepton
nonuniversal character of the local family interactions}

The main point of our considerations in this paper
to study the next chain:

\begin{verse}
The nature of quark and lepton massess $\Longrightarrow$   \\
The quark - lepton family mixing $\Longrightarrow $       \\
A new family dynamics at 1 TeV energy.
\end{verse}

In chapter 2 we have considered the rank eight GUST with gauge symmetry
$G = SU(5)\times U(1)\times (SU(3)\times U(1))_H \subset SO(16)$
($G = SO(10)\times (SU(3)\times U(1))_H \subset SO(16)$.
GUSTs
originating from level-one KMA contain only low-dimensional
representations of the unification group. It is, therefore, difficult to
break the gauge symmetry. In order to solve this problem we have considered
as the observable gauge symmetry the diagonal subgroup $G^{sym}$ of rank 16
group $G\times G\subset SO(16)\times SO(16) (\subset E(8)\times E(8))$.
This construction allows us to break the GUST symmetry to the low energy gauge
group, which includes the $G_H$ family gauge symmetry.
The virtue of the $GUST^{symm}$ considered is that its low-energy spectrum
does not contain particles with exotic charges. This GUST which is based on
the maximal invariant SO(16) subgroup of E(8) leads to the $3_H$ light
and $ 1_H$ heavy families,
and the  GUST construction which is based
on the $E(6)\times SU(3)_H$ $\subset E(8)$ ($SU(3)^4 \subset E(6)$)
gauge symmetry and which predicts only three light quark--lepton famlies.

For examle the model with basis vectors $b_1,\ b_2,\ b_4$ of Model 1
can contain the $E(6)\times SU(3)-$sublattice, which corresponds to the
simle roots:
$$\varepsilon_i-\varepsilon_{i+1}\ ,\ i=1-4,\ 6-8\ ;
\quad \varepsilon_4+\varepsilon_5\ ;
\quad-1/2\,\sum_{i=1}^{8} \varepsilon_i\ ,\ i=1-8 $$
in the ortogonal basis $\varepsilon_i,\ i=1-8$.
Where states from RNS--sector correspond to the first seven roots
and state from $b_2-$sector (Mod.1)
$\psi_{1/2}^{\mu}|0>_L\otimes \prod_{i=1}^{8}\Psi_i^*|0>_R$
corresponds to the last root.
But we must select the $SU(3)-$factor from this model.
We can add the $b_2-$vector of the Model 2 or 3 for this.
However in this case we can not fix the $E(6)\times SU(3)-$lattice
since the lattice destroys or grows. And this way does not lead
to three generations. In this point it's seems more perspective
to work with real world-sheet fermions and with less rank groups.

A variant for unusual nonuniversal family gauge
interactions  of known quarks and leptons could be realized if we
introduce into each generation new heavy quarks (F = U, D ),
and leptons (L, N)
 singlets ( it is possible to consider doublets also)
 under   SU(2)$_L$- and triplets under
\SUTH-groups.( This fermion matter could exist in string spectra.
See the all three models with $SU(3_H) \times SU(3_H)$ family gauge
symmetry).
Let us consider for concreetness a case for charged leptons:
${\Psi}_l= (e,\mu, \tau)$ and ${\Psi}_L$=(E, M,$\cal T$).
Primarily, for simplicity we suggest that  the ordinary
fermions do not take  part in  \SUTH-%
interactions ("white" color states).
Then the interaction is described by  the
relevant part of the SUSY \SUTH- Lagrangian and gets the form

\begin{eqnarray}
{\cal L}_H=g_H \bar{\Psi}_{\cal L} \gamma_{\mu}
\frac{ {\bf {\Lambda^a}_{6x6}} } {2}
 {\Psi}_{\cal L} O_{ab} Z_{\mu}^b  \,\,,
\end{eqnarray}
where

\begin{displaymath}
{\bf {\Lambda^a}_{6x6}} =
\left( \begin{array}{cc}
 S(L\lambda^a  L^+)S& -S(L\lambda^a L^+)C \\
-C(L\lambda^a L^+)S &  C(L\lambda^a L^+)C
\end{array} \right) .
\end{displaymath}

Here we have ${\Psi}_{\cal L} = ({\Psi}_l; {\Psi}_L )$.
The matrix $O_{ab}$  (a,b=1,2,3...8) determines the  relationship
between the bare,  $H_{\mu}^{b}$, and physical, $Z_{\mu}^b$,  gauge
fields. The diagonal 3x3 matrices  S=diag ($s_e, s_{\mu}, s_{\tau}$)
and C=diag ($c_e,c_{\mu},c_{\tau}$)
define the nonuniversal
character for lepton horizontal interactions, as the elements
$s_i$  depend on the lepton masses, like
$s_i \sim \sqrt{m_i/M_0}$ (i=e,$\mu$,$\tau$).
The same suggestion we might accept for local quark family
interactions.

For the family mixing  we might  suggest
the next scheme. The primary 3x3 mass matrix
for the light ordinary fermions is equal to zero :
$M_{ff}^0 \approx  0$.
The 3x3- mass matrix for heavy fermions is approximately
proportional to unite matrix: $M_{FF}^0 \approx M^Y_0 \times 1 $,
where $M^Y_0 \approx 0.5 - 1.0 TeV$ and might be different
for $F_{up}$- , $F_{down}$-  quarks and for $F_L$- leptons.
We assume that the splitting between new heavy fermions in
each class $F_Y$ (Y=up, down, L) is small and,at least in quark sector,
might be described
by the t- quark mass. Such we think that at first
approximation it is possible to neglect by the heavy fermion mixing.
The mixing in the
light sector is completely explained by the coupling
light fermions with the heavy fermions. As a result in of this
coupling the 3x3- mass matrix $ M_{fF}^0 $ could be constructed
by "democratic" way which could lead to the well known mass family
hierarchy:

\begin{displaymath}
{\bf {\it M}_{6x6}^0 } =
\left( \begin{array}{cc}
    M_{ff}^0 & M_{fF}^0 \\
   M_{Ff}^0  &  M_{FF}^0
\end{array} \right) ,
\end{displaymath}
where
\begin{equation}
M_{fF}^0 \approx   M_{fF}^{dem}  \,+ \, M_{fF}^{corr}.
\end{equation}

The diagonalization of the $ M_{fF}^0$- mass matrix
$ X M_{fF}^0 X^+ $
(X = L-, D-, U- mixing matrices)
  gives us the eigen values, which are to define
the family mass
  hierarchy- $ n_1^Y << n_2^Y << n_3^Y $ and the following relations
between the masses of the known light fermions and a new heavy mass
scale:

\begin{equation}
n_i^Y = \sqrt{ m_i M^Y_0},\,\,\, i=1_g,2_g,3_g;\,
\,\,Y=up-, down- fermions.
\end{equation}

 In this "see-saw" mechanism
the common mass scale of new heavy fermions  might be not very far
from the $\sim 1 TeV $ energy, and as a consequence of the last
the mixing angles $ s_i$- might be not too very small.
There is  another interesting relation between the mass scales
$n_i^Y$ might be in this mechanism, at least for the quark case:

\begin{eqnarray}
 n_t/n_c = n_c/n_u =q_H^u, \,\,\,\,\,q_H^u \approx 14 -16, \nonumber\\
 n_b/n_s = n_s/n_d =q_H^d, \,\,\,\,\,q_H^d \approx 4 - 5.  \nonumber
 \end{eqnarray}

As an explicit example of non-universal $SU(3_H) \times SU(3_H)$ local
family interactions could be considered the model 3 (see section 2).

In this approach we get the suppression for quark-lepton flavour
changing processes , like  $\mu$ to e - conversion,
$K \rightarrow \pi + \mu + e $- or $ K \rightarrow \pi + \nu +\nu$-
decays. And as result we can hope to get the very low bound
for the horizontal gauge boson masses ( in some TeV range).

\subsection*{Acknowledgements}
One of us (G.V) would like to thank INFN for financial support and the staff of
the Physics Department in Padova,  especially, Professor C. Voci
 for warm hospitality during his stay in Padova. Also,
he is pleased to the University of Padova and to the Physical Department in
Trieste  and to Professor G. Costa and Professor N. Paver
for hospitality and the financial support. It is a great pleasure
for him to thank  Professor G. Costa, Professor L. Fellin,
Professor J.Gasser,  Professor G. Harrigel, Professor
H. Leutwyller,
Professor P.Minkowski, Professor D. Nanopoulos, Professor N. Paver,
Professor V. Petrov, Professor P. Ramond, Professor P. Sorba,
Professor M. Tonin and  Professor C. Voci for useful
discussions and for the help and support. Many interesting discussions with
colleagues from Theory  Department in Padova INFN Sezione, especially, to
Professor S. Sartory,  during this
work are also acknowledged. Finally, he would like to express his gratitude
to Professor P. Drigo and their colleagues of the Medical
School of the University of Padova.

The research described in this publication was made possible in part
by Grants No RMP000 and RMP300
from the International Science Foundation.

\end{document}